\documentclass[12pt, twoside, english]{article}
\usepackage[utf8]{inputenc}            
\usepackage[T1]{fontenc} 
\usepackage{mathtools}			
\usepackage{amssymb}
\usepackage{a4wide}
\usepackage{graphicx}
\usepackage{fancyhdr}
\usepackage{amsmath, amssymb}
\usepackage{subfigure}
\usepackage{setspace}
\usepackage{verbatim} 
\usepackage[english]{babel}
\usepackage[novbox]{pdfsync} 
\usepackage{caption}
\usepackage{float}
\usepackage{booktabs}

\usepackage{xfrac}

\usepackage{units}

\usepackage{multirow}
\usepackage{tabularx}
\usepackage{array}

\makeatletter
\usepackage{hyperref}

\usepackage{xcolor}

\hypersetup{
	colorlinks = false,
	linktocpage = true,
	allbordercolors = {0.8 0.8 0.8}
}

\addto\captionsenglish{}
\addto\captionsenglish{}
\addto\extrasenglish{}
\addto\extrasenglish{}
\addto\extrasenglish{}
\addto\extrasenglish{}
\addto\extrasenglish{}
\addto\extrasenglish{}
\addto\extrasenglish{}

\newcommand{\lyxaddress}[1]{
\par {\raggedright #1
\vspace{1.4em}
\noindent\par}
}

\numberwithin{equation}{section}

\newcolumntype{L}[1]{>{\raggedright\arraybackslash}m{#1}}
\newcolumntype{M}[1]{>{\centering\arraybackslash}m{#1}}
\newcolumntype{R}[1]{>{\raggedleft\arraybackslash}m{#1}}

\newcommand{\vek}[1]{\mathchoice{\displaystyle\boldsymbol#1}
{\textstyle\boldsymbol#1}{\scriptstyle\boldsymbol#1}
{\scriptscriptstyle\boldsymbol#1}}
\newcommand{\mat}[1]{\mathchoice{\displaystyle\mathbf#1}
{\textstyle\mathbf#1}{\scriptstyle\mathbf#1}
{\scriptscriptstyle\mathbf#1}}

\renewcommand{\d}{ \ensuremath{\mathrm{d} }}

\newcommand{\mult}{\,}

\newcommand{\ti}{ \ensuremath{\mathrm{t} }}

\newcommand{\GradS}{ \ensuremath{\mathrm{Grad}_{\Gamma} }}
\newcommand{\gradS}{ \ensuremath{\mathrm{grad}_{\Gamma} }}
\newcommand{\GradF}{ \ensuremath{\mathrm{Grad}_{\mat{\Upsilon}} }}
\newcommand{\gradF}{ \ensuremath{\mathrm{grad}_{\mat{\Upsilon}} }}
\newcommand{\GradFf}{ \ensuremath{\mathrm{Grad}_{\mat{\Upsilon}_i} }}
\newcommand{\gradFf}{ \ensuremath{\mathrm{grad}_{\mat{\Upsilon}_i} }}
\newcommand{\gradRef}{ \ensuremath{\mathrm{grad}_{\vek{r}} }} 
\newcommand{\divS}{ \ensuremath{\mathrm{div}_{\Gamma} }}
\newcommand{\DivS}{ \ensuremath{\mathrm{Div}_{\Gamma} }}
\newcommand{\divF}{ \ensuremath{\mathrm{div}_{\mat{\Upsilon}} }}
\newcommand{\DivF}{ \ensuremath{\mathrm{Div}_{\mat{\Upsilon}} }}
\newcommand{\divFf}{ \ensuremath{\mathrm{div}_{\mat{\Upsilon}_i} }}
\newcommand{\DivFf}{ \ensuremath{\mathrm{Div}_{\mat{\Upsilon}_i} }}

\makeatother

\usepackage[backend=biber,style=numeric,natbib=true,hyperref=true,doi=true,sorting=none]{biblatex}
\DeclareFieldFormat[article, inbook]{title}{#1} 
\begin{document}

\title{Hyperelastic Membranes with Implicitly Defined, Continuously Embedded Fibers}

\author{Michael Wolfgang Kaiser, Thomas-Peter Fries}
\maketitle

\lyxaddress{\begin{center}
Institute of Structural Analysis\\
Graz University of Technology\\
Lessingstr. 25/II, 8010 Graz, Austria\\
\texttt{www.ifb.tugraz.at}\\
\texttt{michael.kaiser@tugraz.at}\\
\texttt{fries@tugraz.at}
\end{center}}
\vspace{-2cm}
\begin{abstract}
A novel mechanical model and corresponding finite element method for anisotropic, hyperelastic, curved membranes are proposed. Hyperelastic fibers are embedded into the otherwise isotropic membrane, being relevant, for example, in reduced models for biological tissues and textiles. The geometrically nonlinear mechanics is formulated based on first principles of continuum mechanics (finite strain theory). The employed differential operators are formulated in a coordinate-free manner, through a framework known as tangential differential calculus. This enables a (semi-)implicit description of the fiber geometry through the intersection of level sets of some scalar function with the explicitly defined membrane surface. The mechanical model of the implicit fibers is then coupled to the mechanics of the membrane. For the numerical analysis, finite elements are applied such that the resulting scheme is a hybrid between classical Surface FEM and fictitious domain methods. For smooth physical fields, higher-order convergence rates are obtained and confirm the success of the numerical method.\\
\\
\underline{\emph{Keywords}}: Hyperelasticity, Membranes, Fibers, Manifolds, Anisotropy, Composite Materials, Finite Element Method
\end{abstract}
\tableofcontents{}\newpage{}

\section{Introduction}

Composite materials are widely used, e.g., in engineering as lightweight materials and structures, such as textiles, fabrics, and woven composites, and in biomedical applications, see, e.g., \cite{Boisse_1997a,Boisse_2008a,Schulte_2020a,Steigmann_2015a}. They also occur in great variety in nature, for example in the biological tissues of plants, animals, and humans; see, for example, \cite{Holzapfel_2001a,Gasser_2005a,Holzapfel_2025a} for models of (human) artery tissue. The following textbooks and references therein provide an overview on the topic and emphasize the crucial role composites have in the modern world, c.f., \cite{Clyne_2019a,Chawla_2019a,Avk_2013a,Khan_2020a,Hasan_2017a,Krishnasamy_2021a,Paul_2019a}.\\
\\
Different scales of the embedding matrix/bulk materials and sub-structures result in various mechanical models. One crucial aspect to distinguish such models is whether the sub-structures are considered \emph{discretely} or \emph{continuously}, see, e.g., \cite{Madeo_2016a}. In this work, we introduce a novel mechanical model for an anisotropic hyperelastic membrane with continuously embedded fibres with both constituents exhibiting individual hyperelastic behaviour. Furthermore, a modern finite element method is presented to solve the resulting anisotropic membranes numerically. Potential applications include woven fabrics,  paper materials, and tissue modelling and engineering.\\
\\
All fibres (sub-structures) continuously embedded in the membrane are treated \emph{simultaneously} in the mechanical model. This is an extension of previous works by the authors: In \cite{Fries_2023a}, the \emph{Bulk Trace FEM} is introduced for a mechanical model in which curved two-dimensional membranes are solved simultaneously over a three-dimensional bulk domain. Analogously, curved one-dimensional ropes are solved simultaneously over a plane two-dimensional bulk domain. Then, the governing PDEs are formulated for each membrane or rope (manifolds) and later combined in the weak form based on the \emph{coarea formula} \cite{Fries_2023a,Burger_2009a,Dziuk_2013a}. Similar approaches are used in \cite{Kaiser_2024b,Kaiser_2024a,Kaiser_2024c} for curved Timoshenko beams and Reissner--Mindlin shells. In \cite{Fries_2018a,Kaiser_2025a}, this approach was applied to transport problems and incompressible Navier--Stokes flows on curved manifolds. The geometries of the manifolds are defined implicitly, hence, the applied Bulk Trace FEM can be interpreted as a hybrid of the classical Surface FEM and fictitious domain methods, e.g., the Trace FEM \cite{Dziuk_2013a,Dziuk_1988a,Olshanskii_2017a}. Note that the main emphasis in the first Bulk Trace FEM applications, c.f., \cite{Fries_2023a,Kaiser_2024b,Kaiser_2024a,Kaiser_2024c,Kaiser_2025a,Kaiser_2025a,Fries_2023c,Kaiser_2023b}, was to solve the considered PDEs simultaneously on all the considered manifolds (level sets of a level-set function) which are defined over the bulk domain. There, the bulk domain had no mechanical meaning but was only used for the geometry description. A concept where this method is applied to mechanical domains with \emph{continuously embedded sub-structures} (which are described by level sets) was proposed by the authors in \cite{Fries_2023a,Fries_2024c}. Therein, the sub-structures are one-dimensional fibers embedded in two- or three-dimensional bulk domains or two-dimensional membranes embedded in three-dimensional bulk domains. Herein, the one-dimensional fibers are embedded in a curved two-dimensional membrane which itself is defined explicitly in some three-dimensional embedding space, c.f., \cite{Fries_2020a}. The fibers are thus given by the intersection of the membrane with implicitly defined level sets of some scalar function living in the same three-dimensional space in which the membrane is immersed. Further details are described in Sec.~\ref{sec:DiffOp}.\\
\\
To place the proposed model in a broader context of existing models for composite materials with an anisotropic behaviour, we give a short and concise review on this topic. However, due to the vast literature body, a comprehensive survey is beyond the scope of this work. One may classify existing models based on different criteria: (i) The geometric definition and dimensionality of the bulk materials and the embedded fibers (2D/3D, discrete/continuous), (ii) the mechanical modelling of the fibers and the embedding matrix including their coupling, and (iii) different fields of applications.\\
\\
An important application for anisotropic hyperelastic material models are tissues of blood vessels in cardiovascular biomechanics, e.g., the human aorta. Two widely used models are the HGO model \cite{Holzapfel_2000b} and the GOH model \cite{Gasser_2005a}. In these continuum models, collagen fibers are embedded in a three-dimensional matrix material and represented by the (generalized) structure tensor approach. The structure tensor is computed using a single preferred direction of the fibers \cite{Holzapfel_2000a}. Fiber dispersion is a crucial aspect for artery walls and is considered in \cite{Gasser_2005a}. Furthermore, collagen fibers cannot support compression, hence, fibers under compression should be excluded in the model by considering a tension-compression switch, see \cite{Holzapfel_2015a}. For further details of these continuum collagen fiber-reinforced tissue models and their applications in cardiovascular biomechanics, see \cite{Holzapfel_2001a,Holzapfel_2025a,Gasser_2021a}.\\
\\
In this work, the matrix material is a curved two-dimensional, geometrically nonlinear membrane. Therefore, we review next some literature in which fibers are embedded in thin curved matrix material structures, i.e., shells and membranes. In \cite{Klarbring_2017a}, fibers are embedded in a \emph{linear} membrane and an optimization scheme for the fiber orientation design is introduced. The fibers are considered in the material model, similar to the structure tensor approach. Geometrically non-linear shells with embedded beams are introduced in \cite{Duong_2022a,Duong_2022b}. The beams are continuously distributed in the shell surface and considered in the strain energy function via an additive term to the strain energy of the matrix material (shell). A projection of a three-dimensional material model onto two-dimensional membranes where fibers are modelled using the structure tensor approach is shown in \cite{Roohbakhshan_2016a}. Biological membranes with an anisotropic material behaviour due to fiber reinforcement are considered in \cite{Tepole_2015a,Roohbakhshan_2017a}. Note that \emph{biological membranes} or \emph{biomembranes} are thin biological structures, e.g., in cell biomechanics, that can also be subjected to bending. Therefore, from a mechanical perspective, such structures are rather thin \emph{shells}. In this work, we use membranes in the mechanical sense, i.e., bending effects are not considered.\\
\\
Fibers that resist not only tension but also bending and twisting are modeled as beams and considered for continuous fiber reinforcements, e.g., in \cite{Steigmann_2012a,Duong_2022a,Duong_2022b}. For discrete reinforcements, different \emph{coupling methods} are available and have to be addressed in the mechanical modelling and corresponding numerical methods. In \cite{Burman_2018e}, discrete Euler--Bernoulli beams are embedded in Kirchhoff plates. The plate is numerically simulated using a continuous/discontinuous finite element method and the beams are considerd by a Cut FEM approach. Cut FEM is also used for fiber-reinforced composites in \cite{Kerfriden_2020a}. In \cite{Elwi_1989a}, reinforcement elements (reinforcing bars and prestressing tendons) for reinforced concrete structures are considered by inverse mappings in a finite element simulation. Discrete beams are embedded into a shell based on Cosserat continua in \cite{Sky_2024a}. Couplings of discrete beams with three-dimensional volumes using mortar-type methods are introduced in \cite{Steinbrecher_2020a,Steinbrecher_2021a} and of beams with surfaces or boundary surfaces of a three-dimensional body in \cite{Steinbrecher_2025a}. In \cite{Hansbo_2022a}, Nitsches's method is applied for coupling discrete Euler-Bernoulli beams with elastic bulk domains. For a coupling scheme using IGA in numerical analysis for spatially curved Bernoulli beams to surfaces, see \cite{Bauer_2017a}. In \cite{Pechstein_2025a}, direct coupling of a shell to a three-dimensional solid is used for embedding stiff thin-walled structures into soft matrix materials. In this work, the fibers  are added to the matrix material (bulk domain) as homogenized, implicitly defined, continuously embedded sub-structures and cross-coupled through the bulk behaviour. With this approach, sketched by the authors in \cite{Fries_2023a} for fibers embedded in \emph{planar} two-dimensional bulk domains, discontinuities in the physical fields which result from the embedding of \emph{discrete} substructures do not occur, largely simplying the numerical treatment as shown herein.\\
\\
The main contributions of the present work are summarized as:
\begin{itemize}
	\item A novel mechanical model for anisotropic, hyperelastic, and geometrically nonlinear membranes is presented. The anisotropy is considered by embedded fiber families which are geometrically defined implicitly by the intersection of level sets of some level-set function and the explicitly defined membrane surface.
	\item A novel numerical method, based on the Bulk Trace FEM \cite{Fries_2020a}, is introduced for the numerical analysis of the proposed mechanical model.
	\item Numerical examples validate the success of the proposed novelties. For smooth physical fields, higher-order convergence is obtained as expected. The proposed examples may serve as benchmarks for future research.
\end{itemize}
The remainder of this paper is as follows: In Sec.~\ref{sec:DiffOp}, the geometry definitions of the membrane and the fibers are introduced. Furthermore, an overview on later used differential operators is given. The mechanical models for the hyperelastic membrane and the hyperelastic fibers are discussed in Sec.~\ref{sec:MechModelMemb} and Sec.~\ref{sec:MechModelFib}, respectively. This is followed by the definition of the strong and weak forms of the governing equations for the coupled problem in Sec.~\ref{sec:Coupling} which are then discretized in Sec.~\ref{sec:DiscrFEM}. Numerical examples including higher-order convergence studies are presented in Sec.~\ref{sec:NumTCs}. The paper ends with some conclusions and an outlook to further research in Sec.~\ref{sec:CaO}.

\section{Implicitly defined fibers embedded in a membrane} \label{sec:DiffOp}
The \emph{implicitly} defined \emph{fibers} are embedded in a two-dimensional, curved \emph{membrane} which is defined \emph{explicitly}. For the corresponding mechanical models of the fibers and the membrane, the tangential differential calculus (TDC) is used to formulate differential operators. For brevity, we only summarize the general definitions of these herein and refer to previous publications for the fundamentals, c.f., \cite{Delfour_2011a,Fries_2020a,Fries_2023a,Kaiser_2024a}. Furthermore, we focus on a detailed description of the special case of a one-dimensional fiber family $\mat{\Upsilon}_i$, implicitly defined by the level-set function $\phi_i$, which is continuously embedded into a two-dimensional surface $\Gamma$ which itself lives in the three-dimensional embedding space $\mathbb{R}^3$. For now, let us assume that there is \emph{one} fiber family $\mat{\Upsilon}$ such that we can skip the subscript $i$ for brevity. The introduced quantities can later easily be extended to the case of several embedded fiber families.

\subsection{Geometry definition of membrane and fibers} \label{subsec:GeomDiffMF}

\begin{figure}
	\centering
	\includegraphics[width=1.0\textwidth]{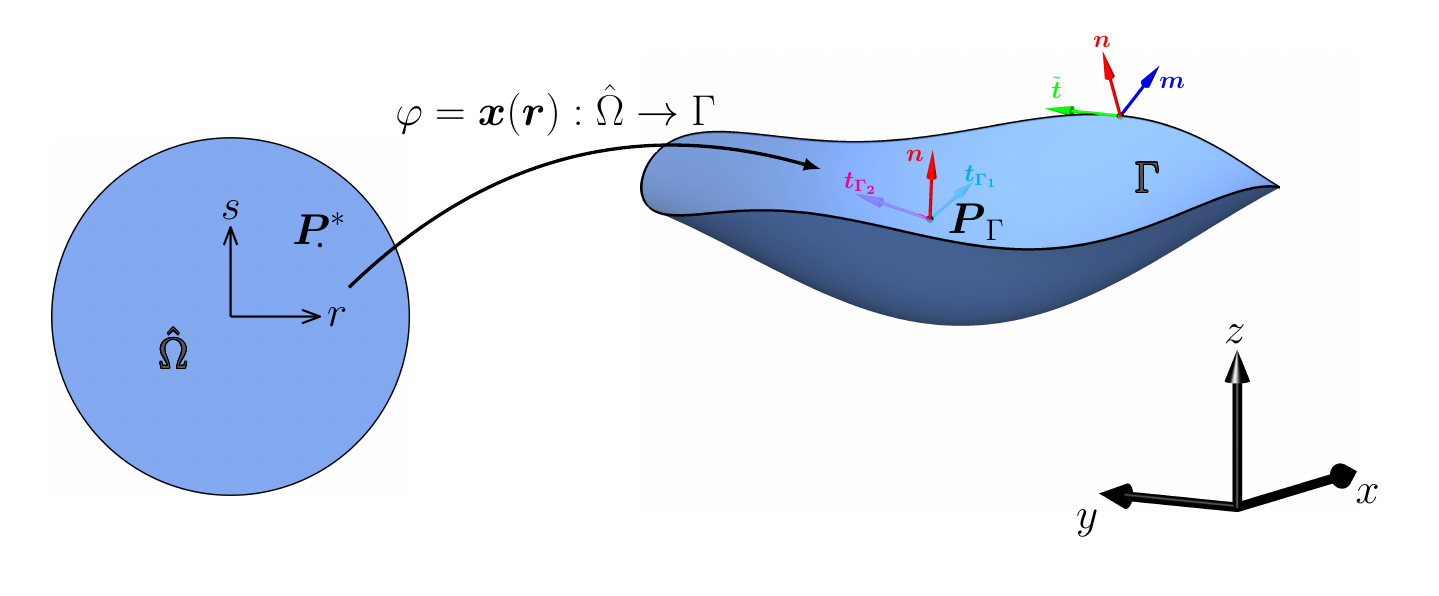}
	\captionsetup{margin=4cc}
	\caption{Explicit geometry description by a map $\varphi$ from the reference domain $\hat{\Omega}$ into the physical domain of the membrane surface $\Gamma$. At point $P_\Gamma$, the vectors $\vek{t}_{\Gamma_1}$, $\vek{t}_{\Gamma_2}$, and $\vek{n}$ are visualized. Furthermore, the local triad $\{\vek{m},\tilde{\vek{t}},\vek{n}\}$ is shown on a point at the boundary $\partial\Gamma$.}
	\label{fig:Mapping}
\end{figure}

\begin{figure}
	\subfigure[$\Gamma$ and $\psi_c$ of $\phi$]{\includegraphics[width=0.5\textwidth]{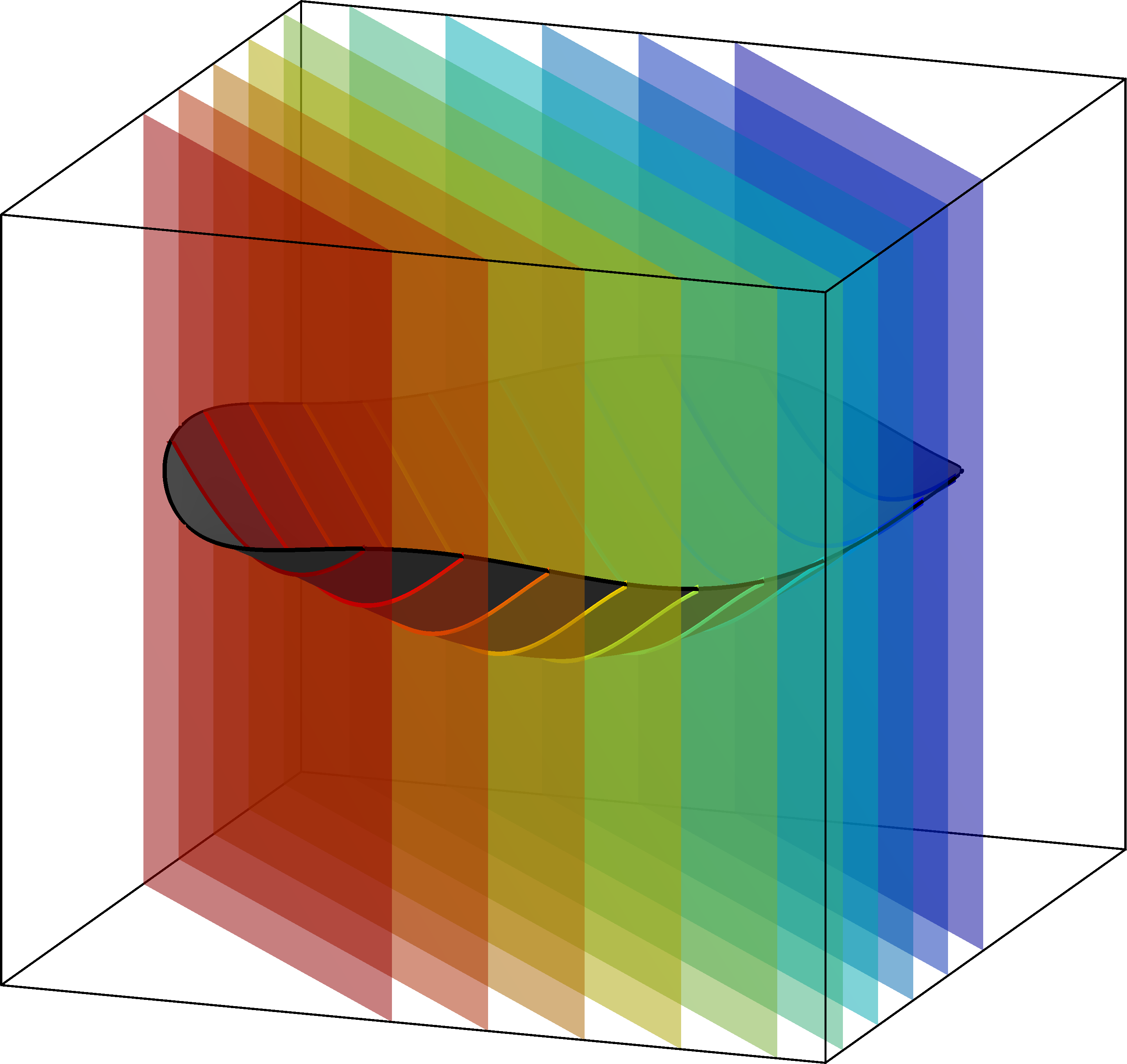}}\qquad
	\subfigure[$\Gamma$ and $\Upsilon_c$]{\includegraphics[width=0.5\textwidth]{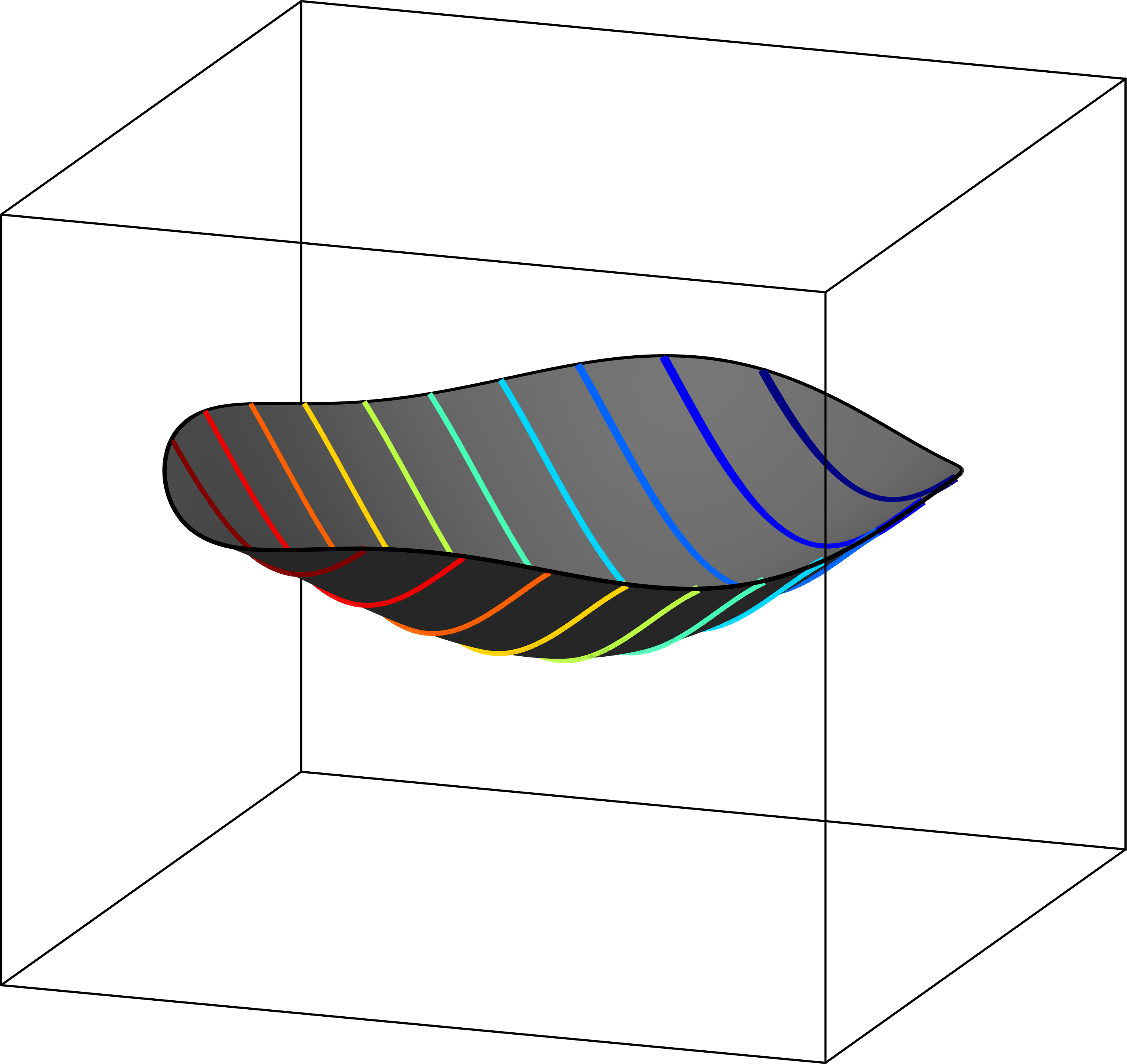}}
	\subfigure[Definition of $\vek{t}$]{\includegraphics[width=0.5\textwidth]{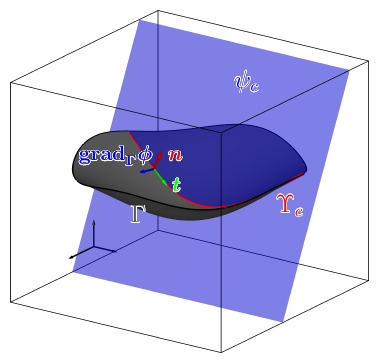}}\qquad
	\subfigure[Definition of $\vek{t}$ (zoomed view)]{\includegraphics[width=0.5\textwidth]{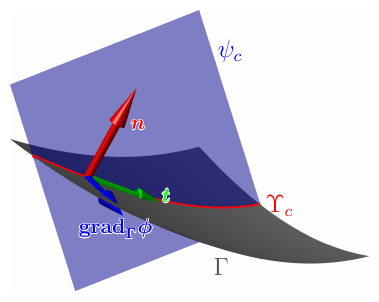}}
	\caption{Geometry definition of the fibers; (a) shows the membrane surface $\Gamma$ in gray and some selected level sets $\psi_c$ of $\phi$ in different colours, (b) shows some fibers $\Upsilon_c$ which result from the intersection of the membrane surface and the level sets in the color code of the level sets in (a); (c) and (d) visualize the tangent vector $\vek{t}$ defined in Eqs.~(\ref{eq:tangVecFib}) and (\ref{eq:normedT}).}
	\label{fig:FiberDef}
\end{figure}

The two-dimensional membrane $\Gamma$ is defined explicitly, e.g., by some map $\varphi: \hat{\Omega}\subset\mathbb{R}^2\rightarrow\Gamma\subset\mathbb{R}^3$, see Fig.~\ref{fig:Mapping}. On the surface $\Gamma$, the \emph{normal vector} $\vek{n}$ and tangential vectors $\vek{t}_{\Gamma_k}$ for $k=\{1,2\}$ are defined as
\begin{equation}
	\mat{j} = \frac{\partial\varphi}{\partial \vek{r}} = \begin{bmatrix}
		\vek{t}_{\Gamma_1}^{\star} & \vek{t}_{\Gamma_2}^{\star}
	\end{bmatrix} \in \mathbb{R}^{3\times 2}; \quad
	\vek{n} = \frac{\vek{t}_{\Gamma_1}^{\star} \times \vek{t}_{\Gamma_2}^{\star}}{\lVert \vek{t}_{\Gamma_1}^{\star} \times \vek{t}_{\Gamma_2}^{\star} \rVert}; \quad \vek{t}_{\Gamma_k} = \frac{\vek{t}_{\Gamma_k}^{\star}}{\lVert \vek{t}_{\Gamma_k}^{\star} \rVert}, \quad k=\{1,2\}. \label{eq:Jacobi_NV}
\end{equation}
Along the boundary $\partial\Gamma$, the normal vector, a tangential vector $\tilde{\vek{t}}$, pointing in direction of $\partial\Gamma$, and a conormal vector defined as 
\begin{equation}
	\vek{m} = \tilde{\vek{t}} \times \vek{n},
\end{equation}
form a local triad, see again Fig.~\ref{fig:Mapping}. It is later used to prescribe boundary conditions. The conormal vector $\vek{m}$ was denoted as $\vek{n}_{\partial\Gamma}$ in \cite{Fries_2020a} and $\vek{q}$ in \cite{Fries_2023a,Kaiser_2024a}.\\
\\
The (tangential) projector\footnote{To avoid confusion later, matrices and tensors are denoted herein with lowercase letters instead of uppercase letters because we will distinguish two mechanical configurations in the remainder of this paper. Then, the undeformed and deformed configurations are notationally distinguished by using uppercase and lowercase letters, respectively; see Sec.~\ref{sec:MechModelMemb}. The identity matrix $\mat{I} \in \mathbb{R}^{(3\times3)}$ is the same in both configurations.} is used to project quantities onto the tangent space of the surface $\Gamma$ and follows as
\begin{equation}
	\mat{p}_{\Gamma} = \mat{I} - \vek{n} \otimes \vek{n},
\end{equation}
with properties: $\mat{p}_{\Gamma}\mult\mat{p}_{\Gamma}=\mat{p}_{\Gamma}$, $\mat{p}_{\Gamma}^{\mathrm{T}} = \mat{p}_{\Gamma}$, $\mat{p}_{\Gamma}\mult\vek{n} = \vek{0}$, and $\vek{v}_\ti = \mat{p}_{\Gamma}\mult\vek{v}$, where $\vek{v}_\ti$ is a vector tangential to the surface (the vector is \emph{in-plane}) and $\vek{v} \in \mathbb{R}^3$ is an arbitrary vector in the three-dimensional space.\\
\\
The embedded fiber family is defined by the \emph{level sets} $\Psi_c$ of the scalar-valued level-set function $\phi(\vek{x})$ and its \emph{intersection} with the surface $\Gamma$. An individual level set is defined as 
\begin{equation}
	\Psi_c = \{\vek{x} \in \mathbb{R}^3: \phi(\vek{x}) = c \in \mathbb{R}\}, \quad \phi_{\mathrm{min}} < c < \phi_{\mathrm{max}}.
\end{equation}
The intersection of the level set $\Psi_c$ with the surface $\Gamma$ results in \emph{one} bounded fiber $\Upsilon_c$ of the fiber family $\mat{\Upsilon}$, i.e., 
\begin{equation}
	\Upsilon_c = \Psi_c \cap \Gamma \rightarrow \Upsilon_c \in \mat{\Upsilon}.
\end{equation}
A tangent vector field $\vek{t}^{\star}$ is related to the fibers 
\begin{equation}
	\vek{t}^{\star} = \vek{n} \times \gradS \, \phi(\vek{x}). \label{eq:tangVecFib}
\end{equation}
Note that the \emph{surface gradient} $\gradS\,\phi$ as defined in Tab.~\ref{tab:DifferentialOperators} is applied in the definition of the tangent vector and that $\vek{t}^{\star}$ is not a unit vector, hence, the unit tangent vector $\vek{t}$ follows as
\begin{equation}
	\vek{t} = \frac{\vek{t}^{\star}}{\lVert \vek{t}^{\star} \rVert}. \label{eq:normedT}
\end{equation}
Fig.~\ref{fig:FiberDef} visualizes the definition of the fibers and their associated tangent vectors. With that definition, an individual fiber can easily be evaluated throughout the modelling and analysis process. However, for the mechanical modelling, the fibers are considered as homogeneous and continuously embedded and, for the later introduced numerical method, the fibers are not discretized individually. Due to the interaction of an implicitly defined level-set function and the explicitly defined surface (of the membrane), the geometry definition of the fibers could also be called as \emph{semi-implicit}.

With the normalized tangent vector of the fiber, the projector, which projects quantities onto the tangent space of the fibers, is defined as
\begin{equation}
	\mat{p}_{\mat{\Upsilon}} = \vek{t} \otimes \vek{t}.
\end{equation}
Note that it has properties analogously to the projector $\mat{p}_{\Gamma}$ given above. The projectors $\mat{p}_{\Gamma}$ and $\mat{p}_{\mat{\Upsilon}}$ are important to define quantities in the TDC w.r.t.~the membrane surface and the fibers, respectively.

\subsection{Definition of differential operators} \label{subsec:DivOpDev}
Next, we define differential operators based on the framework of tangential differential calculus (TDC)
\cite{Delfour_2011a,Fries_2020a,Fries_2023a,Kaiser_2024a}. The surface gradient of a scalar-valued function $f(\vek{r})$ w.r.t.~the explicitly defined membrane surface $\Gamma$ is known as
\begin{equation}
	\gradS \, f(\vek{x}(\vek{r})) = \mat{j}(\vek{r}) \mult \mat{g}^{-1}(\vek{r}) \mult \gradRef \, f(\vek{r}).
\end{equation}
Therein, the metric tensor of the surface is defined as $\mat{g}(\vek{r}) = \mat{j}^{\mathrm{T}}(\vek{r}) \mult \mat{j}(\vek{r})$ and $\gradRef\,\bullet$ is the classical gradient w.r.t.~the reference coordinates. For brevity, we do not provide further details herein; see \cite{Jankuhn_2018a,Schoellhammer_2019a} for a detailed discussion of the surface gradient. In Tab.~\ref{tab:DifferentialOperators}, further differential operators for the curved membrane surface $\Gamma$ and the embedded fibers $\mat{\Upsilon}$ therein are given in a brief and concise manner. For details on their mathematical definition, the reader is referred to previous works, e.g., \cite{Fries_2018a,Jankuhn_2018a,Fries_2020a,Fries_2023a,Kaiser_2024a}. Note that the fibers are one-dimensional structures which are embedded in the two-dimensional surface $\Gamma$, hence, the fibers $\mat{\Upsilon}$ are not (only) implicitly defined one-dimensional curves which live in $\mathbb{R}^3$ as in, e.g., \cite{Fries_2020a,Fries_2023a}, but additionally constrained to live in the curved surface which itself is embedded in $\mathbb{R}^3$, see above. This may be seen as a special case of the implicit geometry definition and the according tangential differential operators formulated using TDC in, e.g., \cite{Fries_2020a, Fries_2023a,Kaiser_2023b, Schoellhammer_2021a}.

\begin{table}[H]
	\centering
	\begin{tabularx}{\textwidth}{L{0.23\textwidth}XX}
		\toprule
		&
		Curved surface $\Gamma$ &
		Fibers $\mat{\Upsilon}$ embedded in $\Gamma$\\
		\midrule
		Tangential projector &
		\multicolumn{1}{c}{$\mat{p}_{\Gamma} = \mat{I} - \vek{n} \otimes \vek{n}$} &
		\multicolumn{1}{c}{$\mat{p}_{\mat{\Upsilon}} = \vek{t} \otimes \vek{t}$}\\
		\midrule
		Tangential gradient of a scalar function &
		\multicolumn{1}{c}{$\gradS \, f = \mat{j} \mult \mat{g}^{-1} \mult \gradRef \, f$} &
		\multicolumn{1}{c}{$\gradF \, f = \mat{p}_{\mat{\Upsilon}} \mult \gradS \, f$}\\
		\midrule
		Directional tangential gradient of a vector function &
		\multicolumn{1}{c}{$\gradS^{\mathrm{dir}} \vek{v} = \begin{bmatrix}
				(\gradS \, v_x)^{\mathrm{T}} \\
				(\gradS \, v_y)^{\mathrm{T}} \\
				(\gradS \, v_z)^{\mathrm{T}} \\
			\end{bmatrix}$} &
		\multicolumn{1}{c}{$\gradF^{\mathrm{dir}} \, \vek{v} = \gradS^{\mathrm{dir}} \vek{v} \mult \mat{p}_{\mat{\Upsilon}}$}\\
		\midrule
		Covariant tangential gradient of a vector function &
		\multicolumn{1}{c}{$\gradS^{\mathrm{cov}} \vek{v} = \mat{p}_{\Gamma} \mult \gradS^{\mathrm{dir}} \vek{v}$} &
		\multicolumn{1}{c}{$\gradF^{\mathrm{cov}} \vek{v} = \mat{p}_{\Upsilon} \mult \gradF^{\mathrm{dir}} \vek{v}$}\\
		\midrule
		Tangential divergence of a vector function &
		\multicolumn{1}{c}{$\begin{aligned}
				\divS \, \vek{v} &= \mathrm{tr}\big(\gradS^{\mathrm{dir}}\vek{v}\big) \\ &= \mathrm{tr}\big(\gradS^{\mathrm{cov}}\vek{v}\big)
			\end{aligned}$} &
		\multicolumn{1}{c}{$\begin{aligned}
				\divF \, \vek{v} &= \mathrm{tr}\big(\gradF^{\mathrm{dir}}\vek{v}\big) \\ &= \mathrm{tr}\big(\gradF^{\mathrm{cov}}\vek{v}\big)
			\end{aligned}$}\\
		\midrule
		Tangential divergence of a tensor function &
		\multicolumn{1}{c}{$\divS \mat{A} = \begin{bmatrix}
				\divS \, (A_{11}, A_{12}, A_{13}) \\
				\divS \, (A_{21}, A_{22}, A_{23}) \\
				\divS \,( A_{31}, A_{32}, A_{33}) \\
			\end{bmatrix}$} &
		\multicolumn{1}{c}{$\divF \mat{A} = \begin{bmatrix}
				\divF \, (A_{11}, A_{12}, A_{13}) \\
				\divF \, (A_{21}, A_{22}, A_{23}) \\
				\divF \,( A_{31}, A_{32}, A_{33}) \\
			\end{bmatrix}$}\\
		\bottomrule
	\end{tabularx}
	\caption{Tangential differential operators for the explicitly defined membrane surface $\Gamma$ and the implicitly defined fiber family $\mat{\Upsilon}$.}
	\label{tab:DifferentialOperators}
\end{table}

\subsection{Integral theorems} \label{subsec:IntTh}

An integral theorem which combines the membrane $\Gamma$ and the embedded fiber family $\mat{\Upsilon}$ is later needed in the formulation of the weak form of the governing equations. Before stating this integral theorem, the divergence theorems for the single two-dimensional membrane and for one family of fibers, i.e., one-dimensional manifolds, on a single two-dimensional surface are given, respectively. We consider a tensor-valued function $\mat{a}$ and a vector-valued function $\vek{w}$ in the following integral theorems.\\
\\
For the membrane, this integral theorem is given as, see, e.g., \cite{Fries_2020a,Schoellhammer_2019a,Fries_2018a}, 
\begin{equation}
	\int_{\Gamma} \vek{w} \cdot \divS\,\mat{a} \,\d\Gamma = - \int_{\Gamma} \gradS^{\mathrm{dir}} \vek{w} : \mat{a} \,\d\Gamma - \int_{\Gamma} \vek{w} \cdot (\mat{a} \mult \divS \,\mat{p}_{\Gamma}) \,\d\Gamma + \int_{\partial\Gamma} \vek{w} \cdot (\mat{a} \mult \vek{m}) \,\d\partial\Gamma. \label{eq:intThSrf}
\end{equation}
Note that the third term on the right hand side considers \emph{curvature} by the relation $\divS \,\mat{p}_{\Gamma} = -\varkappa\mult\vek{n}$, where $\varkappa$ is the mean curvature. For \emph{in-plane} tensors, i.e., $\mat{a} = \mat{p}_{\Gamma} \mult \mat{a} \mult \mat{p}_{\Gamma}$, this term vanishes.\\
\\
The analogous integral theorem for a fiber family $\mat{\Upsilon}$ over a curved surface $\Gamma$ in $\mathbb{R}^3$ is given as
\begin{align}
	\int_{\Gamma} \vek{w} \cdot \divF\,\mat{a} \mult \lVert\vek{t}^{\star}\rVert \,\d\Gamma = &- \int_{\Gamma} \gradF^{\mathrm{dir}}\vek{w} : \mat{a} \mult \lVert\vek{t}^{\star}\rVert \,\d\Gamma - \int_{\Gamma} \vek{w} \cdot (\mat{a} \mult \divF \,\mat{p}_{\mat{\Upsilon}}) \mult \lVert\vek{t}^{\star}\rVert \,\d\Gamma \nonumber \\
	& + \int_{\partial\Gamma} \vek{w} \cdot (\mat{a} \mult (\mat{p}_{\mat{\Upsilon}}\mult\vek{m}))  \mult \lVert\vek{t}^{\star}\rVert \,\d\partial\Gamma. \label{eq:intThFib}
\end{align}
In this integral, \emph{all} fibers over $\Gamma$ are considered simultaneously. Therefore, the \emph{coarea formula} was applied, see, e.g., \cite{Burger_2009a,Fries_2023a,Kaiser_2024a} for further details. The factor $\lVert\vek{t}^{\star}\rVert$ results from the coarea formula. Note that in this integral theorem, there is no (mechanical) interaction between the surface and the fibers. In Eq.~(\ref{eq:intThSrf}), the surface can be interpreted as a bulk domain, without mechanical properties which is only used for geometry definition, analogously to \cite{Fries_2023a,Kaiser_2024a}; later, however, the surface is mechanically modelled as a membrane into which the fibers are embedded. As in Eq.~(\ref{eq:intThSrf}), the third term on the right hand side considers curvature, in this case the curvature of the fibers, and vanishes when the tensor $\mat{a} = \mat{p}_{\mat{\Upsilon}}\mult\mat{a}\mult\mat{p}_{\mat{\Upsilon}}$ is in-plane to the fibers.\\
\\
Finally, the two integral theorems, i.e., Eqs.~(\ref{eq:intThSrf}) and (\ref{eq:intThFib}) are combined to \emph{one} integral theorem for a surface with \emph{embedded} fibers, i.e., the surface and the fibers interact mechanically. For $n_{\mathrm{fib}}$ families of embedded fibers, this integral theorem is defined as
\begin{align}
	\int_{\Gamma} \vek{w} \cdot \big(\divS \, \mat{a} + \sum_{i}^{n_{\mathrm{fib}}} \divFf \, \mat{a} \mult \lVert\vek{t}_i^{\star}\rVert \big) \,\d\Gamma = - &\int_{\Gamma} \mat{a} : \big(\gradS^{\mathrm{dir}} \vek{w} + \sum_{i}^{n_{\mathrm{fib}}} \gradFf^{\mathrm{dir}}\,\vek{w} \mult \lVert\vek{t}_i^{\star}\rVert \big) \,\d\Gamma \nonumber \\
	\underbrace{-\int_{\Gamma} \vek{w} \cdot \big(\mat{a} \mult \big(\divS\,\mat{p}_{\Gamma} + \sum_{i}^{n_{\mathrm{fib}}} \divF \,\mat{p}_{\mat{\Upsilon}_i} \mult \lVert\vek{t}_i^{\star}\rVert\big)\big) \,\d\Gamma}_{\text{curvature term}} +& \int_{\partial\Gamma} \vek{w} \cdot \big(\mat{a} \mult \big(\vek{m} + \sum_{i}^{n_{\mathrm{fib}}}(\mat{p}_{\mat{\Upsilon}_i}\mult\vek{m}) \mult \lVert\vek{t}_i^{\star}\rVert \big)\big)\,\d\partial\Gamma. \label{eq:intThEmbdFib}
\end{align}
For in-plane tensors $\mat{a}_{\Gamma} = \mat{p}_{\Gamma} \mult \mat{a} \mult \mat{p}_{\Gamma}$ and $\mat{a}_{\mat{\Upsilon}} = \mat{p}_{\mat{\Upsilon}}\mult\mat{a}\mult\mat{p}_{\mat{\Upsilon}}$, the curvature term in Eq.~(\ref{eq:intThEmbdFib}) vanishes.

\section{Mechanical model for the hyperelastic membrane} \label{sec:MechModelMemb}

In this section, we introduce the mechanical model for the hyperelastic, isotropic membrane into which the fibers are later embedded. In \cite{Fries_2020a}, a finite strain theory for such membranes governed by St.Venant--Kirchhoff materials is introduced. As usual in finite strain or large displacement theory, the undeformed material and deformed spatial configurations of the membrane are distinguished. Notationally, uppercase letters refer to the undeformed configuration while lowercase letters refer to the deformed configuration. For further details about hyperelasticity and non-linear continuum mechanics, we refer to the classical textbooks, \cite{Holzapfel_2000a,Wriggers_2008a,Ogden_1997a,Belytschko_2000b,Marsden_1994a}.

\subsection{Material and spatial configurations} \label{subsec:MembConfig}

\begin{figure}
	\centering
	\includegraphics[width=1.0\textwidth]{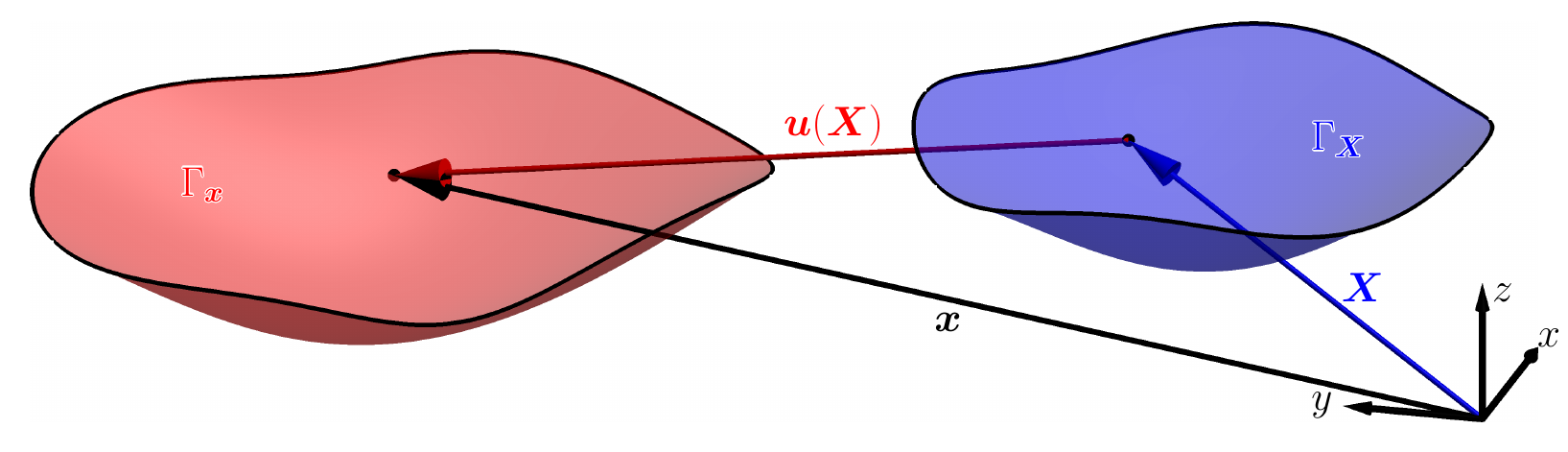}
	\captionsetup{margin=4cc}
	\caption{The displacement of a single membrane: The blue surface represents the undeformed configuration $\Gamma_{\vek{X}}$ and the red surface the deformed configuration $\Gamma_{\vek{x}}$. The vectors are the position vectors $\vek{X}$ and $\vek{x}$ to a point on the membranes and the displacement vector $\vek{u}(\vek{X})$, respectively.}
	\label{fig:MembrKinematics}
\end{figure}

The membrane's surface is represented by the two-dimensional manifolds $\Gamma_{\vek{X}}$ and $\Gamma_{\vek{x}}$ in the undeformed (material) and deformed (spatial) configuration, respectively. The two-dimensional surface represents the mid-surface of a thin membrane which features thickness $\mathrm{T} > 0$. Note that these manifolds are immersed in the three-dimensional space $\mathbb{R}^3$, hence, are of codimension one. In this work, the two-dimensional surfaces are defined explicitly, see Sec.~\ref{subsec:GeomDiffMF} herein and Sec.~2.2 in \cite{Fries_2020a} for further details. The \emph{displacement field} $\vek{u}(\vek{X})$ relates the two configurations by
\begin{equation}
	\vek{x} = \vek{X} + \vek{u}(\vek{X}) \quad \mathrm{with} \, \vek{X} \in \Gamma_{\vek{X}} \subset \mathbb{R}^3 \, \mathrm{and} \, \vek{x} \in \Gamma_{\vek{x}} \subset \mathbb{R}^3. \label{eq:disp}
\end{equation}
With the displacement $\vek{u}(\vek{X})$, the \emph{in-plane} surface deformation gradient is
\begin{equation}
	\mat{F}_{\Gamma} = \mat{P}_{\Gamma} + \GradS^{\mathrm{dir}} \vek{u}. \label{eq:DefGradMemb}
\end{equation}
Note that, in the remainder of this paper, we often omit the subscripts $\vek{X}$ or $\vek{x}$ for $\Gamma$ when used as an index of a certain quantity written by uppercase or lowercase letters. For example, we write $\mat{P}_{\Gamma} = \mat{P}_{\Gamma_{\vek{X}}}(\vek{X})$ and $\mat{p}_{\Gamma} = \mat{p}_{\Gamma_{\vek{x}}}(\vek{x})$ for the tangential projector. For the gradient, an uppercase `G' in $\GradS$ is the gradient w.r.t.~the undeformed configuration while a lowercase `g' indicates the gradient $\gradS$ w.r.t.~the deformed configuration, respectively. Fig.~\ref{fig:MembrKinematics} visualizes the kinematic situation for some individual membrane.

The inverse of the surface deformation gradient is defined as 
\begin{equation}
	\mat{F}_{\Gamma}^{-1} = \mat{p}_{\Gamma} - \gradS^{\mathrm{dir}} \vek{u}. \label{eq:InvDefGradMemb}
\end{equation}
It is important to note that the \emph{in-plane} deformation gradient $\mat{F}_{\Gamma}$ is of \emph{rank two}, hence, $\mat{F}_{\Gamma}^{-1}$ is \emph{not} the standard \emph{inverse} of a tensor but computed by Eq.~(\ref{eq:InvDefGradMemb}). Furthermore the transposed quantities of the deformation gradient and its inverse are defined as
\begin{align}
	\mat{F}_{\Gamma}^{\mathrm{T}} &= \mat{P}_{\Gamma} + \big(\GradS^{\mathrm{dir}} \vek{u} \big)^{\mathrm{T}},\\
	\mat{F}_{\Gamma}^{-\mathrm{T}} &= \mat{p}_{\Gamma} - \big(\gradS^{\mathrm{dir}} \vek{u} \big)^{\mathrm{T}}.
\end{align} 
Note the following properties of the surface deformation gradient: (i) $\mat{F}_{\Gamma}^{-1} \mult \mat{F}_{\Gamma} = \mat{P}_{\Gamma}$, (ii) $\mat{F}_{\Gamma} \mult \mat{F}_{\Gamma}^{-1} = \mat{p}_{\Gamma}$, (iii) $\mat{F}_{\Gamma} = \mat{p}_{\Gamma} \mult \mat{F}_{\Gamma}  \mult \mat{P}_{\Gamma}$, and (iv) $\mat{F}_{\Gamma}^{-1} = \mat{P}_{\Gamma} \mult \mat{F}_{\Gamma}^{-1}  \mult \mat{p}_{\Gamma}$.
 
\subsection{Strain tensors} \label{subsec:StrainTenMemb}
Strain tensors are defined based on the deformation gradient and are important quantities in hyperelasticity. For the membrane considered herein, the \emph{right} Cauchy--Green tensor is defined as
\begin{equation}
	\mat{C}_{\Gamma} = \mat{F}_{\Gamma}^{\mathrm{T}} \mult \mat{F}_{\Gamma}, \label{eq:rCGTmemb}
\end{equation}
and measures strain in the material coordinates \cite{Holzapfel_2000a}. An \emph{eigenvalue} decomposition of the right Cauchy--Green tensor leads to the first and second principal stretches $\lambda_1^{\Gamma}$ and $\lambda_2^{\Gamma}$, respectively. Note that $\mat{C}_{\Gamma}$ is of rank two and has, therefore, only two non-zero eigenvalues. \emph{Incompressible} membranes are defined by the volume ratio $J_{\Gamma} = \lambda_1^{\Gamma} \mult \lambda_2^{\Gamma} \mult \lambda_3^{\Gamma} = 1$, hence, the third principal stretch $\lambda_3^{\Gamma}$, associated with the normal vector, is defined as $\lambda_3^{\Gamma} = \nicefrac{1}{(\lambda_1^{\Gamma} \mult \lambda_2^{\Gamma})}$. For the relation of the membrane thicknesses in the deformed and undeformed configurations, this leads to $\mathrm{t} = \lambda_3^{\Gamma} \mult \mathrm{T}$. Using the eigenvalues, i.e., the squared first and second principal stretches, the right Cauchy--Green tensor can be rewritten as
\begin{equation}
	\mat{C}_{\Gamma} = \sum_{k=1}^{2} \left(\lambda_k^{\Gamma}\right)^2 \mult \bar{\vek{T}}_{\Gamma_k} \otimes \bar{\vek{T}}_{\Gamma_k}.
\end{equation}
The eigenvectors $\bar{\vek{T}}_{\Gamma_k}$, associated to the eigenvalues $(\lambda_k^{\Gamma})^2$, are tangential to the manifold, hence, $\bar{\vek{T}}_{\Gamma_k} \cdot \vek{N} = 0$ holds.\\
\\
For the deformed configuration, the \emph{left} Cauchy--Green tensor is relevant and defined as
\begin{equation}
	\mat{b}_{\Gamma} = \mat{F}_{\Gamma} \mult \mat{F}_{\Gamma}^{\mathrm{T}}. \label{eq:lCGTmemb}
\end{equation}
The inverses of the Cauchy--Green tensors are defined as
\begin{align}
	\mat{C}_{\Gamma}^{-1}& = \mat{F}_{\Gamma}^{-1} \mult \mat{F}_{\Gamma}^{-\mathrm{T}}, \\
	\mat{b}_{\Gamma}^{-1}& = \mat{F}_{\Gamma}^{-\mathrm{T}} \mult \mat{F}_{\Gamma}^{-1}.
\end{align}
The Green--Lagrange strain tensor, defined w.r.t.~the undeformed configuration, is
\begin{equation}
	\mat{E}_{\Gamma} = \frac{1}{2} \big(\mat{C}_{\Gamma} - \mat{P}_{\Gamma}\big), \label{eq:GLSTmemb}
\end{equation}
and the Euler--Almansi strain tensor, defined w.r.t.~the deformed configuration,
\begin{equation}
	\mat{e}_{\Gamma} = \frac{1}{2} \big(\mat{p}_{\Gamma} - \mat{b}_{\Gamma}^{-1}\big). \label{eq:EASTmemb}
\end{equation}
Note that all of these strain tensors are in-plane, i.e., tangential tensors. For the Green--Lagrange and Euler--Almansi strain tensors, there holds  (i) $\mat{E}_{\Gamma} = \mat{P}_{\Gamma} \mult \mat{E}_{\Gamma} \mult \mat{P}_{\Gamma}$, (ii) $\mat{e}_{\Gamma} = \mat{p}_{\Gamma} \mult \mat{e}_{\Gamma} \mult \mat{p}_{\Gamma}$, (iii) $\mat{e}_{\Gamma} = \mat{F}_{\Gamma}^{-\mathrm{T}} \mult \mat{E}_{\Gamma} \mult \mat{F}_{\Gamma}^{-1}$, and (iv) $\mat{E}_{\Gamma} = \mat{F}_{\Gamma}^{\mathrm{T}} \mult \mat{e}_{\Gamma} \mult \mat{F}_{\Gamma}$. 

\subsection{Strain energy function and stress tensors} \label{subsec:SEF-STMemb}
The \emph{stored internal energy function} is formulated based on the principal stretches. For incompressible membranes, a useful starting point is the definition of some general Ogden material in $\mathbb{R}^3$, see Eq.~(6.119) on p.~236 in \cite{Holzapfel_2000a}, which is reduced to the membrane case as
\begin{equation}
	W_{\Gamma}(\lambda_1^{\Gamma},\lambda_2^{\Gamma}) = \sum_{p=1}^{N} \frac{\mu_p}{\alpha_p} \mult\big((\lambda_1^{\Gamma})^{\alpha_p}+(\lambda_2^{\Gamma})^{\alpha_p}+(\lambda_1^{\Gamma}\mult \lambda_2^{\Gamma})^{-\alpha_p}-3\big), \label{eq:EngFctMemb}
\end{equation}
with $p = \{1, 2,\ldots,N\}$ and dimensionless constants $\alpha_p$. The classical shear modulus from linear elasticity is related to the material coefficients by $2\mult\mu = \sum_{p} \frac{\mu_p}{\alpha_p}$. For the definition of the stress tensor, the derivative of the strain energy function with respect to the principal stretches is needed, defined as
\begin{align}
	\frac{\partial W}{\partial \lambda_1^{\Gamma}} &=\sum_{p=1}^{N} \mu_p \mult\big( (\lambda_1^{\Gamma})^{\alpha_p-1} - (\lambda_1^{\Gamma})^{-\alpha_p-1}\mult (\lambda_2^{\Gamma})^{-\alpha_p}\big),\\
	\frac{\partial W}{\partial \lambda_2^{\Gamma}} &=\sum_{p=1}^{N} \mu_p \mult\big( (\lambda_2^{\Gamma})^{\alpha_p-1} - (\lambda_1^{\Gamma})^{-\alpha_p}\mult (\lambda_2^{\Gamma})^{-\alpha_p-1}\big).
\end{align}
Furthermore, two important classes of materials are derived from this general definition of the strain energy function. First, for $N=2$, $\alpha_1 = 2$, and $\alpha_2 = -2$ follows the strain energy function for \emph{incompressible Mooney--Rivlin materials} as
\begin{equation}
	W(\lambda_1^{\Gamma},\lambda_2^{\Gamma}) = \frac{\mu_1}{2}\mult \big((\lambda_1^{\Gamma})^2 + (\lambda_2^{\Gamma})^2 + (\lambda_1^{\Gamma} \mult \lambda_2^{\Gamma})^{-2} - 3\big) - \frac{\mu_2}{2} \mult \big((\lambda_1^{\Gamma})^{-2} + (\lambda_2^{\Gamma})^{-2} + (\lambda_1^{\Gamma} \mult \lambda_2^{\Gamma})^2 - 3\big).
\end{equation}
The respective derivatives, which are used in the definition of the stress tensors later, are
\begin{align}
	\frac{\partial W}{\partial \lambda_1^{\Gamma}} &= \mu_1 \mult \big(\lambda_1^{\Gamma} - (\lambda_1^{\Gamma})^{-3} \mult (\lambda_2^{\Gamma})^{-2}\big) + \mu_2 \mult \big((\lambda_1^{\Gamma})^{-3} - \lambda_1^{\Gamma}\mult (\lambda_2^{\Gamma})^2\big), \\
	\frac{\partial W}{\partial \lambda_2^{\Gamma}} &= \mu_1 \mult \big(\lambda_2^{\Gamma} - (\lambda_1^{\Gamma})^{-2} \mult (\lambda_2^{\Gamma})^{-3}\big) + \mu_2 \mult \big((\lambda_2^{\Gamma})^{-3} - (\lambda_1^{\Gamma})^2 \mult \lambda_2^{\Gamma}\big).
\end{align}
For $N=1$, $\alpha_1 = 2$ and $\mu_1 = \mu$, there follows from Eq.~(\ref{eq:EngFctMemb}) the strain energy function for membranes composed of \emph{incompressible Neo-Hooke materials} as
\begin{equation}
	W(\lambda_1^{\Gamma},\lambda_2^{\Gamma}) = \frac{\mu}{2} \mult \big((\lambda_1^{\Gamma})^2 + (\lambda_2^{\Gamma})^2 + (\lambda_1^{\Gamma} \mult \lambda_2^{\Gamma})^{-2} - 3).
\end{equation}
Again, we need the derivatives with respect to $\lambda_1^{\Gamma}$ and $\lambda_2^{\Gamma}$, respectively, which follow as
\begin{align}
	\frac{\partial W}{\partial \lambda_1^{\Gamma}} &= \mu_1 \mult \left(\lambda_1 - (\lambda_1^{\Gamma})^{-3}\mult(\lambda_2^{\Gamma})^{-2}\right), \\
	\frac{\partial W}{\partial \lambda_2^{\Gamma}} &= \mu_1 \mult \left(\lambda_2 - (\lambda_1^{\Gamma})^{-2} \mult (\lambda_2^{\Gamma})^{-3}\right).
\end{align}
Next, we define stress tensors. The second Piola--Kirchhoff stress tensor is defined as
\begin{equation}
	\mat{S}_{\Gamma} = \sum_{k=1}^{2} \frac{1}{\lambda_k^{\Gamma}} \mult \frac{\partial W_{\Gamma}}{\partial \lambda_k^{\Gamma}} \mult \bar{\vek{T}}_{\Gamma_k} \otimes \bar{\vek{T}}_{\Gamma_k}. \label{eq:2PKmemb}
\end{equation}
The first Piola--Kirchhoff stress tensor is given as
\begin{equation}
	\mat{K}_{\Gamma} = \mat{F}_{\Gamma} \mult \mat{S}_{\Gamma} = \sum_{k=1}^{2} \frac{\partial W_{\Gamma}}{\partial \lambda_k^{\Gamma}} \mult \bar{\vek{t}}_{\Gamma_k} \otimes \bar{\vek{T}}_{\Gamma_k}. \label{eq:1PKmemb}
\end{equation}
The Cauchy stress tensor for the membrane is
\begin{equation}
	\vek{\sigma}_{\Gamma} = \frac{1}{J_{\Gamma}} \mult \mat{F}_{\Gamma} \mult \mat{S}_{\Gamma} \mult \mat{F}_{\Gamma}^{\mathrm{T}} = \sum_{k=1}^{2} \frac{\lambda_k^{\Gamma}}{J_{\Gamma}} \mult \frac{\partial W_{\Gamma}}{\partial \lambda_k^{\Gamma}} \mult \bar{\vek{t}}_{\Gamma_k} \otimes \bar{\vek{t}}_{\Gamma_k}. \label{eq:CSmemb}
\end{equation}
Note that for \emph{incompressible} membranes, $J_{\Gamma} = \lambda_1^{\Gamma} \mult \lambda_2^{\Gamma} \mult \lambda_3^{\Gamma} = 1$. Furthermore, note the relations between the stress tensors as (i) $\mat{S}_{\Gamma} = \mat{P}_{\Gamma} \mult \mat{S}_{\Gamma} \mult \mat{P}_{\Gamma}$, (ii) $\vek{\sigma}_{\Gamma} = \mat{p}_{\Gamma} \mult \vek{\sigma}_{\Gamma} \mult \mat{p}_{\Gamma}$, and (iii)  $\mat{K}_{\Gamma} = \mat{K}_{\Gamma} \mult \mat{P}_{\Gamma} = \mat{p}_{\Gamma} \mult \mat{K}_{\Gamma}$.

\subsection{Equilibrium of the membrane in strong form} \label{subsec:EqMemb}
The equilibrium in finite strain theory is formulated in the deformed configuration as
\begin{equation}
	\divS\left(t\mult\vek{\sigma}_{\Gamma}(\vek{x})\right) = -t\mult\vek{f}_{\Gamma}(\vek{x}) \quad \forall \vek{x} \in \Gamma_{\vek{x}}, \label{eq:sfDefMemb}
\end{equation}
where $t = \lambda_3^{\Gamma} \mult T$ and $\vek{f}_{\Gamma}(\vek{x})\in\mathbb{R}^3$ are body forces. Note the relation of stresses between the deformed and undeformed configuration, known as \cite{Fries_2020a}
\begin{equation}
	T \mult \DivS\,\mat{K}_{\Gamma}(\vek{X}) = \lambda_1^{\Gamma} \mult \lambda_2^{\Gamma} \mult \divS \left(\lambda_3^{\Gamma}\mult T\mult\vek{\sigma}_{\Gamma}(\vek{x})\right)\!.
\end{equation}
The equilibrium can then be formulated in the \emph{undeformed} configuration as
\begin{equation}
	T \mult \DivS \mat{K}_{\Gamma}(\vek{X}) = -T\vek{F}_{\Gamma}(\vek{X}) \quad \forall \vek{X} \in \Gamma_{\vek{X}}. \label{eq:sfUndefMemb}
\end{equation}
The BVP modelling the mechanics of hyperelastic membranes in the undeformed configuration is now given by (i) the \emph{kinematics} by the strain tensor $\mat{E}_{\Gamma}$ from Eq.~(\ref{eq:GLSTmemb}), (ii) the \emph{constitutive equations} by the stress tensor $\mat{K}_{\Gamma}$ from Eq.~(\ref{eq:1PKmemb}), and (iii) by the \emph{equilibrium} from Eq.~(\ref{eq:sfUndefMemb}). This BVP can be reformulated into one (vector-valued) field equation for the sought displacements $\vek{u}$. The boundary conditions are stated as follows: The boundary of the undeformed membrane is split into two non-overlapping parts, i.e., the \emph{Dirichlet} part $\partial\Gamma_{\vek{X}}^{\mathrm{D}}$ and the \emph{Neumann} part $\partial\Gamma_{\vek{X}}^{\mathrm{N}}$, respectively. This leads to the boundary conditions
\begin{align}
	\vek{u}(\vek{X}) &= \hat{\vek{G}}(\vek{X}) \quad \mathrm{on} \quad \partial\Gamma_{\vek{X}}^{\mathrm{D}},\label{eq:DirBCsMemb}\\
	T\mult\mat{K}_{\Gamma}(\vek{X}) \mult \vek{M}(\vek{X}) &= \hat{\vek{H}}_{\Gamma}(\vek{X}) \quad \mathrm{on} \quad \partial\Gamma_{\vek{X}}^{\mathrm{N}}, \label{eq:NeuBCsMemb}
\end{align}
where $\hat{\vek{G}}(\vek{X})$ are prescribed displacements and $\hat{\vek{H}}_{\Gamma}(\vek{X})$ are prescribed tractions.

\section{Mechanical model for fibers embedded in a surface} \label{sec:MechModelFib}

In this section, we introduce the mechanical model for fibers of one hyperelastic, isotropic fiber family $\mat{\Upsilon}$. The fibers are to be embedded in the membrane later, analogously to the previous section. Note that the fibers could also be considered without their embedding as substructures into the membrane but as independent structures which are defined over the surface bulk domain. This approach would be similar to \cite{Fries_2023a}, where the mechanical models for ropes and membranes are solved simultaneously over a bulk domain without mechanical properties, only needed for geometrical purposes. In this work, however, the fiber model described in this section is later coupled to some bulk domain featuring hyperelastic material behaviour. Again, as in the section before, the material and spatial configurations of the fibers are distinguished, using the same notational conventions.

\subsection{Material and spatial configurations} \label{subsec:FiberConfig}

Following the outline of the previous section, there follows for the deformation gradient of the fibers and its inverse:

\begin{align}
	\mat{F}_{\mat{\Upsilon}} &= \mat{P}_{\mat{\Upsilon}} + \GradF^{\mathrm{dir}} \vek{u}, \label{eq:DefGradFib}\\
	\mat{F}_{\mat{\Upsilon}}^{-1} &= \mat{p}_{\mat{\Upsilon}} - \gradF^{\mathrm{dir}} \vek{u}, \label{eq:InvDefGradFib}\\
	\mat{F}_{\mat{\Upsilon}}^{\mathrm{T}} &= \mat{P}_{\mat{\Upsilon}} + \big(\GradF^{\mathrm{dir}} \vek{u} \big)^{\mathrm{T}},\\
	\mat{F}_{\mat{\Upsilon}}^{-\mathrm{T}} &= \mat{p}_{\mat{\Upsilon}} - \big(\gradF^{\mathrm{dir}} \vek{u} \big)^{\mathrm{T}}.
\end{align} 
Similarly, they feature the properties:: (i) $\mat{F}_{\mat{\Upsilon}}^{-1} \mult \mat{F}_{\mat{\Upsilon}} = \mat{P}_{\mat{\Upsilon}}$, (ii) $\mat{F}_{\mat{\Upsilon}} \mult \mat{F}_{\mat{\Upsilon}}^{-1} = \mat{p}_{\mat{\Upsilon}}$, (iii) $\mat{F}_{\mat{\Upsilon}} = \mat{p}_{\mat{\Upsilon}} \mult \mat{F}_{\mat{\Upsilon}}  \mult \mat{P}_{\mat{\Upsilon}}$, and (iv) $\mat{F}_{\mat{\Upsilon}}^{-1} = \mat{P}_{\mat{\Upsilon}} \mult \mat{F}_{\mat{\Upsilon}}^{-1}  \mult \mat{p}_{\mat{\Upsilon}}$.

\subsection{Strain tensors} \label{subsec:StrainTenFib}
We follow the same procedure as for the mebranes above. As such, the right Cauchy--Green tensor for the fibers is defined as
\begin{equation}
	\mat{C}_{\mat{\Upsilon}} = \mat{F}_{\mat{\Upsilon}}^{\mathrm{T}} \mult \mat{F}_{\mat{\Upsilon}}. \label{eq:rCGTfib}
\end{equation}
It has rank one and, therefore, an \emph{eigenvalue} decomposition leads to the fiber stretch $\lambda_1^{\vek{\Upsilon}}$. The other two eigenvalues are zero. For \emph{incompressible} fibers, the volume ratio follows as $J_{\mat{\Upsilon}} = \lambda_1^{\mat{\Upsilon}} \mult \lambda_2^{\mat{\Upsilon}} \mult \lambda_3^{\mat{\Upsilon}} = 1$, hence, the second and third principal stretches are computed as $\lambda_2^{\mat{\Upsilon}} = \lambda_3^{\mat{\Upsilon}} = \nicefrac{1}{\sqrt{\lambda_1^{\mat{\Upsilon}}}}$. With the first eigenvalue, i.e., the squared fiber stretch, the right Cauchy--Green tensor can be rewritten as
\begin{equation}
	\mat{C}_{\mat{\Upsilon}} = \left(\lambda_1^{\mat{\Upsilon}}\right)^2 \mult \bar{\vek{T}} \otimes \bar{\vek{T}}.
\end{equation}
The eigenvector $\bar{\vek{T}}$ is associated to the squared fiber stretch and is tangential to the fibers.\\
\\
For the various strain tensors related to the fiber family $\mat{\Upsilon}$, we find analogously to Sec.~\ref{subsec:StrainTenMemb} for the membranes:
\begin{align}
	&\text{left Cauchy--Green tensor:} & \mat{b}_{\mat{\Upsilon}} &= \mat{F}_{\mat{\Upsilon}} \mult \mat{F}_{\mat{\Upsilon}}^{\mathrm{T}}, \label{eq:lCGTfib}\\
	&\text{inverse of the right Cauchy--Green tensor:} & \mat{C}_{\mat{\Upsilon}}^{-1}& = \mat{F}_{\mat{\Upsilon}}^{-1} \mult \mat{F}_{\mat{\Upsilon}}^{-\mathrm{T}}, \\
	&\text{inverse of the left Cauchy--Green tensor:} & \mat{b}_{\mat{\Upsilon}}^{-1}& = \mat{F}_{\mat{\Upsilon}}^{-\mathrm{T}} \mult \mat{F}_{\mat{\Upsilon}}^{-1},\\
	&\text{Green--Lagrange strain tensor:} & \mat{E}_{\mat{\Upsilon}} &= \frac{1}{2} \big(\mat{C}_{\mat{\Upsilon}} - \mat{P}_{\mat{\Upsilon}}\big), \label{eq:GLSTfib}\\
	&\text{Euler--Almansi strain tensor:}& \mat{e}_{\mat{\Upsilon}} &= \frac{1}{2} \big(\mat{p}_{\mat{\Upsilon}} - \mat{b}_{\mat{\Upsilon}}^{-1}\big). \label{eq:EASTfib}	 
\end{align}
For the strain tensors, the following properties hold: (i) $\mat{E}_{\mat{\Upsilon}} = \mat{P}_{\mat{\Upsilon}} \mult \mat{E}_{\mat{\Upsilon}} \mult \mat{P}_{\mat{\Upsilon}}$, (ii) $\mat{e}_{\mat{\Upsilon}} = \mat{p}_{\mat{\Upsilon}} \mult \mat{e}_{\mat{\Upsilon}} \mult \mat{p}_{\mat{\Upsilon}}$, (iii) $\mat{e}_{\mat{\Upsilon}} = \mat{F}_{\mat{\Upsilon}}^{-\mathrm{T}} \mult \mat{E}_{\mat{\Upsilon}} \mult \mat{F}_{\mat{\Upsilon}}^{-1}$, and (iv) $\mat{E}_{\mat{\Upsilon}} = \mat{F}_{\mat{\Upsilon}}^{\mathrm{T}} \mult \mat{e}_{\mat{\Upsilon}} \mult \mat{F}_{\mat{\Upsilon}}$.

\subsection{Strain energy function and stress tensors} \label{subsec:SEF-STFib}
The \emph{stored internal energy function} is formulated based on the fiber stretch, analogously to Sec.~\ref{subsec:SEF-STMemb} for membranes. For incompressible, isotropic, and hyperelastic fibers, the strain energy function for Ogden materials is defined as
\begin{equation}
	W_{\mat{\Upsilon}}(\lambda_1^{\mat{\Upsilon}}) = \sum_{p=1}^{N} \frac{\mu_p}{\alpha_p} \mult\big((\lambda_1^{\mat{\Upsilon}})^{\alpha_p}+2\mult(\lambda_1^{\mat{\Upsilon}})^{-\alpha_p/2}-3\big). \label{eq:EngFctFib}
\end{equation}
The derivative of this strain energy function, as later used in the definition of stress tensors, follows as
\begin{equation}
	\frac{\partial W_{\mat{\Upsilon}}}{\partial \lambda_1^{\mat{\Upsilon}}} = \sum_{p=1}^{N} \mu_p \mult \big((\lambda_1^{\mat{\Upsilon}})^{\alpha_1-1} - (\lambda_1^{\mat{\Upsilon}})^{-\alpha_1/2-1}\big).
\end{equation}
Analogously to Sec.~\ref{subsec:SEF-STMemb}, the strain energy function for the general Ogden material for fibers may, for example, be simplified to incompressible Mooney--Rivlin or Neo-Hooke materials if desired. For Mooney--Rivlin materials, the strain energy function and its derivative with respect to the fiber stretch is
\begin{align}
	W_{\mat{\Upsilon}}(\lambda_1^{\mat{\Upsilon}}) &= \frac{\mu_1}{2}  \mult \left((\lambda_1^{\mat{\Upsilon}})^2 + 2 \mult (\lambda_1^{\mat{\Upsilon}})^{-1} - 3\right) - \frac{\mu_2}{2} \mult \left((\lambda_1^{\mat{\Upsilon}})^{-2} + 2 \mult \lambda_1^{\mat{\Upsilon}} - 3\right), \\
	\frac{\partial W_{\mat{\Upsilon}}}{\partial \lambda_1^{\mat{\Upsilon}}} &= \mu_1 \mult \left(\lambda_1^{\mat{\Upsilon}} - (\lambda_1^{\mat{\Upsilon}})^{-2}\right) + \mu_2 \mult \left((\lambda_1^{\mat{\Upsilon}})^{-3} - 1\right).
\end{align} 
For Neo-Hooke materials, there follows
\begin{align}
	W_{\mat{\Upsilon}}(\lambda_1^{\mat{\Upsilon}}) &= \frac{\mu}{2}  \mult \left((\lambda_1^{\mat{\Upsilon}})^2 + 2 \mult (\lambda_1^{\mat{\Upsilon}})^{-1} - 3\right), \\
	\frac{\partial W_{\mat{\Upsilon}}}{\partial \lambda_1^{\mat{\Upsilon}}} &= \mu \mult \left(\lambda_1^{\mat{\Upsilon}} - (\lambda_1^{\mat{\Upsilon}})^{-2}\right).
\end{align} 
With these energy functions and their derivatives, the following stress tensors of the fibers result as
\begin{align}
	&\text{second Piola–Kirchhoff:} & \mat{S}_{\mat{\Upsilon}} &= \frac{1}{\lambda_1^{\mat{\Upsilon}}} \mult \frac{\partial W_{\mat{\Upsilon}}}{\partial \lambda_1^{\mat{\Upsilon}}} \mult \bar{\vek{T}} \otimes \bar{\vek{T}}, \label{eq:2PKfib}\\
	&\text{first Piola–Kirchhoff:} & \mat{K}_{\mat{\Upsilon}} &= \mat{F}_{\mat{\Upsilon}} \mult \mat{S}_{\mat{\Upsilon}} = \frac{\partial W_{\mat{\Upsilon}}}{\partial \lambda_1^{\mat{\Upsilon}}} \mult \bar{\vek{t}} \otimes \bar{\vek{T}}, \label{eq:1PKfib}\\
	&\text{Cauchy:} & \vek{\sigma}_{\mat{\Upsilon}} &= \frac{1}{J_{\mat{\Upsilon}}} \mult \mat{F}_{\mat{\Upsilon}} \mult \mat{S}_{\mat{\Upsilon}} \mult \mat{F}_{\mat{\Upsilon}}^{\mathrm{T}} = \frac{\lambda_1^{\mat{\Upsilon}}}{J_{\mat{\Upsilon}}} \mult \frac{\partial W_{\mat{\Upsilon}}}{\partial \lambda_1^{\mat{\Upsilon}}} \mult \bar{\vek{t}} \otimes \bar{\vek{t}}. \label{eq:CSfib}
\end{align}
They feature similar relations as described above for the membrane. 

\subsection{Equilibrium of the fiber family in strong form} \label{subsec:EqFib}
The equilibrium in finite strain theory for \emph{one single} fiber $\Upsilon_c$ of the fiber family $\mat{\Upsilon}$ is formulated in the deformed configuration as
\begin{equation}
	\mathrm{div}_{\Upsilon_c}\left(\lambda_2^{\Upsilon_c}\mult\lambda_3^{\Upsilon_c}\mult A\mult\vek{\sigma}_{\Upsilon_c}(\vek{x})\right) = -\lambda_2^{\Upsilon_c}\mult\lambda_3^{\Upsilon_c}\mult A\mult\vek{f}_{\Upsilon_c}(\vek{x}) \quad \forall \vek{x} \in \Upsilon_c. \label{eq:sfDef1Fib}
\end{equation}
The differential operators for one fiber follow directly from the definitions given in Sec.~\ref{sec:DiffOp}. $A$ is the cross section area of the fiber in the undeformed configuration and is related to the cross section area in the deformed configuration by $a = \lambda_2^{\Upsilon_c}\mult\lambda_3^{\Upsilon_c}\mult A$. For the \emph{fiber family}, i.e., \emph{all} fibers over the surface $\Gamma$, the equilibrium in strong form with respect to the deformed configuration follows as
\begin{equation}
	\mathrm{div}_{\mat{\Upsilon}}\left(\lambda_2^{\mat{\Upsilon}}\mult\lambda_3^{\mat{\Upsilon}}\mult A \mult\vek{\sigma}_{\mat{\Upsilon}}(\vek{x})\right) \! \mult \lVert \vek{t}^{\star}(\vek{x}) \rVert = -\lambda_2^{\Upsilon_c}\mult\lambda_3^{\Upsilon_c}\mult A\mult\vek{f}_{\mat{\Upsilon}}(\vek{x})\mult \lVert \vek{t}^{\star}(\vek{x}) \rVert \quad \forall \vek{x} \in \Gamma_{\vek{x}}. \label{eq:sfDefFib}
\end{equation}
Herein, the term $\lVert \vek{t}^{\star}(\vek{x}) \rVert$ is related to the coarea formula from the integral theorem in Eq.~(\ref{eq:intThEmbdFib}) but is applied in the strong form already because instead of considering all fibers over the surface individually, we want to describe the whole fiber family at once. Note that all fibers $\Upsilon_c$ which belong to the fiber family $\mat{\Upsilon}$ have the same cross section area $A$. With the relation $A\mult\DivF\,\mat{K}(\vek{X})\mult\lVert \vek{T}^{\star}(\vek{X}) \rVert = \lambda_1^{\mat{\Upsilon}} \mult \divS \left(\lambda_2^{\mat{\Upsilon}} \mult \lambda_3^{\mat{\Upsilon}}\mult A\mult\vek{\sigma}_{\mat{\Upsilon}}(\vek{x})\right)\mult\lVert \vek{t}^{\star}(\vek{x}) \rVert$ of the stresses between the deformed and undeformed configuration, the equilibrium is formulated in the \emph{undeformed} configuration as
\begin{equation}
	A\mult\DivF \mat{K}_{\mat{\Upsilon}}(\vek{X}) \mult \lVert \vek{T}^{\star}(\vek{X}) \rVert = -A\mult\vek{F}_{\mat{\Upsilon}}(\vek{X}) \mult \lVert \vek{T}^{\star}(\vek{X}) \rVert \quad \forall \vek{X} \in \Gamma_{\vek{X}}. \label{eq:sfUndefFib}
\end{equation}
The BVP modelling the family of hyperelastic fibers $\mat{\Upsilon}$ over a surface $\Gamma$ is completed by boundary conditions defined as
\begin{align}
	\vek{u}(\vek{X}) &= \hat{\vek{G}}(\vek{X}) \quad \mathrm{on} \quad \partial\Gamma_{\vek{X}}^{\mathrm{D}}, \label{eq:DirBCsFib}\\
	A\mult\mat{K}_{\mat{\Upsilon}}(\vek{X}) \mult \vek{M}(\vek{X}) \mult \lVert \vek{T}^{\star}(\vek{x}) \rVert &= \hat{\vek{H}}_{\mat{\Upsilon}}(\vek{X}) \quad \mathrm{on} \quad \partial\Gamma_{\vek{X}}^{\mathrm{N}}. \label{eq:NeuBCsFib}
\end{align}
The union of the set of all boundary points of the fibers is equal to the boundary contour of the surface, i.e., $\partial \Gamma = \cup_{c \in \Phi} \,\partial\Upsilon_c$ with $\Phi = (\phi_{\mathrm{min}},\phi_{\mathrm{max}})$.

\section{Coupled mechanical model in strong and weak form}\label{sec:Coupling}

\begin{figure}
	\centering
	\includegraphics[width=1.0\textwidth]{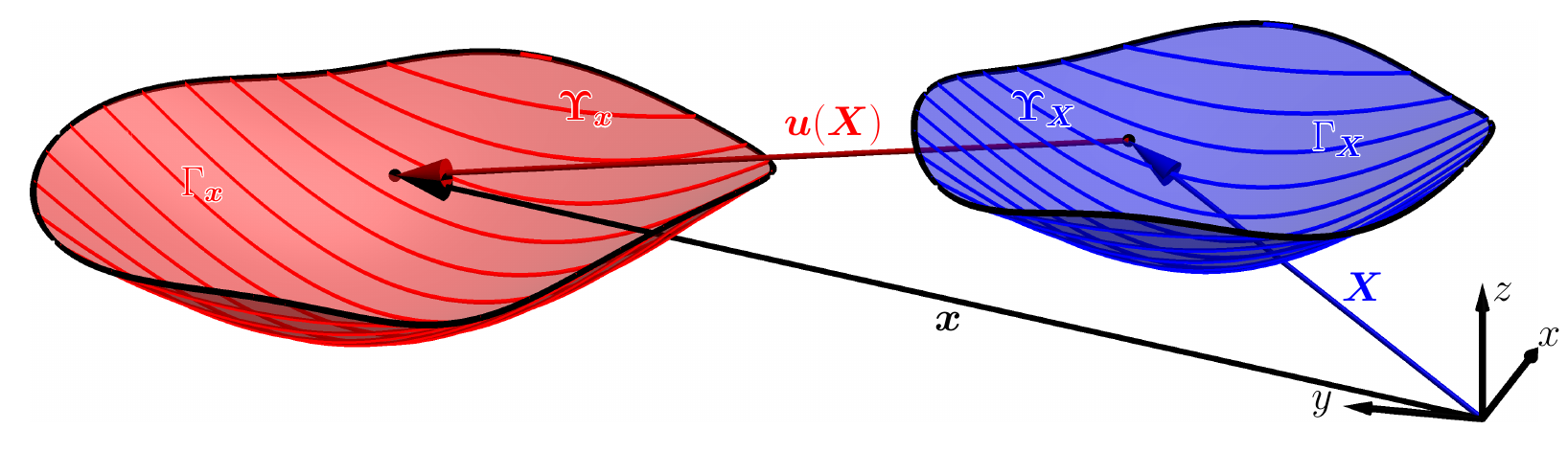}
	\captionsetup{margin=4cc}
	\caption{Visualization of the kinematic situation for fibers embedded in a membrane: Some fibers $\Upsilon_c$ of the fiber family $\mat{{\Upsilon}}$ are plotted within the surface. The situation in the deformed configuration is shown in red, while the undeformed configuration is visualized in blue.}
	\label{fig:CoupledKinematics}
\end{figure}

In this section, the \emph{coupled} mechanical model for fibers embedded in a membrane is introduced first and then, the weak form is derived which is later needed in the finite element formulation. As already described in the introduction, the homogeneously distributed fibers are cross-coupled with the membrane. This results in an innovative concept for anisotropic curved membranes. Fig.~\ref{fig:CoupledKinematics} visualizes the kinematic situation.\\
\\
The equilibrium in strong form for some membrane $\Gamma$ with $n_{\mathrm{fib}}$ embedded fiber families $\mat{\Upsilon}_i$ results by combining of Eqs.(\ref{eq:sfDefMemb}) and (\ref{eq:sfDefFib}) as
\begin{align}
	\divS\left(t\mult\vek{\sigma}_{\Gamma}(\vek{x})\right) & +\sum_{i}^{n_{\mathrm{fib}}}\divFf\left(\lambda_2^{\mat{\Upsilon}_i}\mult\lambda_3^{\mat{\Upsilon}_i}\mult\tilde{A}\mult\vek{\sigma}_{\mat{\Upsilon}_i}(\vek{x})\right) \! \mult \lVert \vek{t}_i^{\star}(\vek{x}) \rVert = \nonumber\\
	-t\mult\vek{f}_{\Gamma}(\vek{x}) & -\sum_{i}^{n_{\mathrm{fib}}}\lambda_2^{\mat{\Upsilon}_i}\mult\lambda_3^{\mat{\Upsilon}_i}\mult\tilde{A}\mult\vek{f}_{\mat{\Upsilon}_i}(\vek{x})\mult \lVert \vek{t}_i^{\star}(\vek{x}) \rVert \quad \forall \vek{x} \in \Gamma_{\vek{x}}. \label{eq:sfDefFCoupled}
\end{align}
Note that, for brevity, we assume a fiber cross section $\tilde{A} = B \mult T$ with a width  of unity, i.e., $B=1$, and the membrane thickness $T$ in the reference configuration for the continuously embedded fibers. To choose $\tilde{A} = T$ is justified by compatibility between the continuously embedded fibers and the membrane. Thereby, the addends in Eq.~(\ref{eq:sfDefFCoupled}) feature consistent units. The second and third principal stretches of the fibers, i.e., $\lambda_2^{\mat{\Upsilon}_i}$ and $\lambda_3^{\mat{\Upsilon}_i}$, respectively, are considering the cross section area stretch in the equilibrium equation with respect to the undeformed configuration.\\
\\
The equilibrium expressed in quantities of the undeformed configuration, which is used in the finite element formulation introduced below, results as
\begin{align}
	T\mult\DivS \mat{K}_{\Gamma}(\vek{X}) &+ \sum_{i}^{n_{\mathrm{fib}}} \tilde{A}\mult\DivFf \mat{K}_{\mat{\Upsilon}_i}(\vek{X}) \mult \lVert \vek{T}_i^{\star}(\vek{X}) \rVert	
	 = \nonumber\\
	 -T\mult\vek{F}_{\Gamma}(\vek{X}) &- \sum_{i}^{n_{\mathrm{fib}}} \tilde{A}\mult\vek{F}_{\mat{\Upsilon}_i}(\vek{X}) \mult \lVert \vek{T}_i^{\star}(\vek{X}) \rVert \quad \forall \vek{X} \in \Gamma_{\vek{X}}. \label{eq:sfUndefCoupled}
\end{align}
The field equations of the BVP modelling the mechanics of fiber-embedded membranes to be fulfilled for all $\vek{X} \in \Gamma_{\vek{X}}$ is now given by (i) the kinematics, i.e., the strain tensors $\mat{E}_{\Gamma}(\vek{X})$ in Eq.~(\ref{eq:GLSTmemb}) and  $\mat{E}_{\mat{\Upsilon}_i}(\vek{X})$ in Eq.~(\ref{eq:GLSTfib}), (ii) the constitutive equations, i.e, the stress tensors $\mat{K}_{\Gamma}(\vek{X})$ in Eq.~(\ref{eq:1PKmemb}) and $\mat{K}_{\mat{\Upsilon}}(\vek{X})$ in Eq.~(\ref{eq:1PKfib}), the equilibrium in Eq.~(\ref{eq:sfUndefCoupled}), and (iv) the boundary conditions $\hat{\vek{G}}(\vek{X})$, i.e., Eqs.~(\ref{eq:DirBCsMemb}) and (\ref{eq:DirBCsFib}), and $\hat{\vek{H}}(\vek{X}) = \hat{\vek{H}}_{\Gamma}(\vek{X}) + \hat{\vek{H}}_{\vek{\Upsilon}_i}(\vek{X})$ from Eqs.~(\ref{eq:NeuBCsMemb}) and (\ref{eq:NeuBCsFib}).\\
\\
For the continuous weak form, we introduce the following function spaces 
\begin{align}
	\mathcal{S}_{\vek{u}} &= \big\{\vek{v} \in \left[\mathcal{H}^1(\Gamma_{\vek{X}})\right]^3:\vek{v} = \hat{\vek{G}}\,\,\mathrm{on}\,\, \partial \Gamma_{\vek{X}}^{\mathrm{D}}\big\},\\
	\mathcal{V}_{\vek{u}} &= \big\{\vek{v} \in \left[\mathcal{H}^1(\Gamma_{\vek{X}})\right]^3:\vek{v} = \vek{0}\,\,\mathrm{on}\,\, \partial \Gamma_{\vek{X}}^{\mathrm{D}}\big\}.
\end{align}
Herein, $\mathcal{H}^1$ is the Sobolev space of functions with square-integrable first derivatives. With the integral theorems of Eqs.~(\ref{eq:intThSrf}) to (\ref{eq:intThEmbdFib}), we can formulate the weak form of the BVP w.r.t.~the deformed configuration. For fiber families $\vek{\Upsilon}_i$ with $i\in\{1,2,\ldots,n_\mathrm{fib}\}$, the task is to find $\vek{u} \in \mathcal{S}_{\vek{u}}$ such that for all $\vek{w} \in \mathcal{V}_{\vek{u}}$, there holds in the undeformed configuration,
\begin{align}
	&\int_{\Gamma_{\vek{X}}}\!\!\!T\mult(\GradS^{\mathrm{dir}}\,\vek{w}: \mat{K}_{\Gamma}(\vek{u})) +\sum_{i}^{n_{\mathrm{fib}}}\tilde{A}\mult\GradFf^{\mathrm{dir}}\,\vek{w}:\mat{K}_{\mat{\Upsilon}_i}(\vek{u})) \! \mult \lVert \vek{T}_i^{\star}(\vek{X}) \rVert \, \d\Gamma = \nonumber\\
	&\int_{\Gamma_{\vek{X}}}\!\!\! T\mult\vek{w} \cdot \vek{F}_{\Gamma}(\vek{X}) +\sum_{i}^{n_{\mathrm{fib}}}\tilde{A}\mult\vek{w} \cdot \vek{F}_{\mat{\Upsilon}_i}(\vek{X})\mult \lVert \vek{T}_i^{\star}(\vek{X}) \rVert\,\d\Gamma + \int_{\partial\Gamma_{\vek{X}}^\mathrm{N}}\!\!\! \vek{w} \cdot \hat{\vek{H}}\,\d\partial\Gamma. \label{eq:wfUndefFCoupled}
\end{align}
This weak form is used in the finite element implementation, discussed in the next section.

\section{Discretization and finite element implementation} \label{sec:DiscrFEM}
In this section, we state the discretized weak form and briefly discuss implementational aspects in the context of anisotropic two-dimensional, curved membranes with embedded fibers in the applied Bulk Trace FEM framework. For more technical details on the general consideration of the substructures (herein fibers) in a bulk domain, we refer to section 4.3 in our previous work \cite{Fries_2023a}.

\begin{figure}
	\centering
	\includegraphics[width=0.5\textwidth]{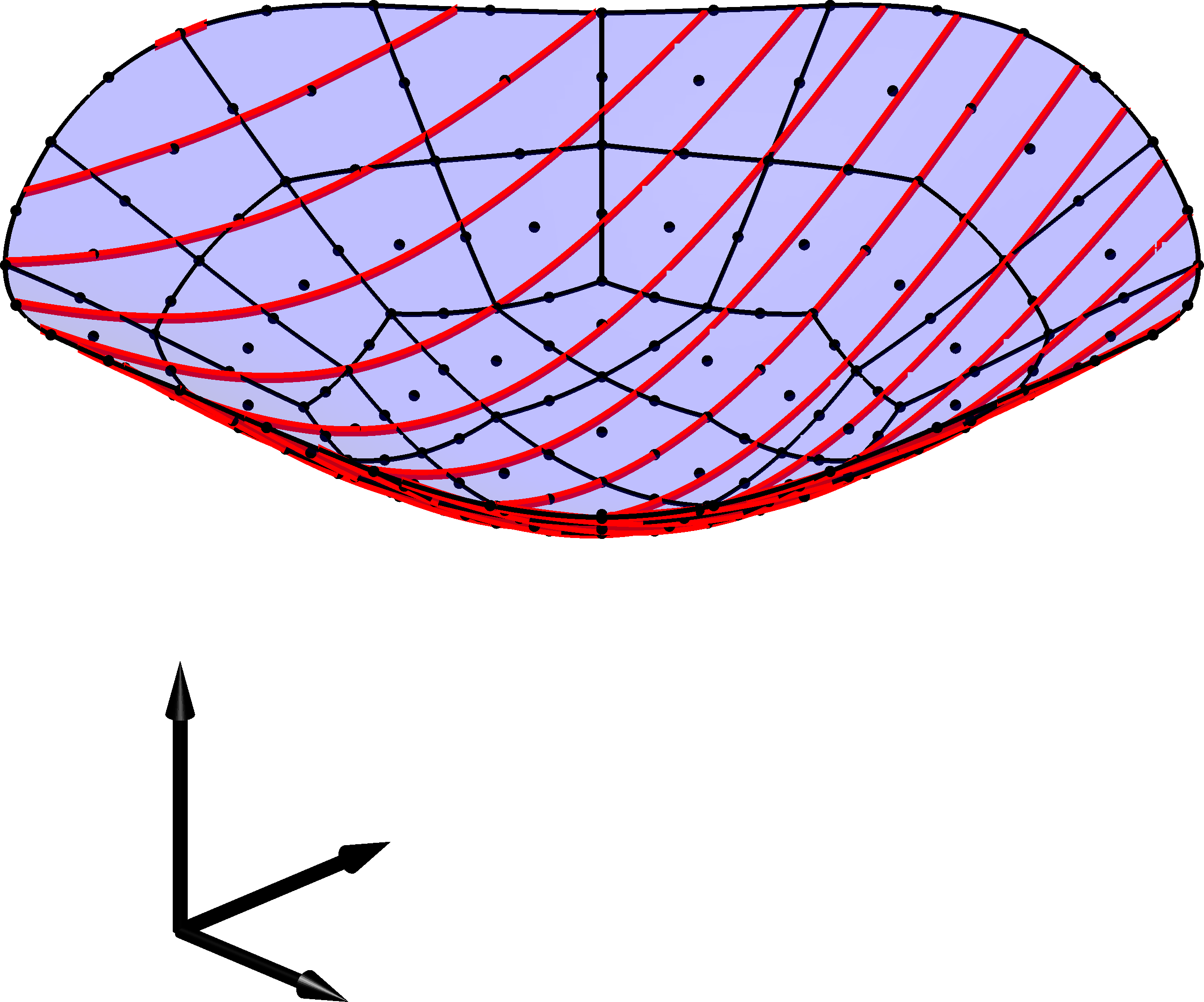}
	\captionsetup{margin=4cc}
	\caption{Some arbitrarily selected fibers $\Upsilon_c$ of a fiber family $\mat{\Upsilon}$ embedded in a membrane surface which is discretized by second-order  quadrilateral elements of Lagrange type. Note that the fibers do not have to be aligned with the element edges.}
	\label{fig:FibreMesh}
\end{figure}

The membrane's surface $\Gamma_{\vek{X}}$ in the undeformed configuration is discretized \emph{explicitly} by two-dimensional quadrilateral or triangular, higher-order Lagrange-type elements of order $p$. The embedded fibers are completely independent of the elements, i.e., not aligned with element edges, see Fig.~\ref{fig:FibreMesh}. The nodal coordinates of the elements used in the discretization are labelled $\vek{X}_i$ with $i \in \{1,2,\ldots n_q\}$ and $n_q$ being the total number of nodes in the mesh. The resulting \emph{nodal basis functions} $B_i(\vek{X})$ span a $C^0$-continuous finite element space as
\begin{equation}
	\mathcal{Q}_{\Gamma_{\vek{X}}}^h := \bigg\{v^h \in C^0(\Gamma_{\vek{X}}^h):v^h=\sum_{i=1}^{n_q} B_i(\vek{X}) \mult \hat{v}_i \,\mathrm{with}\,\hat{v}_i \in \mathbb{R} \bigg\} \subset \mathcal{H}^1(\Gamma_{\vek{X}}^h).
\end{equation}
Based on this, the discrete function spaces for test and trial functions are
\begin{align}
	\mathcal{S}_{\vek{u}}^h &= \big\{\vek{v}^h \in \left[\mathcal{Q}_{\Gamma_{\vek{X}}}^h\right]^3:\vek{v}^h = \hat{\vek{G}}\,\,\mathrm{on}\,\, \partial \Gamma_{\vek{X}}^{\mathrm{D},h}\big\},\\
	\mathcal{V}_{\vek{u}}^h &= \big\{\vek{v}^h \in \left[\mathcal{Q}_{\Gamma_{\vek{X}}}^h\right]^3:\vek{v}^h = \vek{0}\,\,\mathrm{on}\,\, \partial \Gamma_{\vek{X}}^{\mathrm{D},h}\big\}.
\end{align}
The \emph{discrete} formulation of Eq.~(\ref{eq:wfUndefFCoupled}) results as: Given $n_{\mathrm{fib}}$ fiber families $\mat{\Upsilon}_i$, body forces $\vek{F}_{\Gamma}^h  \in \mathbb{R}^3$ on $\Gamma_{\vek{X}}$ for the membrane and $\vek{F}_{\mat{\Upsilon}_i}^h \in \mathbb{R}^3$ on $\Gamma_{\vek{X}}$ for the fibers, tractions $\hat{\vek{H}}^h \in \mathbb{R}^3$ (combined for the membrane and the fibers), find the displacement field $\vek{u}^h \in \mathcal{S}_{\vek{u}}^h$ such that for all test functions $\vek{w}^h \in \mathcal{V}_{\vek{u}}^h$, there holds in $\Gamma_{\vek{X}}^h$
\begin{align}
	\int_{\Gamma_{\vek{X}}^h}\!\!\!T\mult(\GradS^{\mathrm{dir}}\,\vek{w}^h &: \mat{K}_{\Gamma}(\vek{u}^h)) +\sum_{i}^{n_{\mathrm{fib}}}\tilde{A}\mult\GradFf^{\mathrm{dir}}\,\vek{w}^h :\mat{K}_{\mat{\Upsilon}_i}(\vek{u}^h)) \! \mult \lVert \vek{T}_i^{\star}\rVert \, \d\Gamma = \nonumber\\
	\int_{\Gamma_{\vek{X}}^h}\!\!\! T\mult\vek{w}^h \cdot \vek{F}_{\Gamma}^h &+\sum_{i}^{n_{\mathrm{fib}}}\tilde{A}\mult\vek{w}^h \cdot \vek{F}_{\mat{\Upsilon}_i}^h\mult \lVert \vek{T}_i^{\star}\rVert\,\d\Gamma + \int_{\partial\Gamma_{\vek{X}}^\mathrm{N,h}}\!\!\! \vek{w}^h \cdot \hat{\vek{H}}^h\,\d\partial\Gamma. \label{eq:wfDiscUndefFCoupled}
\end{align}
Note that although also geometric quantities (such as the surface, normal vectors etc.) and differential operators (such as gradients and divergence) refer to the discrete mesh, we omit the superscript $h$ for these quantities. For example, we do not write $\vek{T}_i^{\star,h}$ or $\GradFf^{\mathrm{dir},h}\,\bullet$ but rather $\vek{T}_i^{\star}$ or $\GradFf^{\mathrm{dir}}\,\bullet$. The sought displacement field $\vek{u}^h(\vek{X})$ is found by solving the non-linear system of equations for the $n_{\mathrm{DOF}} = 3 \mult n_q$ degrees of freedom by applying the Newton-Raphson method, as usual in the finite strain theory.

\section{Numerical examples}\label{sec:NumTCs}

Some numerical test cases of anisotropic membranes, modelled and simulated using the proposed methods, are shown in this section. Herein, these test cases are rather academic and real world applications are beyond the scope of this paper and subject to further research, see Sec.~\ref{sec:CaO}. To confirm the success of the proposed methods, different error measures are evaluated. The first one is the so-called \emph{residual error} $\varepsilon_{\mathrm{res}}$, which was successfully used in previous works for PDEs on manifolds, e.g., \cite{Fries_2020a,Schoellhammer_2019b,Schoellhammer_2021a,Fries_2023a,Kaiser_2024a}. For this error measure, the error in the equilibrium in strong form is integrated over the $n_{\mathrm{el}}$ elements of the deformed, discretized membrane, i.e.,
\begin{equation}
	\varepsilon_{\mathrm{res}} = \sqrt{\sum_{i=1}^{n_{\mathrm{el}}} \int_{\Gamma_{\vek{x}}^{\mathrm{el}}} \mathfrak{r}(\vek{u}) \cdot \mathfrak{r}(\vek{u}) \,\d\Gamma}, \label{eq:resError}
\end{equation}
where the residual is defined as
\begin{equation}
	\mathfrak{r}(\vek{u}) = \divS\left(t\mult\vek{\sigma}_{\Gamma}\right) +\sum_{i}^{n_{\mathrm{fib}}}\divFf\left(\lambda_2^{\mat{\Upsilon}_i}\mult\lambda_3^{\mat{\Upsilon}_i}\mult\tilde{A}\mult\vek{\sigma}_{\mat{\Upsilon}_i}\right) \! \mult \lVert \vek{t}_i^{\star} \rVert +
	t\mult\vek{f}_{\Gamma} +\sum_{i}^{n_{\mathrm{fib}}}\lambda_2^{\mat{\Upsilon}_i}\mult\lambda_3^{\mat{\Upsilon}_i}\mult\tilde{A}\mult\vek{f}_{\mat{\Upsilon}_i}\mult \lVert \vek{t}_i^{\star} \rVert, \label{eq:residuum}
\end{equation}
using the strong form of the coupled equilibrium from Eq.~(\ref{eq:sfDefFCoupled}). For the analytical solution, this error vanishes. Note that the element-wise integration in Eq.~(\ref{eq:resError}) is necessary due to the second-order derivatives in the residual, hence, element boundaries where the used $C^0$-continuous shape functions feature jumps are excluded in the evaluation of this error measure. The expected order of convergence is $\mathcal{O}(p-1)$ due to the second-order derivatives in the residual.\\
\\
Another useful error measure is the \emph{stored-energy error}, see, e.g., \cite{Zienkiewicz_2013a,Zienkiewicz_2014a,Fries_2023a,Neumeyer_2025a}, defined as
\begin{equation}
	\varepsilon_{\mathfrak{e}}(\vek{u}) = \lvert W_{tot}(\vek{u}) - W_{tot}(\vek{u}^h) \rvert.
\end{equation}
Using the strain energy functions, the total energy of the coupled model is evaluated as
\begin{equation}
	W_{tot}(\vek{u}) = \int_{\Gamma_{\vek{X}}} T \mult W_{\Gamma}(\lambda_1^{\Gamma},\lambda_2^{\Gamma}) + \sum_{i}^{n_{\mathrm{fib}}} \mult \tilde{A}\mult W_{\mat{{\Upsilon}}}(\lambda_1^{\mat{\Upsilon}})\mult\lVert\vek{T}_i^{\star}\rVert \, \d\Gamma,
\end{equation}
where $T$ is the thickness of the undeformed membrane and $\vek{T}_i^\star$ represents the tangent vectors along the fibers, see Eq.~(\ref{eq:normedT}). If no analytical solution is available for the reference energy, either an overkill approximation obtained from a very fine discretization with higher-order elements or a manufactured solution may be used.

\subsection{Membrane with different embedded fiber families} \label{subsec:TC1}
\begin{figure}
	\subfigure[fiber family $\mat{\Upsilon}_{1}$]{\includegraphics[width=0.3\textwidth]{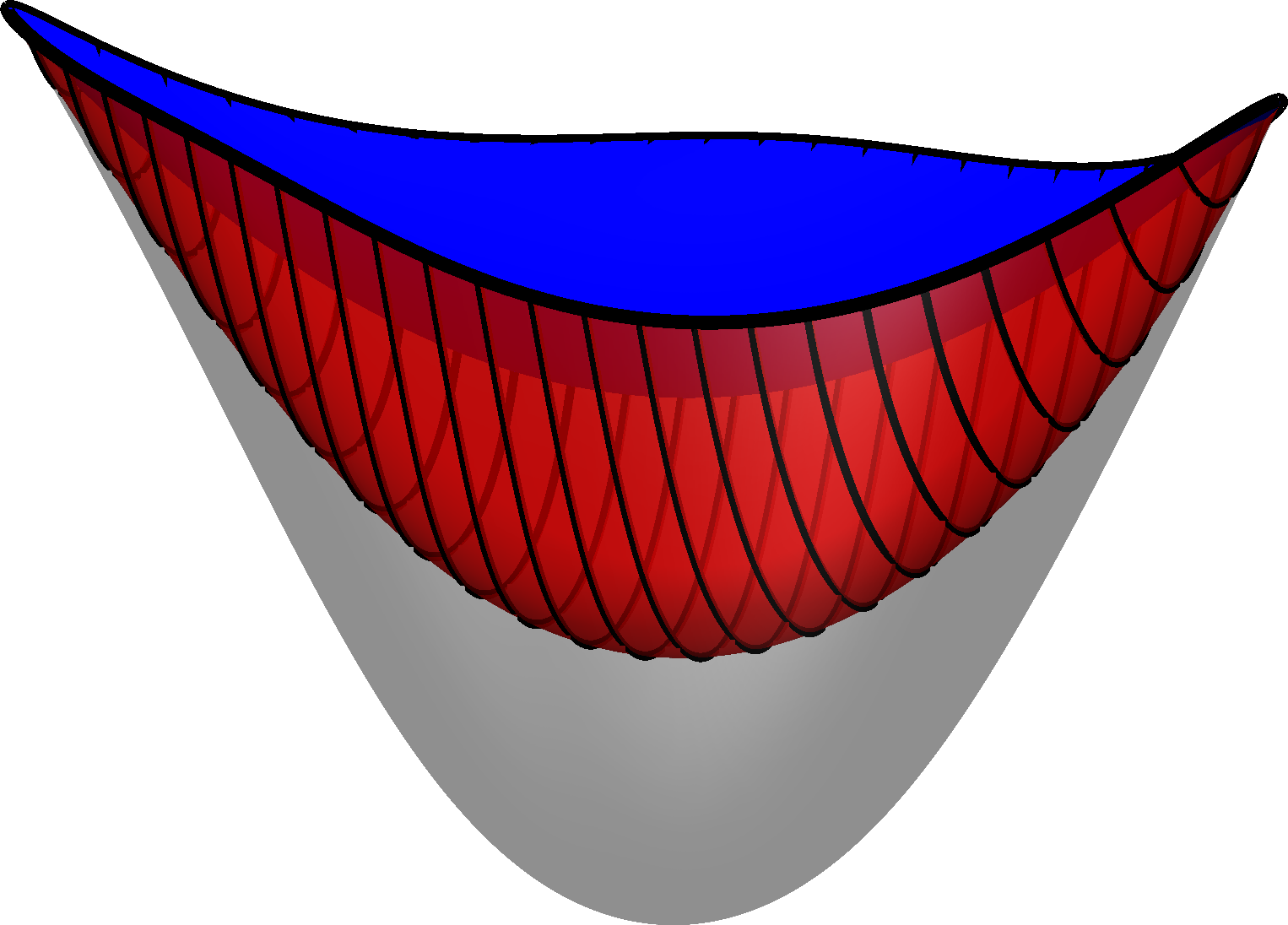}}\hfill
	\subfigure[fiber family $\mat{\Upsilon}_{2}$]{\includegraphics[width=0.3\textwidth]{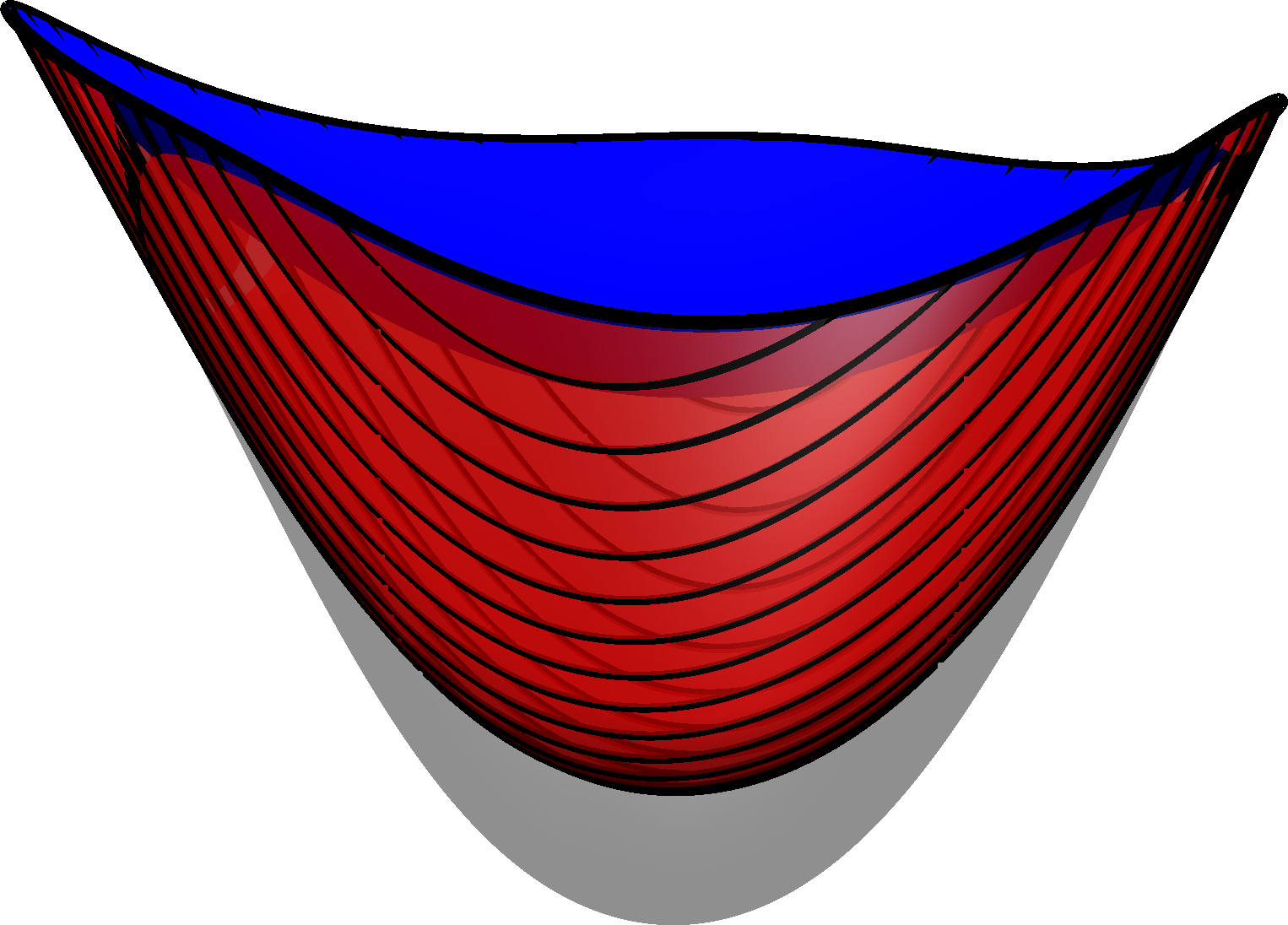}}\hfill
	\subfigure[$\mat{\Upsilon}_{1}$ and $\mat{\Upsilon}_{2}$ together]{\includegraphics[width=0.3\textwidth]{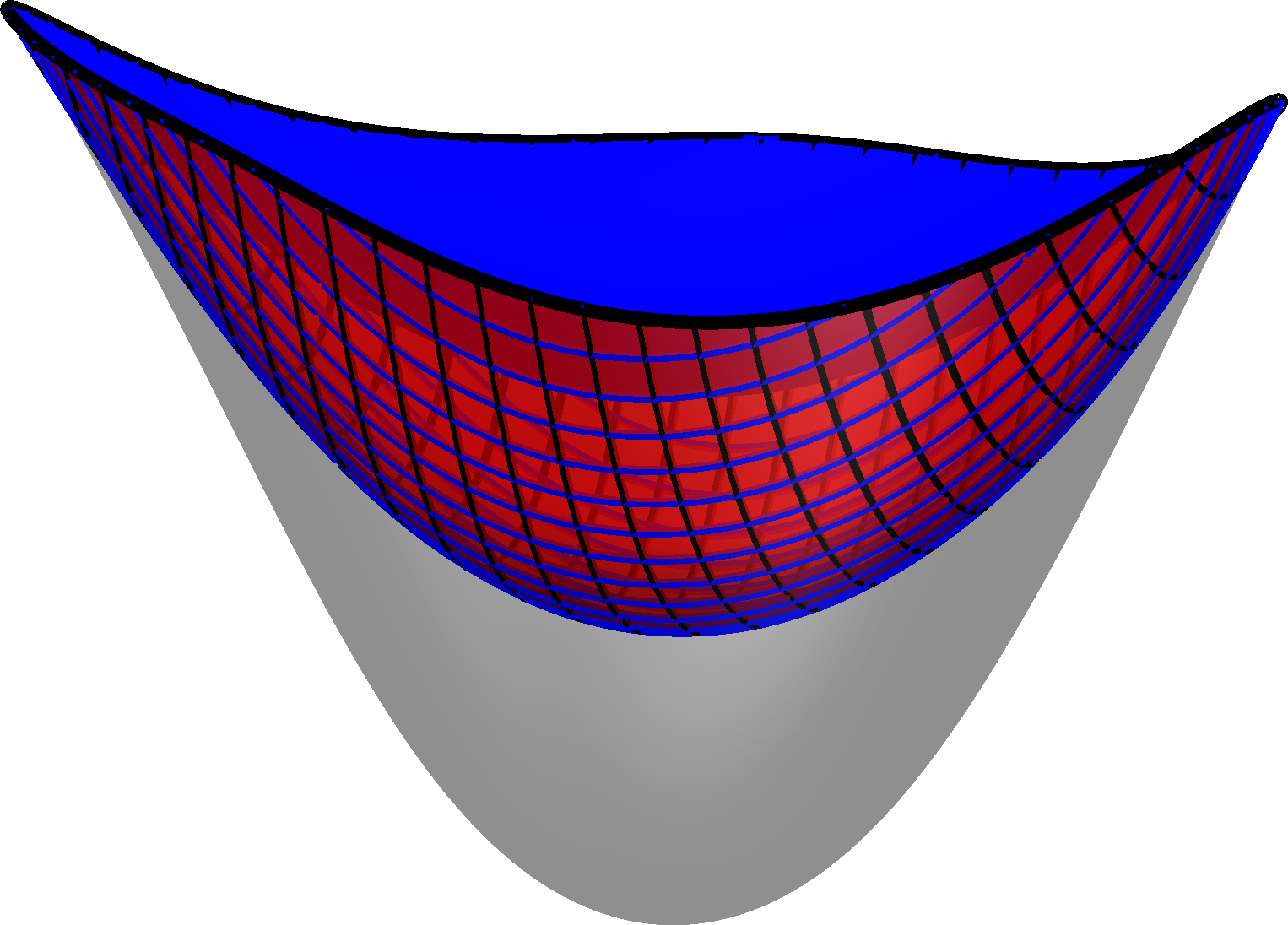}}
	\subfigure[deformation magn. $\mat{\Upsilon}_{1}$]{\includegraphics[width=0.3\textwidth]{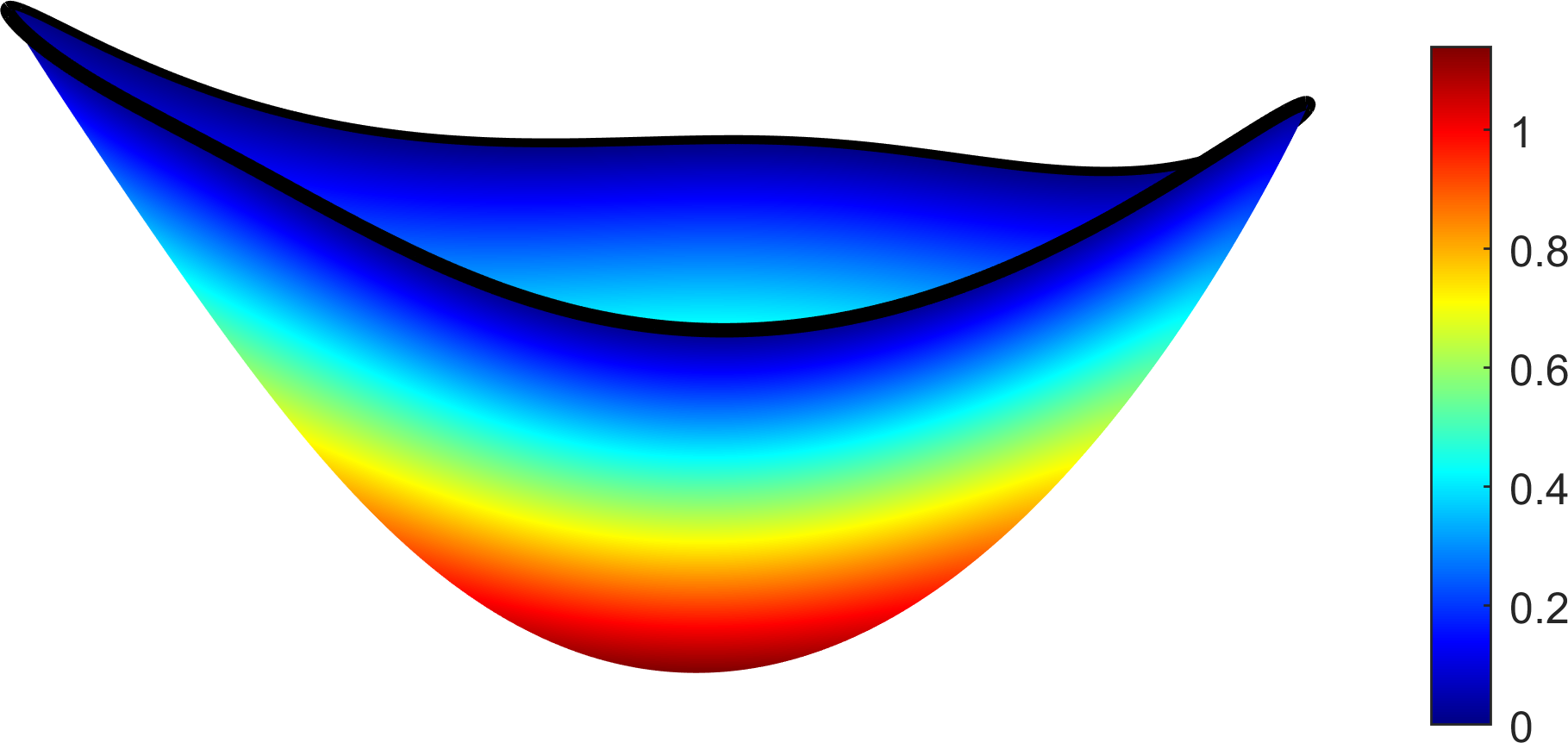}}\hfill
	\subfigure[deformation magn. $\mat{\Upsilon}_{2}$]{\includegraphics[width=0.3\textwidth]{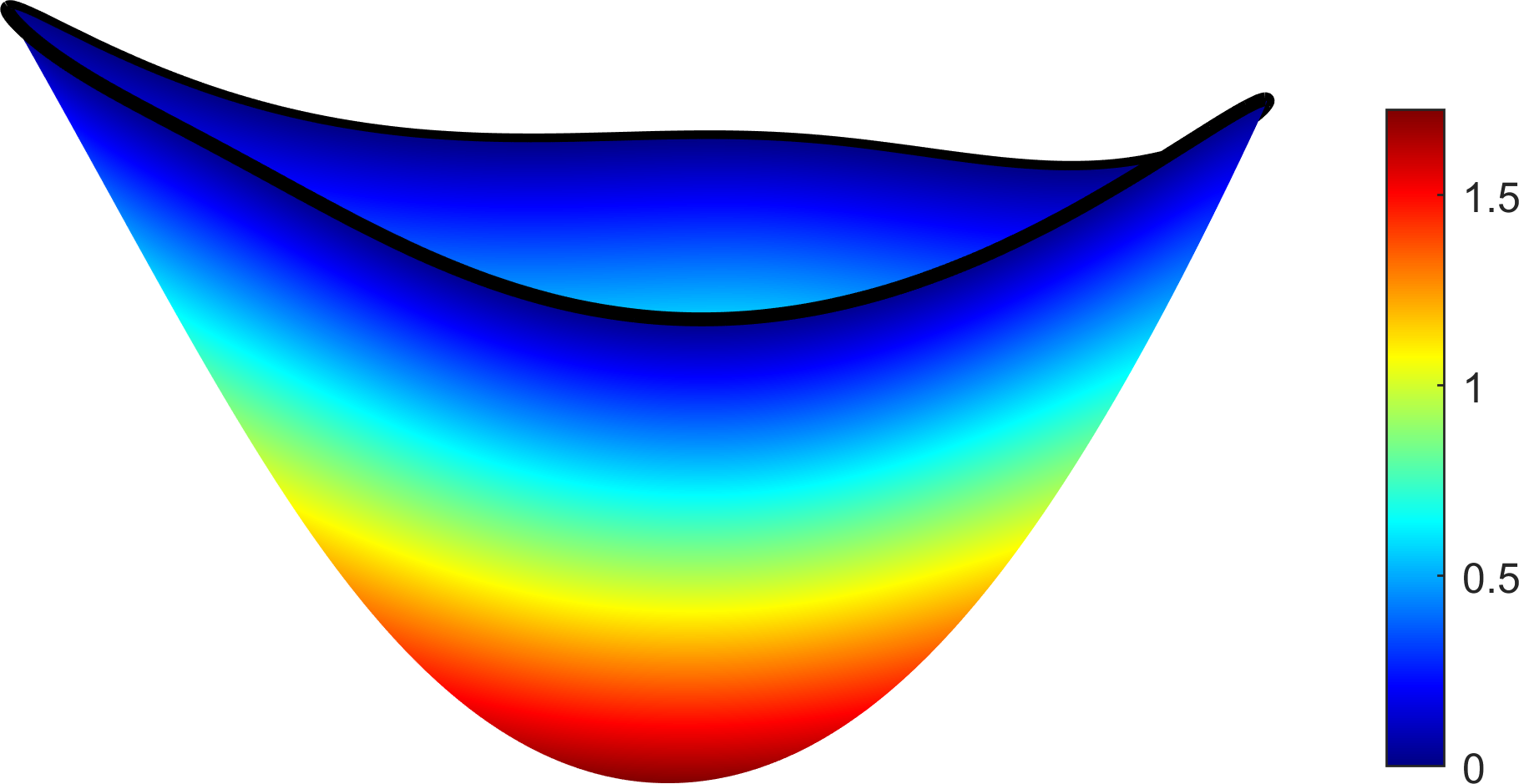}}\hfill
	\subfigure[def. magn. $\mat{\Upsilon}_{1}$ and $\mat{\Upsilon}_{2}$]{\includegraphics[width=0.3\textwidth]{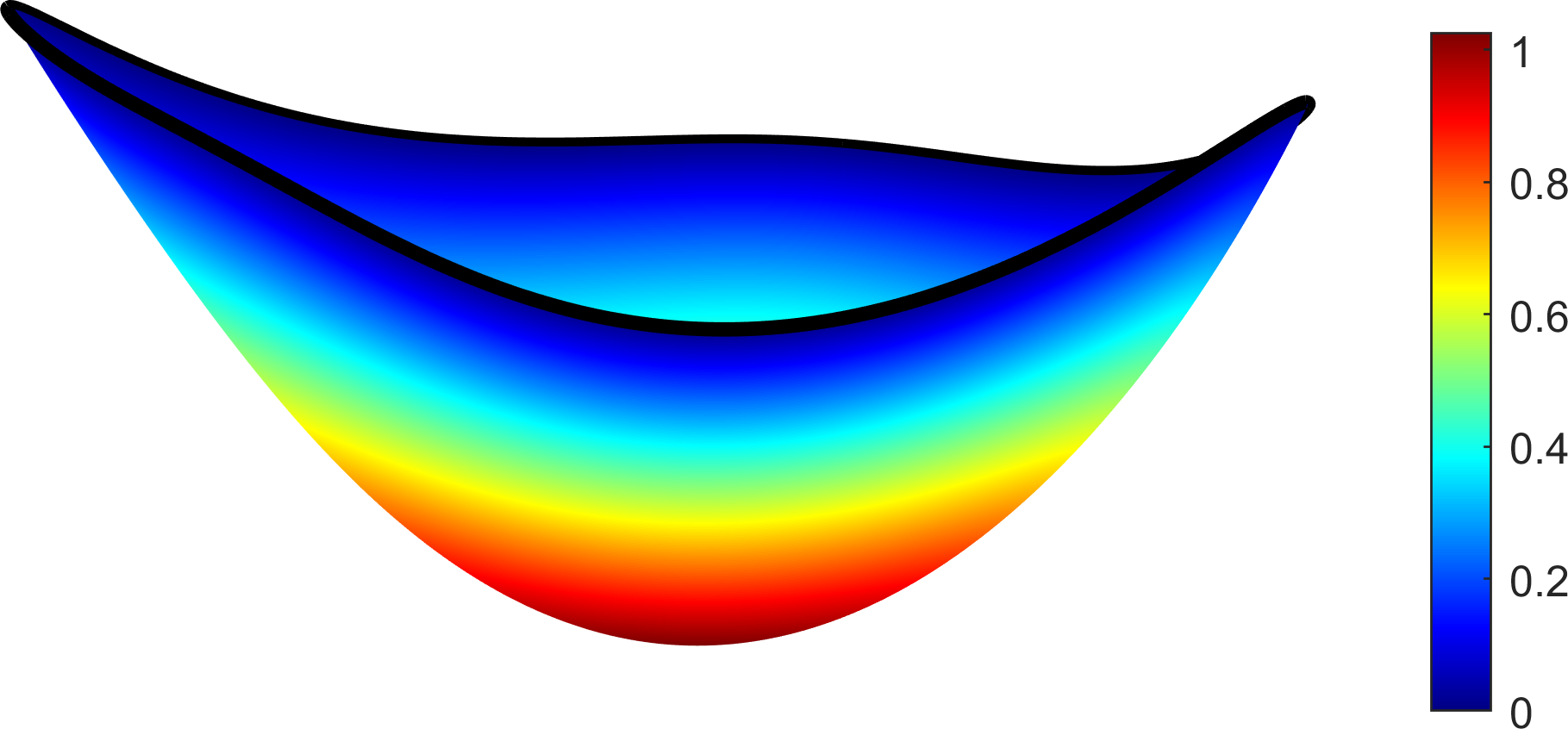}}
	\subfigure[von Mises stress $\mat{\Upsilon}_{1}$]{\includegraphics[width=0.3\textwidth]{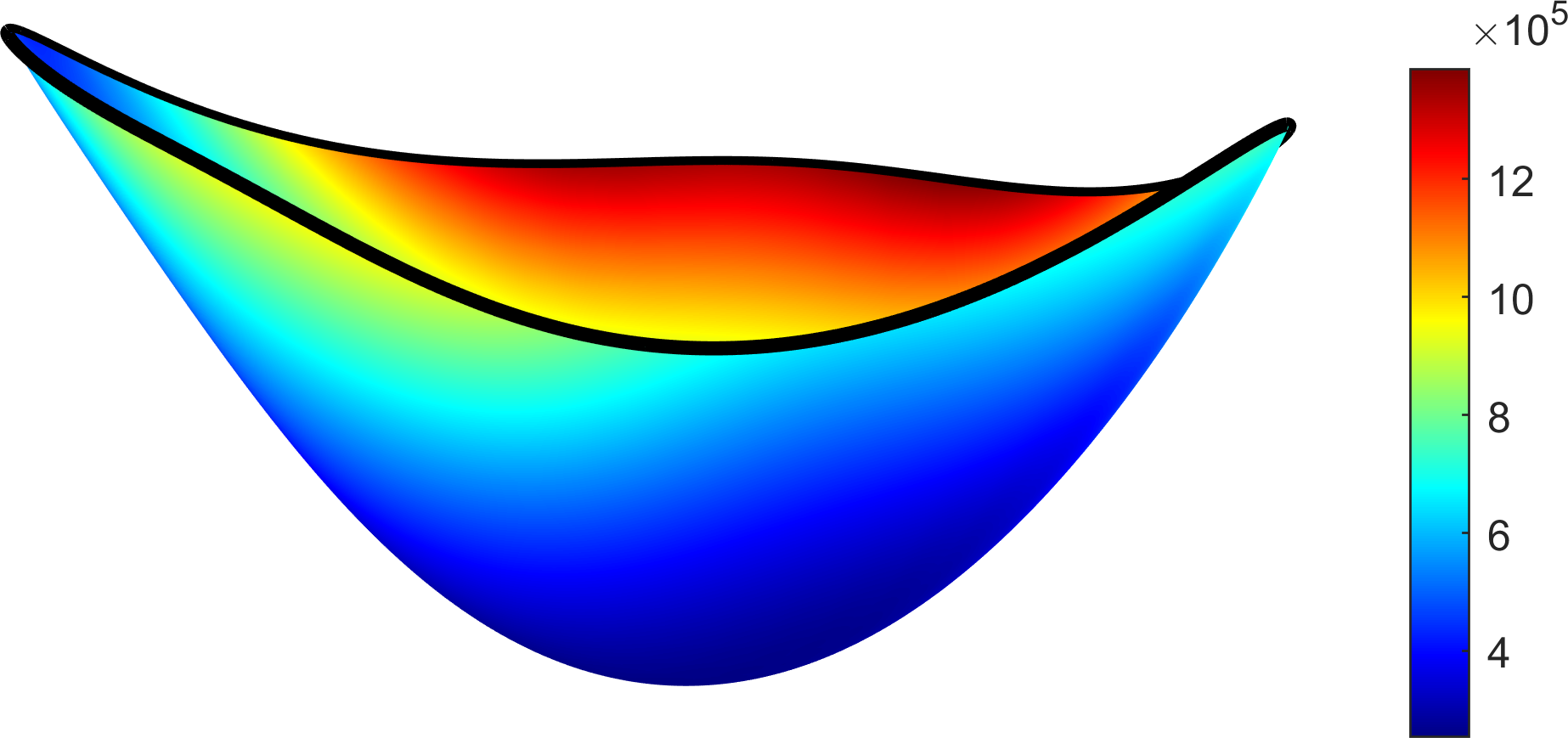}}\hfill
	\subfigure[von Mises stress $\mat{\Upsilon}_{2}$]{\includegraphics[width=0.3\textwidth]{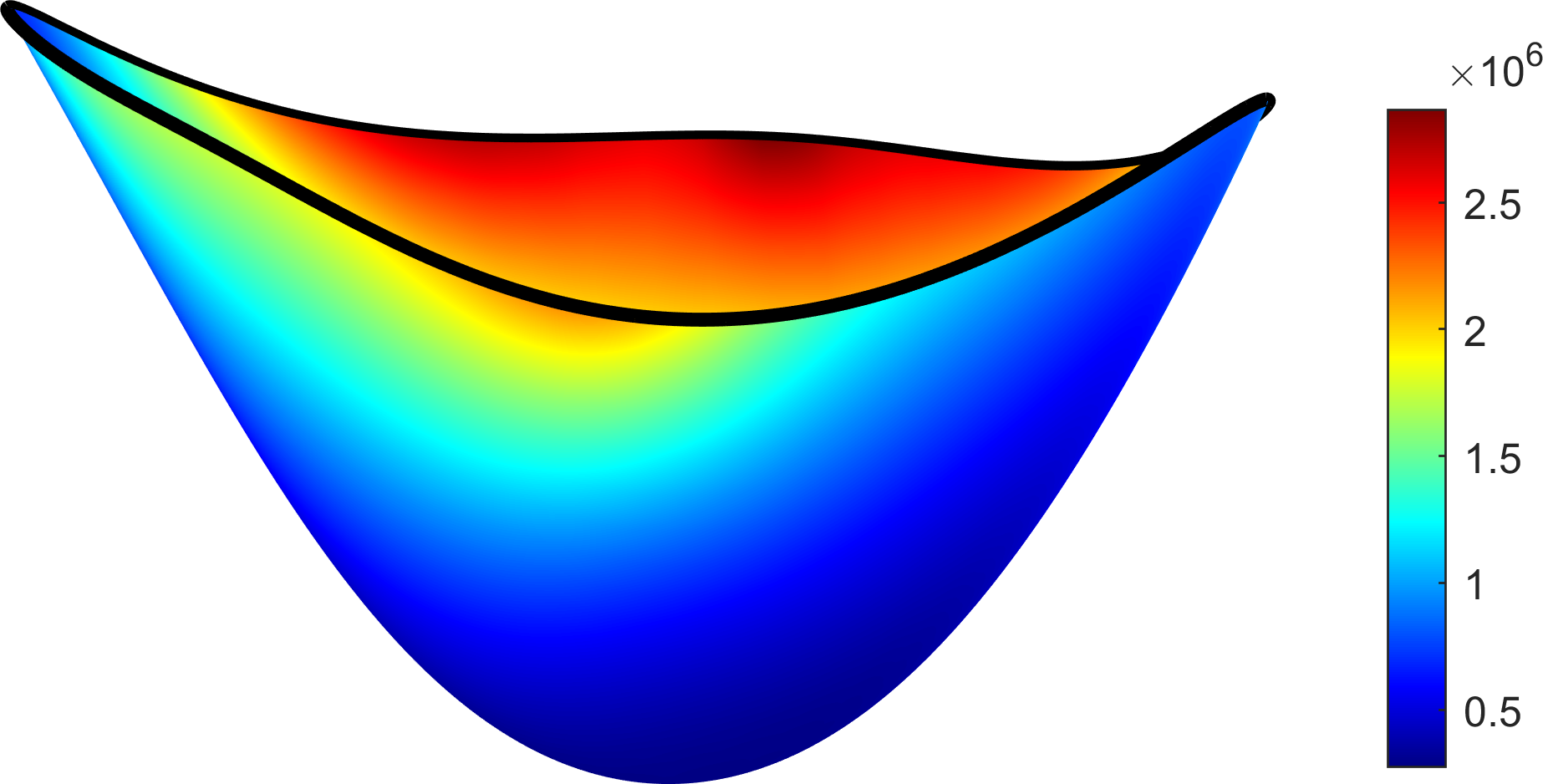}}\hfill
	\subfigure[von Mises stress $\mat{\Upsilon}_{1} + \mat{\Upsilon}_{2}$]{\includegraphics[width=0.3\textwidth]{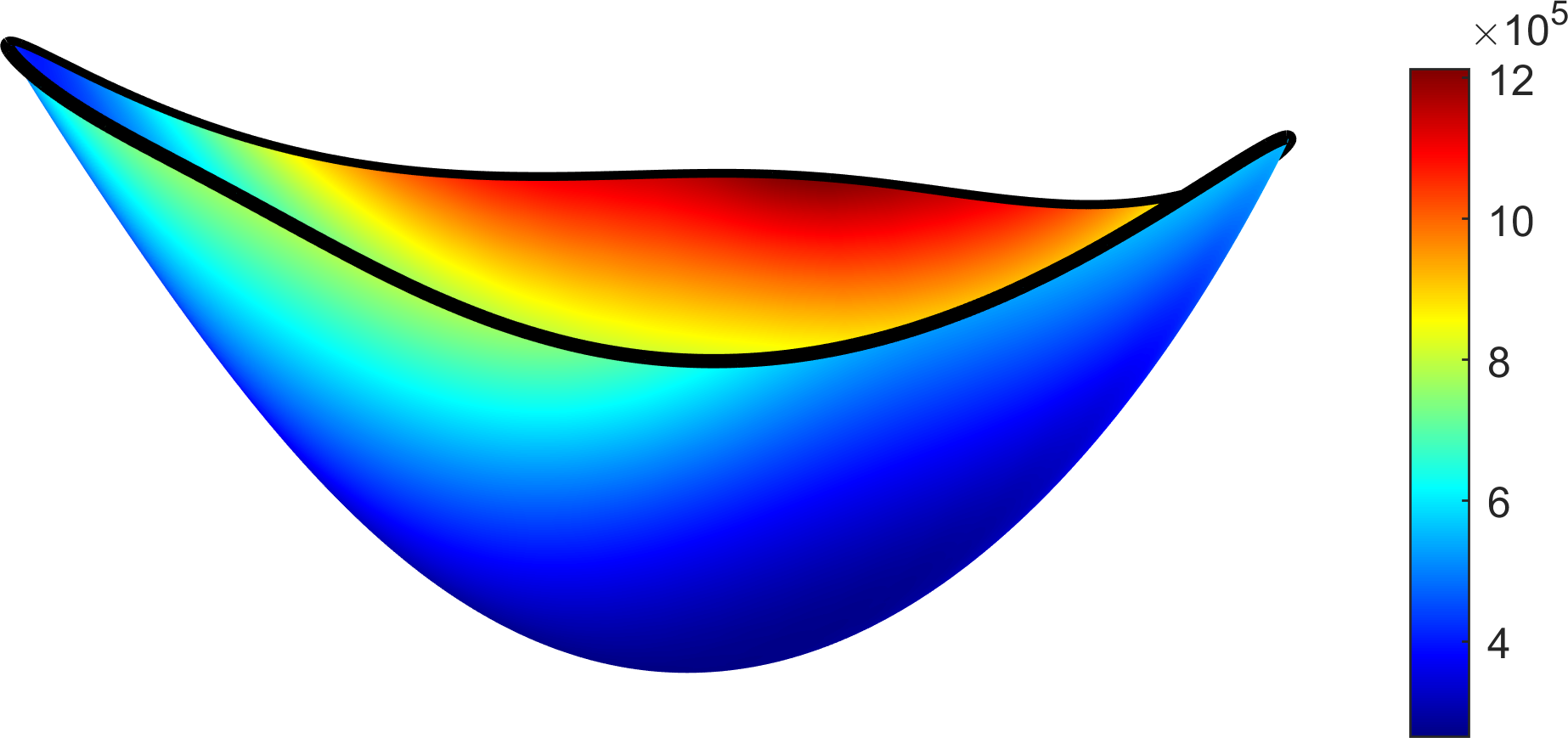}}
	\caption{Results of test case 1, where the first column, i.e., figures (a), (d), and (g), shows the results for the membrane with embedded fiber family 1, the second column, i.e., figures (b), (e), and (h), shows the results for the membrane with embedded fiber family 2, and the third column, i.e., figures (c), (f), and (i), shows the results for the fiber families 1 and 2 embedded together.}
	\label{fig:TC1-ResVis}
\end{figure}

\begin{figure}
	\subfigure[fiber family $\mat{\Upsilon}_{1}$]{\includegraphics[width=0.3\textwidth]{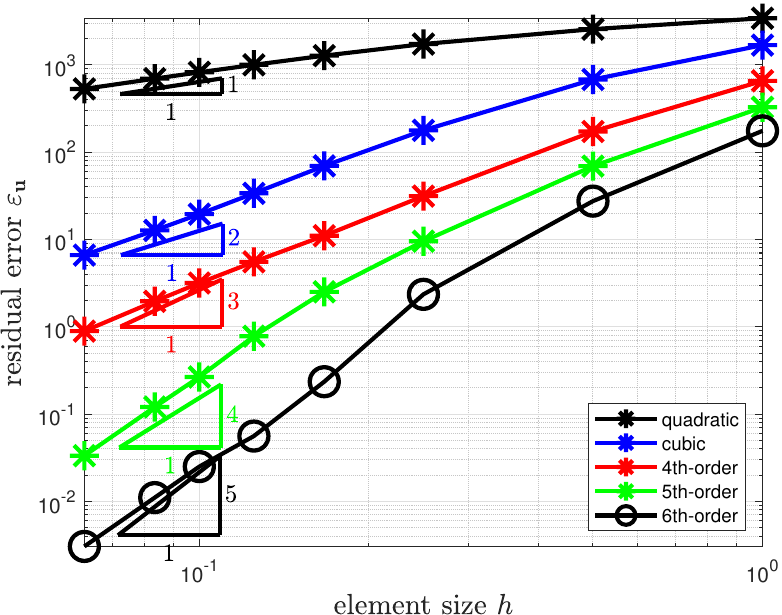}}\hfill
	\subfigure[fiber family $\mat{\Upsilon}_{2}$]{\includegraphics[width=0.3\textwidth]{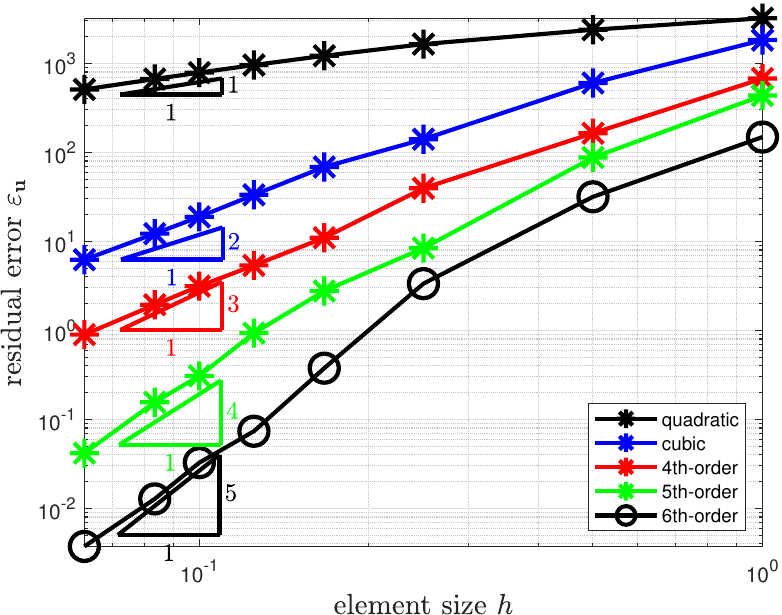}}\hfill
	\subfigure[$\mat{\Upsilon}_{1}$ and $\mat{\Upsilon}_{2}$ together]{\includegraphics[width=0.3\textwidth]{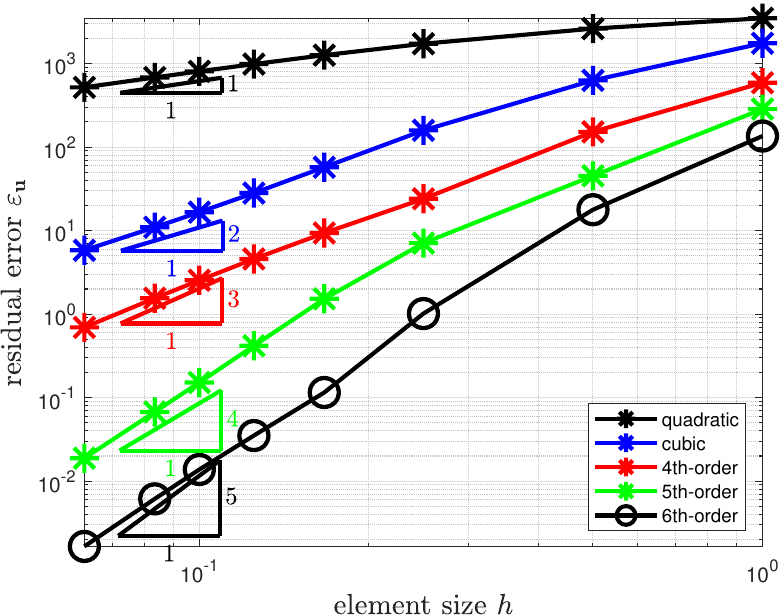}}
	\subfigure[fiber family  $\mat{\Upsilon}_{1}$]{\includegraphics[width=0.3\textwidth]{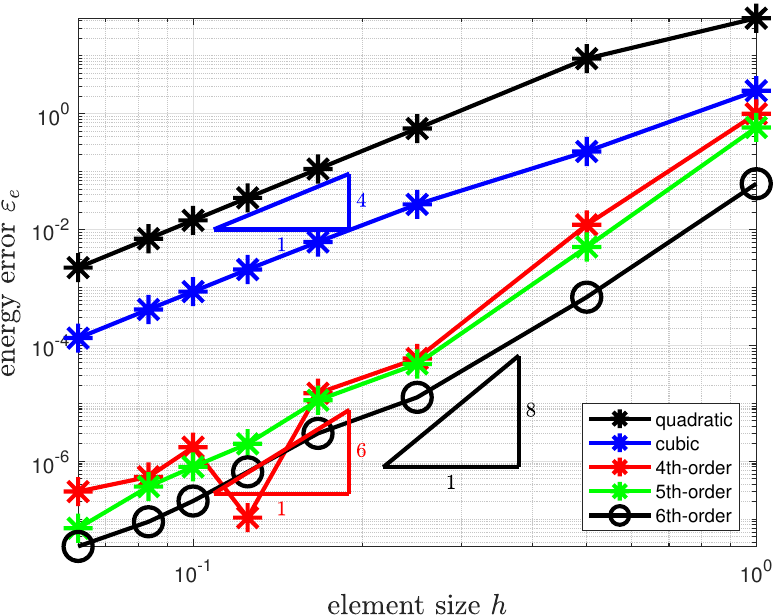}}\hfill
	\subfigure[fiber family  $\mat{\Upsilon}_{2}$]{\includegraphics[width=0.3\textwidth]{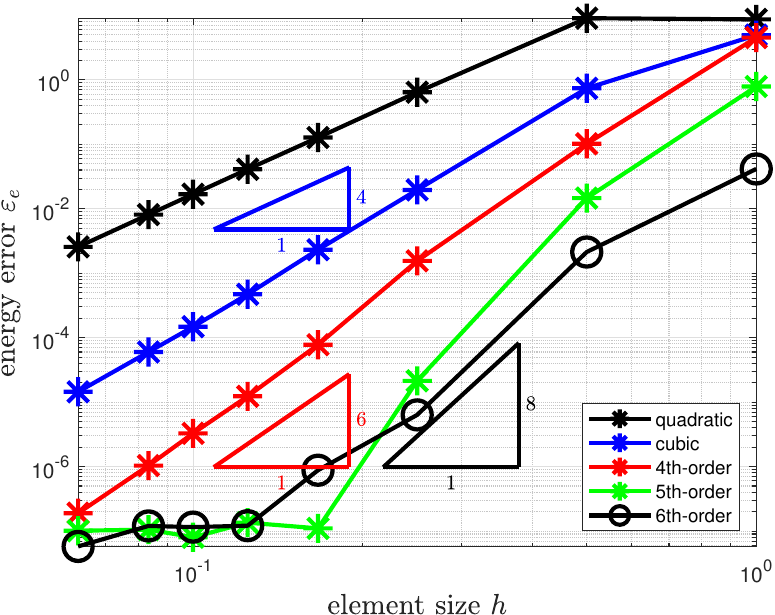}}\hfill
	\subfigure[$\mat{\Upsilon}_{1}$ and $\mat{\Upsilon}_{2}$ together]{\includegraphics[width=0.3\textwidth]{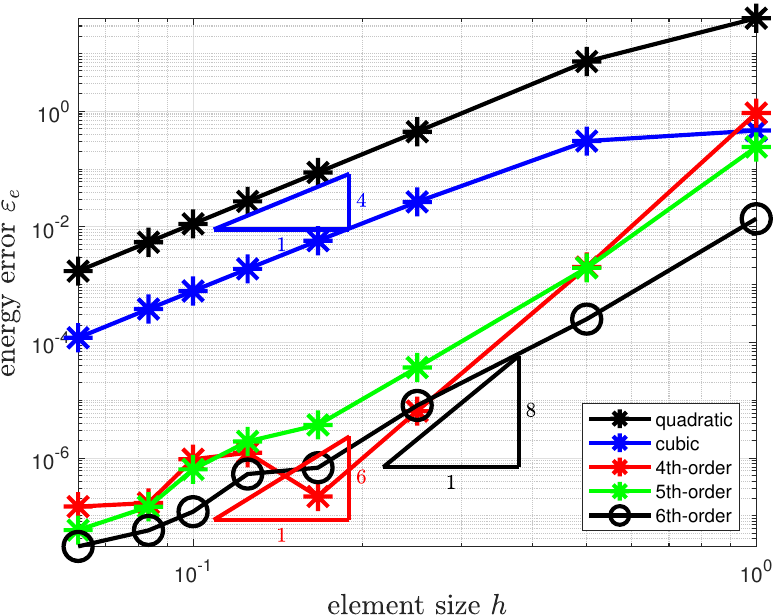}}
	\caption{Convergence results in (a) to (c) the residual errors and in (d) to (f) the stored-energy errors for a membrane with different embedded fiber families.}
	\label{fig:TC1-ResConv}
\end{figure}

The membrane geometry $\Gamma$ of the first example results from mapping a unit circle in two-dimensions to a curved surface by
\begin{equation*}
	\varphi = \vek{x}(\vek{r}) = \begin{cases}
		X(\vek{r})= 2 + r - 2 \mult s - 0.2 \mult r \mult s + 0.75 \mult \sin(2 \mult r+0.3),\\
		Y(\vek{r})= 1 + r + s + 0.5 \mult r \mult s + 0.5 \mult \cos(r+1.5 \mult s),\\
		Z(\vek{r},X(\vek{r}))= -0.3 + 0.5 \mult r^2 + 0.75 \mult s + \sin(r \mult s) + 0.2\mult(X-2)^2,
	\end{cases}
\end{equation*}
where $r$ and $s$ are the two-dimensional coordinates in which the original unit circle, centered at the origin of the $rs$-coordinate system, is defined. The membrane thickness is defined as $T = 0.01$ m. We consider three different situations based on the two fiber families $n_{\mathrm{fib}} = \{1,2\}$. In this example, the level sets are parallel to coordinate axis, i.e., $\mat{\Upsilon}_{1} = X$ and $\mat{\Upsilon}_{2} = Y$, respectively. The membrane is an incompressible multi-term Ogden material with $N=3$ and the fibers are governed by the incompressible Neo-Hooke material law. The material parameters, based on Example 6.6 in \cite{Holzapfel_2000a} where a spherical rubber balloon is considered, are defined as:

\begin{table}[htb!]
	\centering
	\begin{tabular}{l l}
	three-parameter Ogden material: & $(\mu_1,\mu_2,\mu_3) = (0.63, 0.0012, -0.01)$ MPa,\\
	 & $(\alpha_1,\alpha_2,\alpha_3) = (1.3, 5, -2)$,\\
	 Neo-Hooke: & $\mu = 0.4225$ MPa.
	\end{tabular}
\end{table}
The $Z$-component of the load vector $\vek{f}_{\Gamma}$ is $f_Z = -845$ kN, acting downwards in vertical direction, while the other two components and $\vek{f}_{\mat{{\Upsilon}}_i}$ are set to zero. The whole boundary is treated as a Dirichlet boundary with prescribed zero-displacements. The membrane is discretized by quadrilateral Lagrange-type elements of order $p$. Useful start vectors for the Newton-Raphson iteration may be generated using simplified hyperelastic materials such as St.~Venant--Kirchhoff solids. Fig.~\ref{fig:TC1-ResVis} shows the results for the three different types of fiber families. Figures (a) to (c) show the undeformed configuration in blue, the deformed configuration in red and some embedded fibers in black, and, the deformed configuration of the membrane without any embedded fibers in gray for comparison. Figures (d) to (f) show the Euclidean magnitude of the deformation vector and figures (g) to (d) show the von Mises stress on the deformed membrane with embedded fibers. The von Mises stress is computed by
\begin{equation}
	\sigma_M = \sqrt{\sigma_{\Gamma,11}^2 + \sigma_{\Gamma,22}^2 + \sigma_{\Gamma,33}^2 - 	\sigma_{\Gamma,11}\mult\sigma_{\Gamma,22} - \sigma_{\Gamma,11} \mult \sigma_{\Gamma,33} - \sigma_{\Gamma,22} \mult \sigma_{\Gamma,33}
	+ 3 \mult (\sigma_{\Gamma,12}^2  + \sigma_{\Gamma,13}^2  + \sigma_{\Gamma,23}^2)}, \label{eq:vonMisesStress}
\end{equation}
where $\sigma_{\Gamma,ij},\,(i,j)\in\{1,2,3\}$ are the components of the membrane's Cauchy stress tensor.\\
\\
Fig.~\ref{fig:TC1-ResConv} shows the convergence rates for the residual errors $\varepsilon_{\mathrm{res}}$ and for the stored-energy error $\varepsilon_{\mathfrak{e}}$, respectively. The reference value for the stored elastic energy is $\mathfrak{e}_{\mathrm{ref}} = 3.398467671137\cdot10^4$ J for fiber family 1, $\mathfrak{e}_{\mathrm{ref}} = 5.77652818747\cdot10^4$ J for fiber family 2, and $\mathfrak{e}_{\mathrm{ref}} = 2.963740520732\cdot10^4$ J for fiber families 1 and 2 embedded together. Convergence of the residual errors is optimal, i.e., $\mathcal{O}(p-1)$. Results for the energy error are seen in Fig.~\ref{fig:TC1-ResConv}(d) to (f) for the three different scenarios. Convergence of $\mathcal{O}(p+1)$ is expected, for even orders  $\mathcal{O}(p+2)$ was reported in literature, e.g., \cite{Fries_2023a} for further details. These convergence rates are confirmed here; only for the very high order case of $p=6$ the results appear slightly suboptimal.

\subsection{Spherical cap membrane subjected to internal pressure} \label{subsec:TC2}
\begin{figure}
	\subfigure[deformation]{\includegraphics[width=0.3\textwidth]{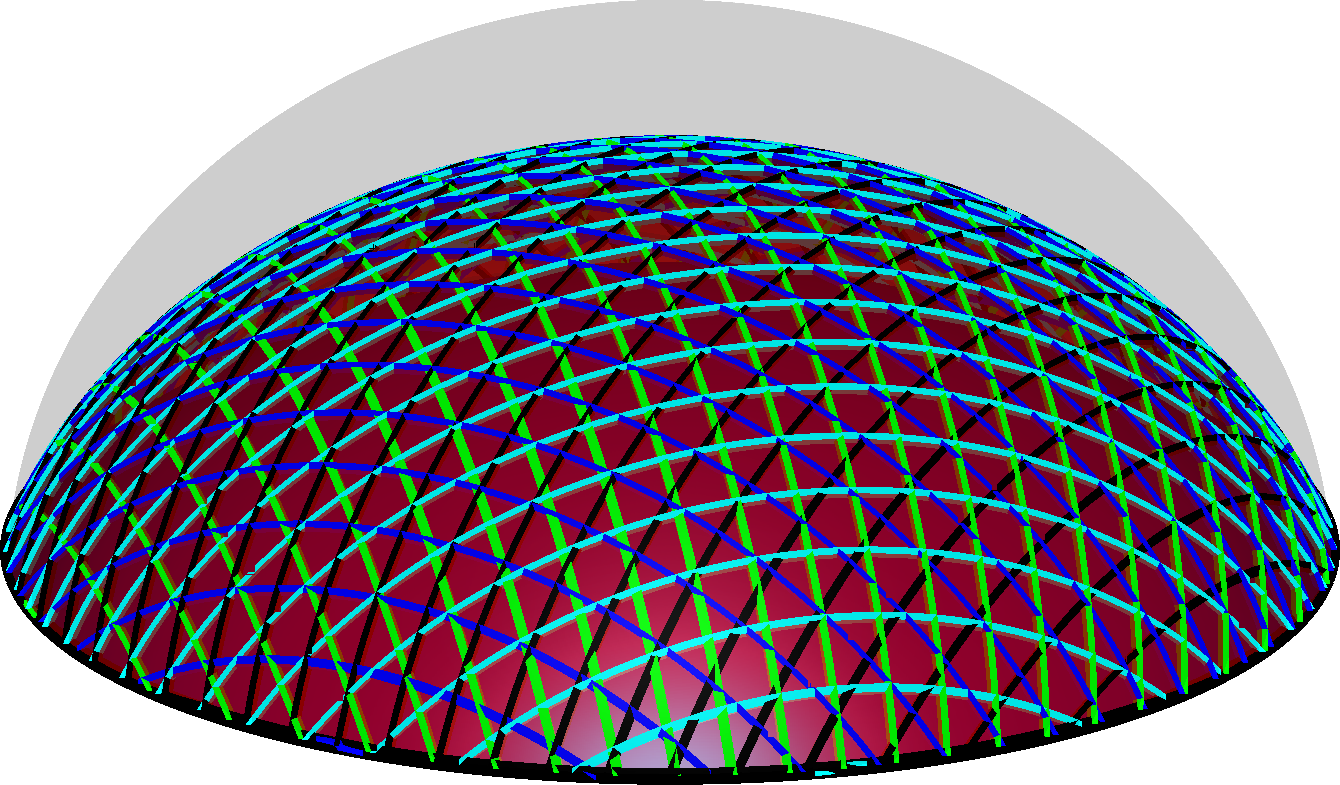}}\hfill
	\subfigure[deformation magnitude]{\includegraphics[width=0.3\textwidth]{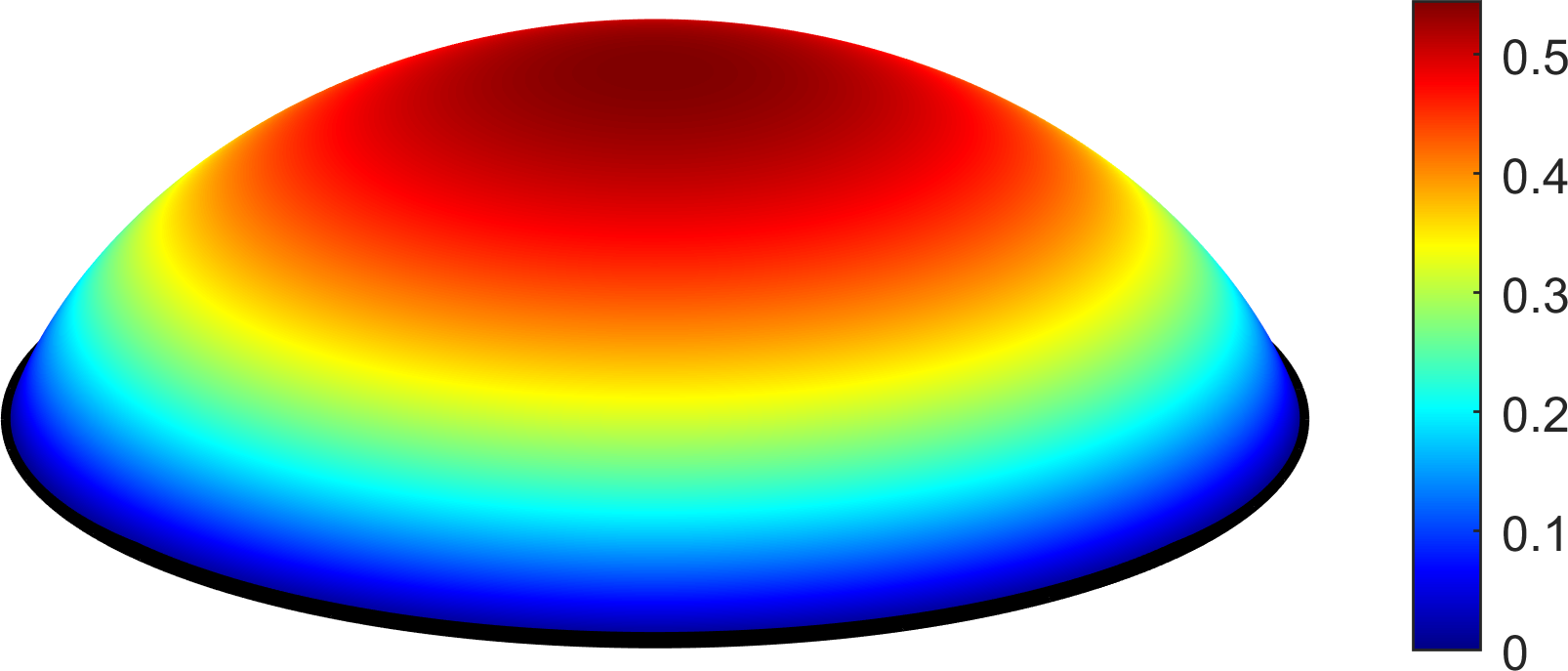}}\hfill
	\subfigure[von Mises stress]{\includegraphics[width=0.3\textwidth]{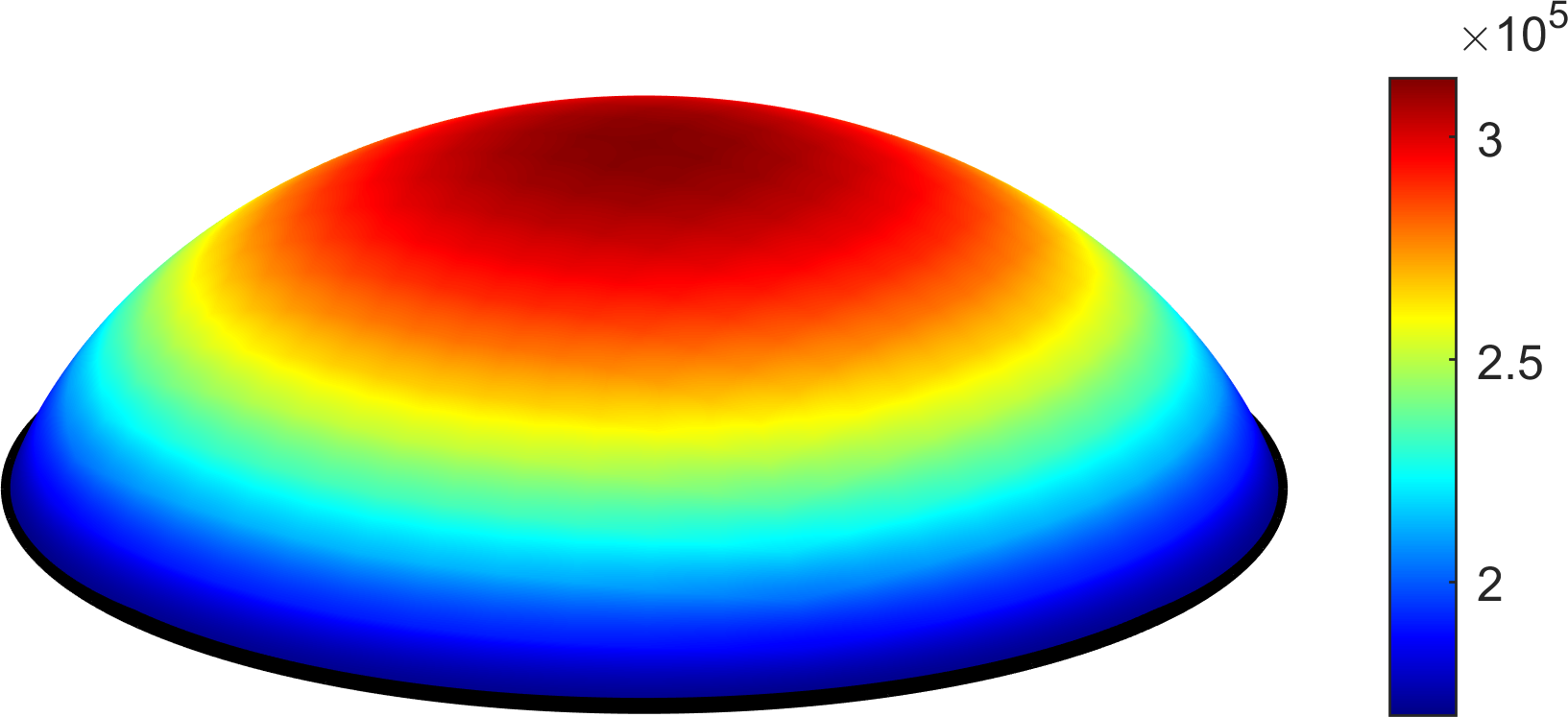}}
	\caption{Results of test case 2: (a) shows the deformed configuration in red and some arbitrarily selected fibers of the four fiber families in black, green, blue, and cyan. The gray surface in (a) is the deformed membrane without the reinforcing fibers. (b) shows the Euclidean norm of the deformation and (c) the von Mises stress.}
	\label{fig:TC2-ResVis}
\end{figure}

\begin{figure}
	\subfigure[residual error]{\includegraphics[width=0.5\textwidth]{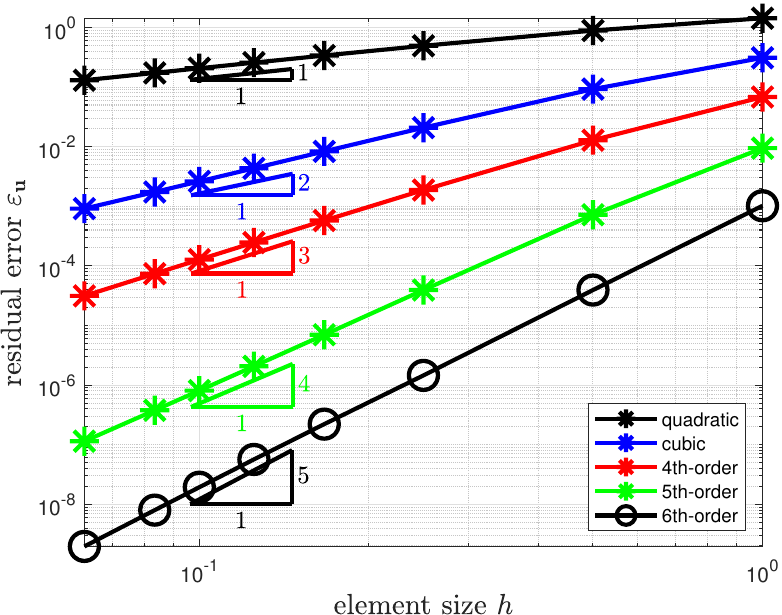}}\hfill
	\subfigure[stored-energy error]{\includegraphics[width=0.5\textwidth]{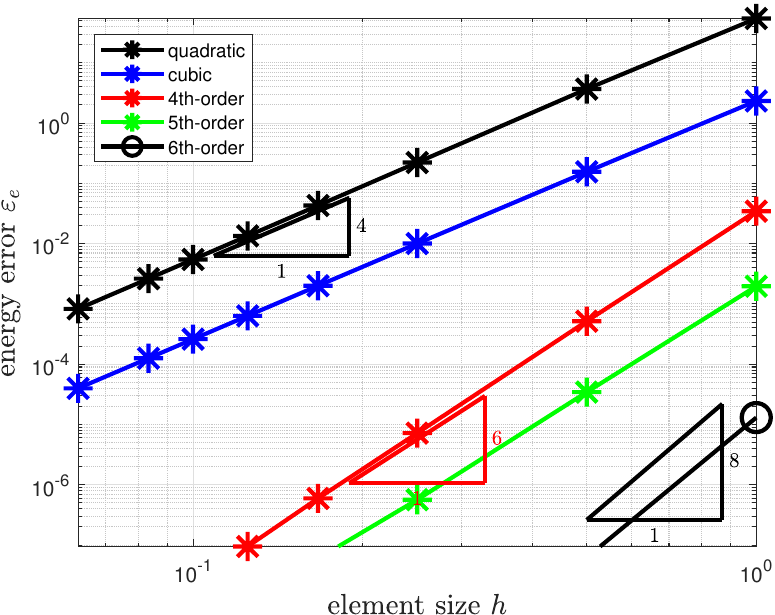}}	
	\caption{Results of test case 2: (a) shows the residual errors and (b) the stored-energy errors.}
	\label{fig:TC2-ResConv}
\end{figure}

In this test case, a spherical cap of thickness $T=0.01$ m is subjected to internal pressure. Therefore, the loading is given as
\begin{equation}
	\vek{f}_{\Gamma}(\vek{x}) = \frac{p}{T} \mult \vek{n},
\end{equation}
with the constant internal pressure $p = 3$ kPa which acts on the inner membrane surface in the deformed configuration. This introduces a new source of non-linearity in the loading which has to be considered properly in the Newton--Raphson scheme of the solver. $\vek{f}_{\mat{{\Upsilon}}_i}$ is set to zero. Again, the whole boundary is treated as a Dirichlet boundary with prescribed zero-displacements. The geometry of the membrane is the upper cap of a sphere, centered at the origin, with radius $r=4\,$m and a polar angle $\beta = 40^{\circ}$ from the $Z$-axis. The material of the membrane is modelled with the Mooney--Rivlin material law. The applied parameters are $\mu_1 = 0.240\,$MPa and $\mu_2 = -0.1825\,$MPa. There are four families of fibers embedded in the membrane defined via the level-set functions $\phi_1 = X$, $\phi_2 = Y$, $\phi_3 = X+Y$, and $\phi_4 = X-Y$. The material parameters for the fibers are the same as for the test case before in Sec.~\ref{subsec:TC1}.\\
\\
Fig.~\ref{fig:TC2-ResVis} shows visualizations of the results. In Fig.~\ref{fig:TC2-ResVis}(a), the deformed configuration is plotted as a red surface and selected fibers of the four fiber families are plotted using different colours. Furthermore, the gray surface is the deformed configuration of the isotropic membrane without fibers. The Euclidean magnitude of the deformation and the von Mises stress, evaluated by Eq.~(\ref{eq:vonMisesStress}), is plotted over the deformed configuration for the reinforced membrane in Figs.~\ref{fig:TC2-ResVis}(b) and (c), respectively. Fig.~\ref{fig:TC2-ResConv} shows the convergence rates for the residual error and the stored-energy error, respectively. The reference value for the stored elastic energy is $\mathfrak{e}_{\mathrm{ref}} = 1.130293929170\cdot10^4$ J. Optimal convergence rates are obtained for both error types.
 
\subsection{Cylindrical membrane subjected to internal pressure} \label{subsec:TC3}
In this section, we show two examples for cylindrical membranes which are subjected to internal pressure. The difference between the two examples are the boundary conditions as outlined below.
\subsection*{Example 3a: Strong enforcement of Dirichlet boundary conditions}
\begin{figure}
	\centering
	\includegraphics[width=0.6\textwidth]{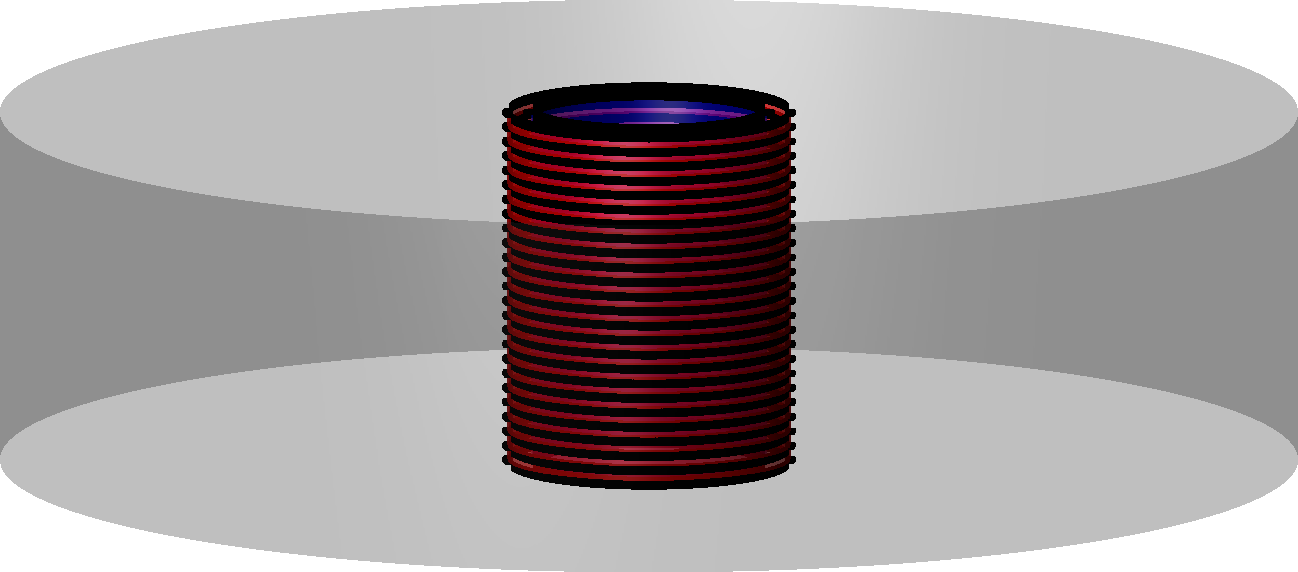}
	\caption{Results of test case 3a: The undeformed configuration in light blue, the deformed configuration in red and some arbitrarily selected fibers in black. The gray cylinder is the deformed membrane without any embedded fibers.}
	\label{fig:TC3a-ResVis}
\end{figure}
\begin{figure}
	\subfigure[residual error]{\includegraphics[width=0.5\textwidth]{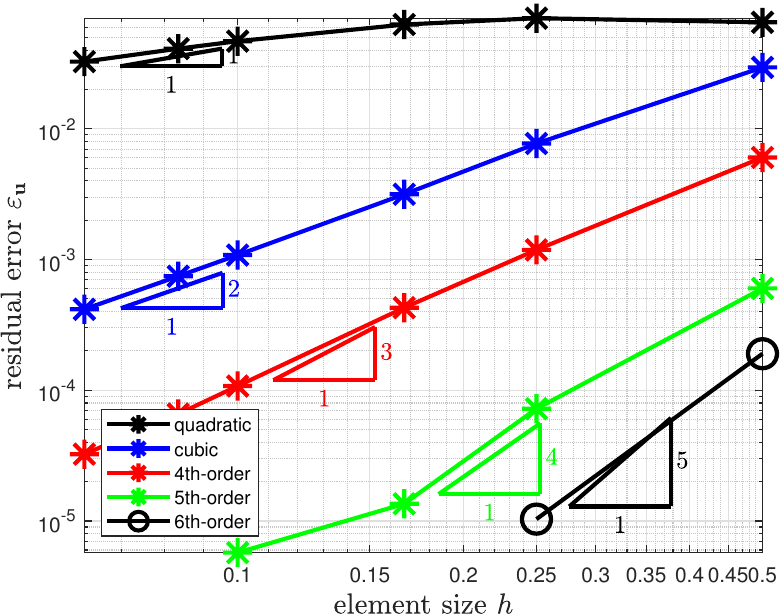}}\hfill
	\subfigure[stored-energy error]{\includegraphics[width=0.5\textwidth]{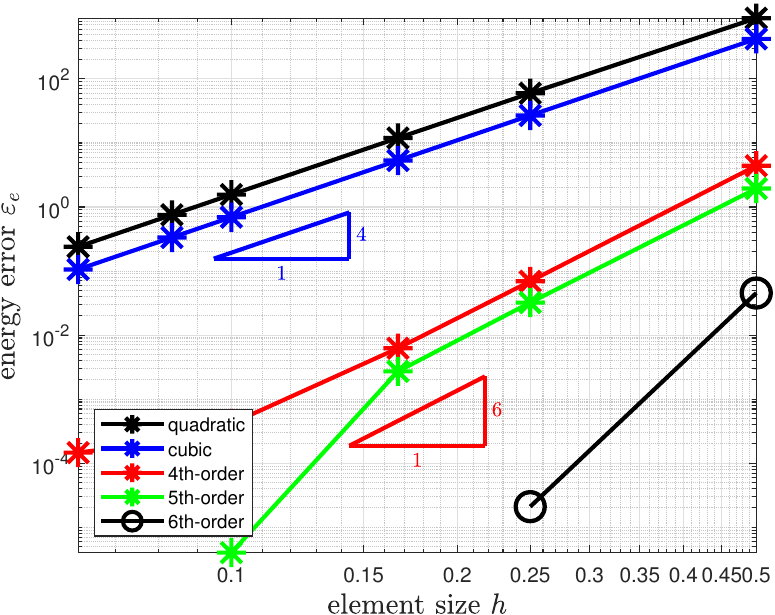}}
	\caption{Results of test case 3a: (a) the residual error and (b) the stored-energy error.}
	\label{fig:TC3a-ResConv}
\end{figure}
The axis of the cylinder is the $Z$-axis, the radius is $r=2$ m, the height of the cylinder is $h=6$ m, and the thickness of the membrane is set to unity, i.e., the tensile stiffness is governed by the material parameters. The level-set function which was used for the fiber definition is $\phi = Z$. The material parameters of the membrane and the fibers are the same as in test case 1, see Sec.~\ref{subsec:TC1}. The internal pressure is set to $p=200\,$kPa. Dirichlet boundary conditions are prescribed at the bottom and the top of the cylinder. That is, at the lower and upper nodes of the cylinder mesh, we prescribe zero-displacements in the $Z$-direction and in the circumferential direction. Fig.~\ref{fig:TC3a-ResVis} visualizes the result of this test case. The deformed configuration is shown in red and some arbitrarily selected fibers in black. The gray cylinder is the deformed membrane without any embedded fibers. Note the big difference of the deformation between an isotropic and an anisotropic, i.e., fiber-reinforced, membrane. The obtained uniform displacement in radial direction is $u_r = 0.40921\,$m for the anisotropic case.  The residual error in Fig.~\ref{fig:TC3a-ResConv}(a) and the stored-energy error in Fig.~\ref{fig:TC3a-ResConv}(b) show the expected convergence properties, hence, confirm the success of the proposed methods. The reference value for the stored elastic energy is $\mathfrak{e}_{\mathrm{ref}} = 3.993797057964$ MJ.

\subsection*{Example 3b: Weak enforcement of Dirichlet boundary conditions}
\begin{figure}
	\centering
	\subfigure[deformation]{\includegraphics[width=0.32\textwidth]{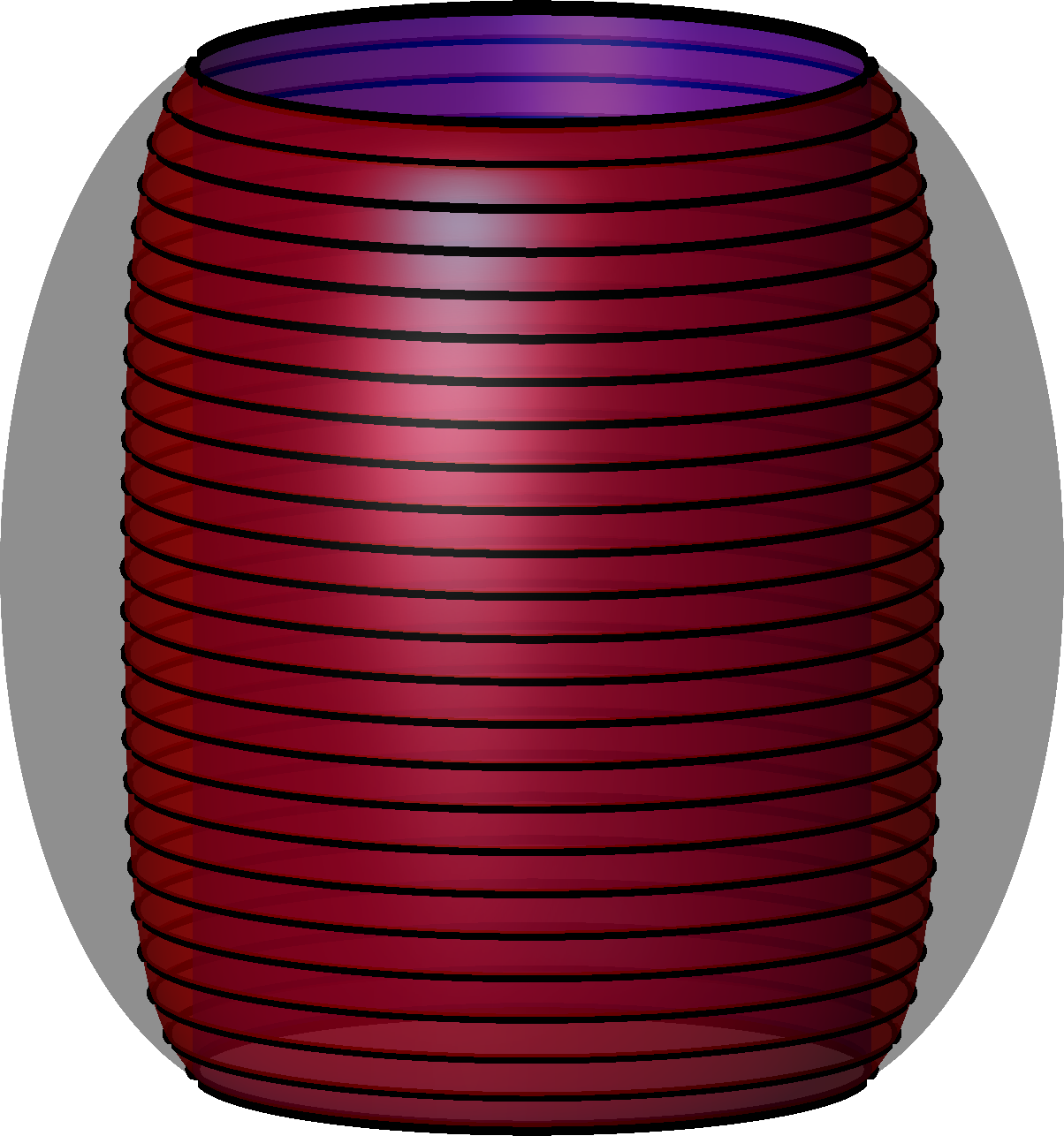}}\hfill
	\subfigure[residual error]{\includegraphics[width=0.45\textwidth]{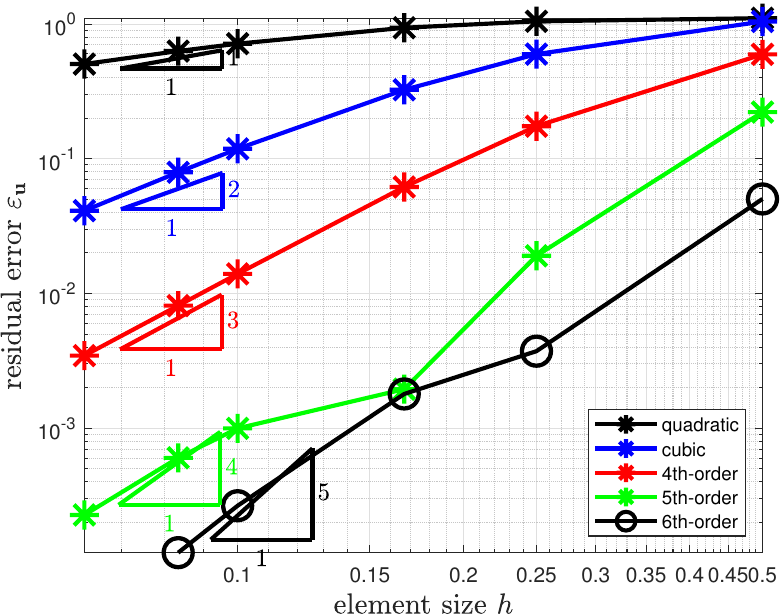}}\linebreak
	\subfigure[deformation magnitude]{\includegraphics[width=0.32\textwidth]{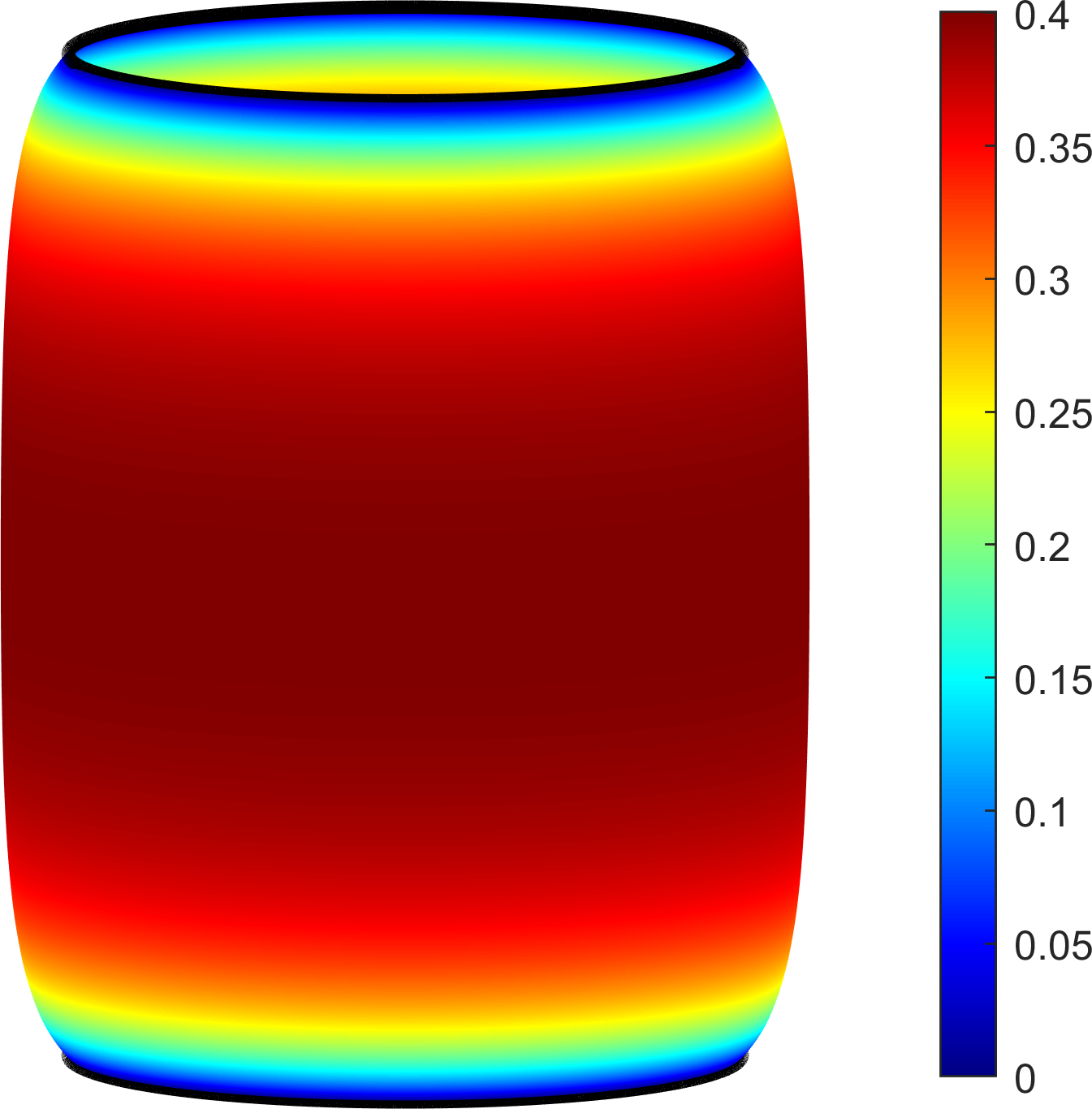}}\hfill
	\subfigure[von Mises stress]{\includegraphics[width=0.32\textwidth]{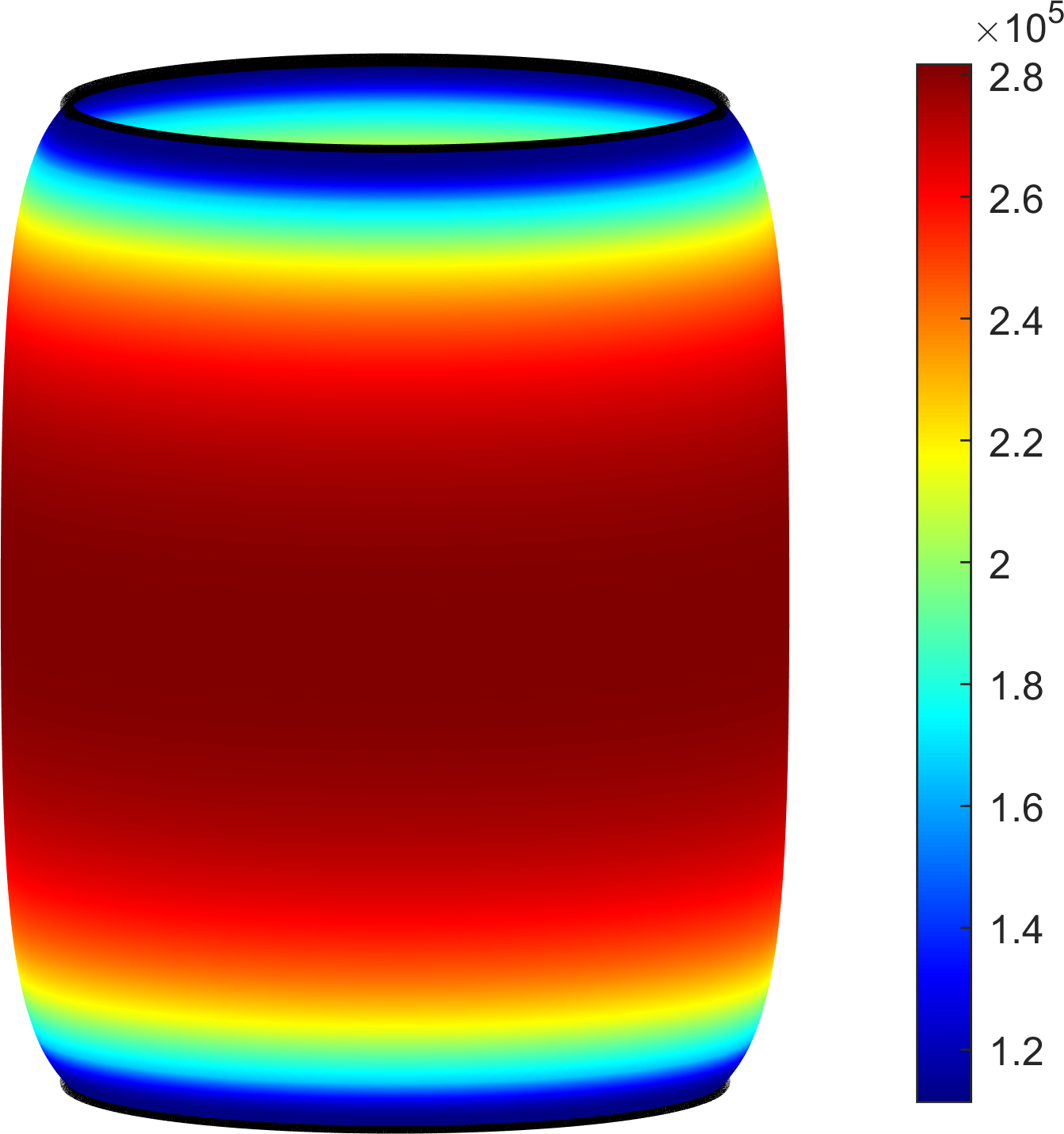}}
	\caption{Results of test case 3b: (a) shows the deformed configuration in red and some arbitrarily selected fibers in black, (b) shows the convergence in the residual error, (c) the Euclidean norm of the deformation, and (d) the von Mises stress of the membrane.}
	\label{fig:TC3b-ResVis}
\end{figure}
The overall definitions for the cylindrical membrane and the fibers, respectively, i.e, the geometry, the material parameters, and the loading, are the same as in Example 3a above. The difference to the former test case is that at all boundary nodes, all-zero displacements are prescribed. We use Lagrange multipliers to weakly enforce zero-components of the displacements in normal, tangential, and co-normal directions (w.r.t.~the deformed configuration). Let us introduce the Jacobi matrix from the Newton--Raphson iteration scheme as $\bar{\mat{K}}$, then this matrix is modified to $\begin{bmatrix}
	\bar{\mat{K}} & \mat{C}^{\mathrm{T}} \\ \mat{C} & \mat{0}
\end{bmatrix}$, with the constraint matrix $\mat{C}$ resulting from the Lagrange multiplier approach; see \cite{Kaiser_2023a} for the application of Lagrange multipliers for Dirichlet boundary conditions of curved beams formulated using TDC, which is similar to this work. Fig.~\ref{fig:TC3b-ResVis} shows the results for this example, i.e, the deformation, the convergence in the residual error, the deformation magnitude, and the von Mises stress, evaluated by Eq.~(\ref{eq:vonMisesStress}). 

\subsection{Comparison to discrete fiber embeddings} \label{subsec:TC4}
\begin{figure}
	\centering
	\includegraphics[width=0.4\textwidth]{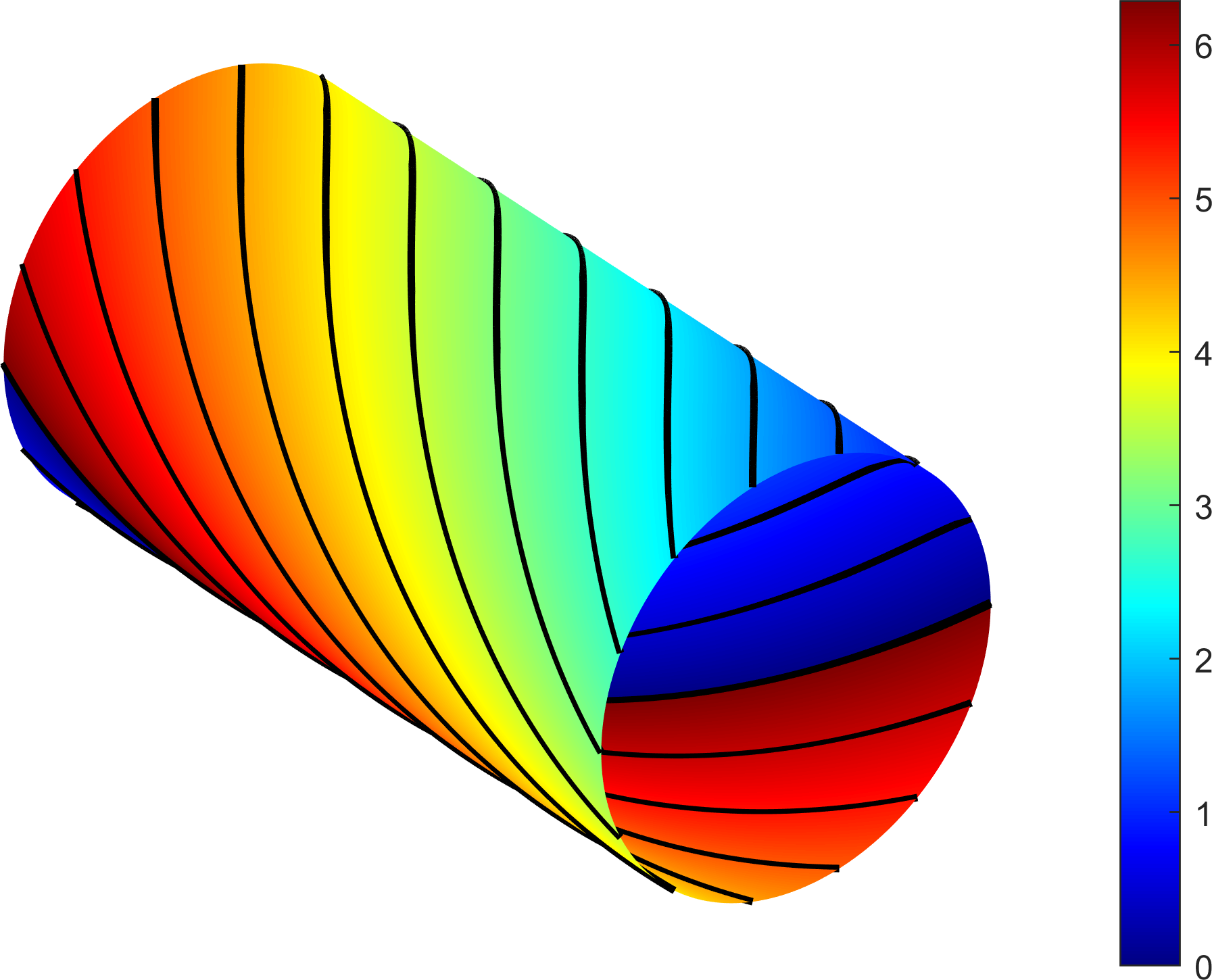}
	\caption{The geometry of the cylinder and the embedded fibers. The color indicates the level-set values.}
	\label{fig:TC4-GeomVis}
\end{figure}

\begin{figure}
	\subfigure[deformation]{\includegraphics[width=0.5\textwidth]{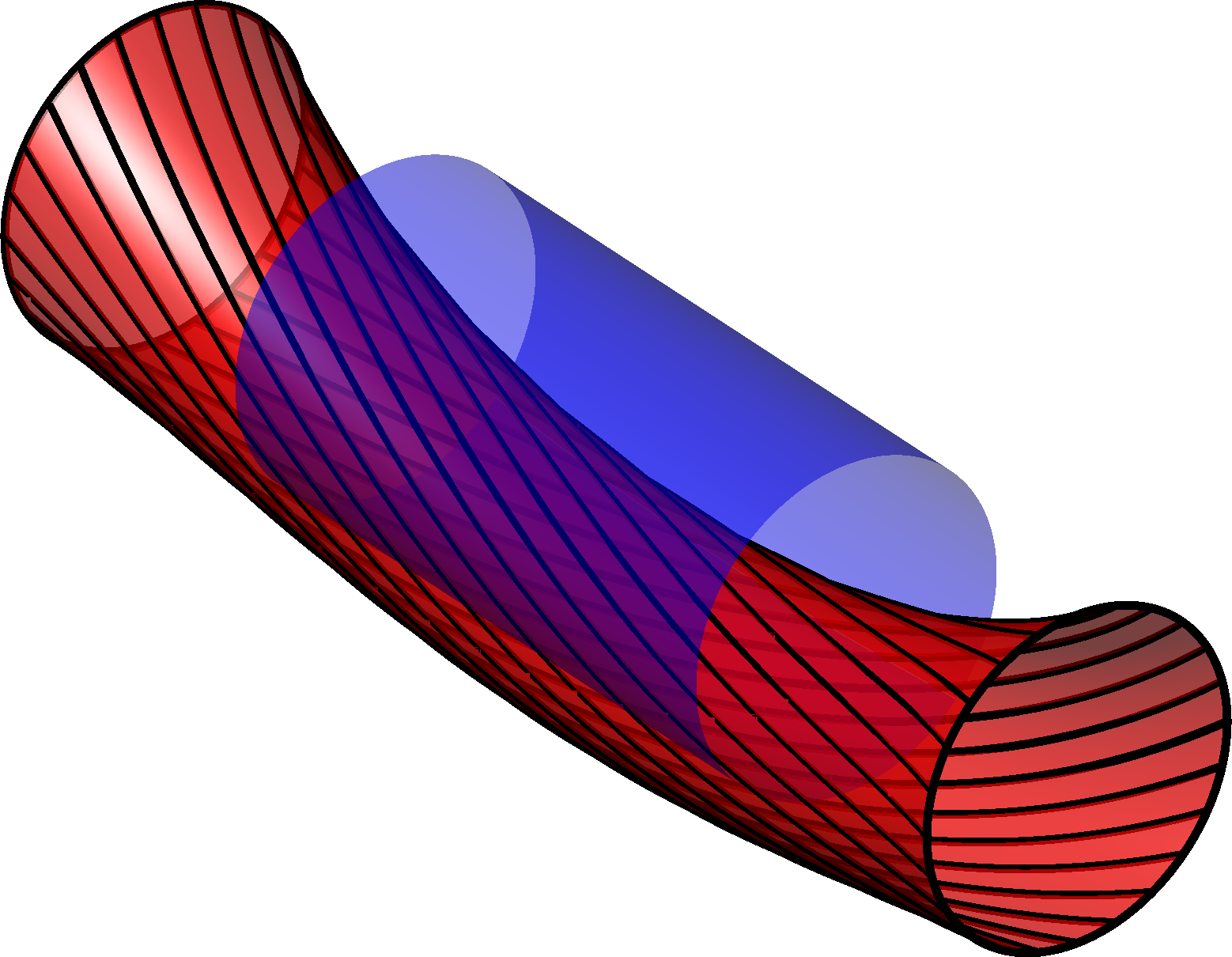}}\hfill
	\subfigure[von Mises stress]{\includegraphics[width=0.5\textwidth]{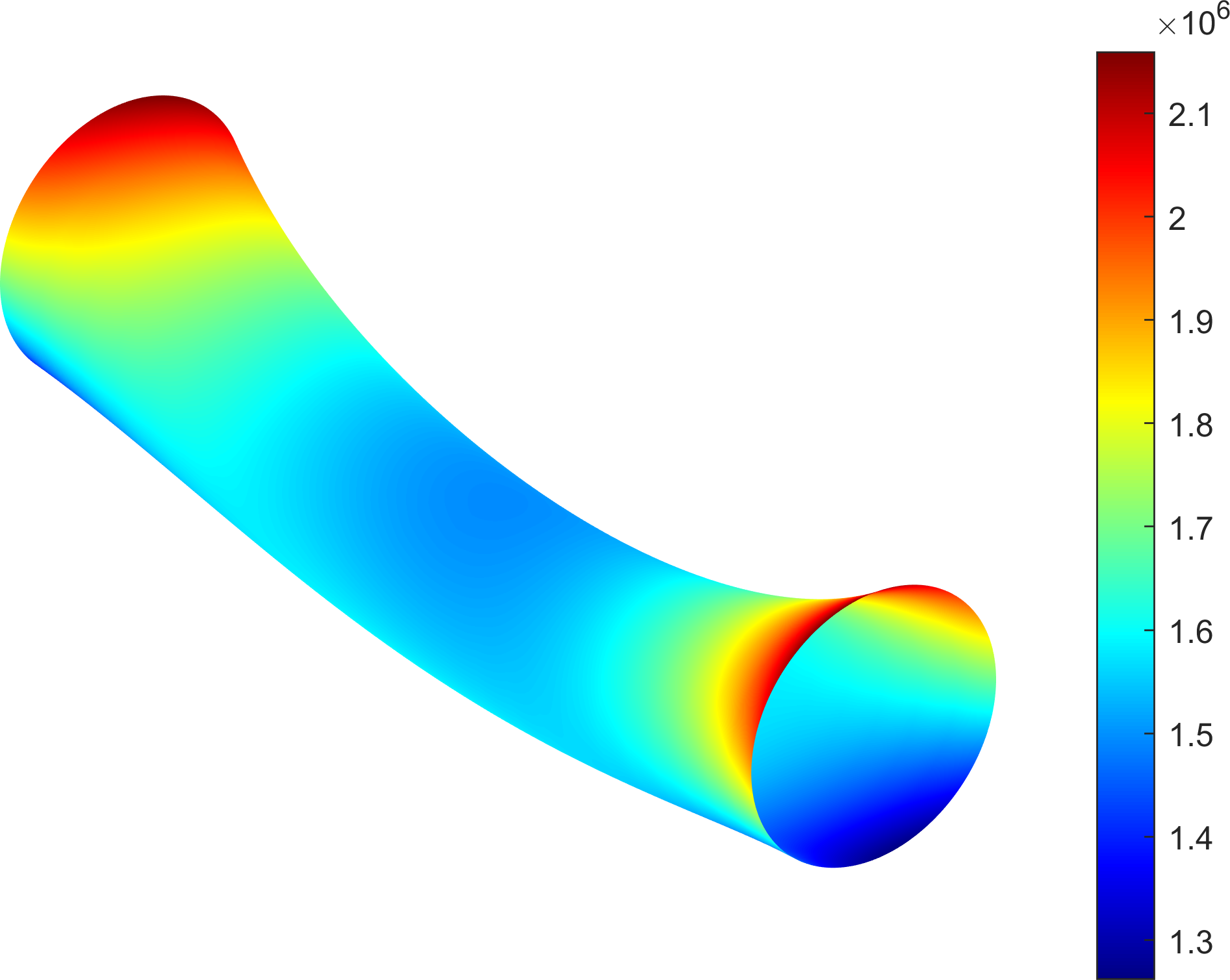}}
	\caption{(a) shows the deformed configuration in red together with some arbitrarily selected fibers in black. The blue surface is the undeformed configuration. (b) depicts the von Mises stress plotted on the deformed configuration.}
	\label{fig:TC4-ResVis}
\end{figure}

\begin{figure}
	\centering
	\subfigure[Cylinder with sprialing fibers]{\includegraphics[width=0.4\textwidth]{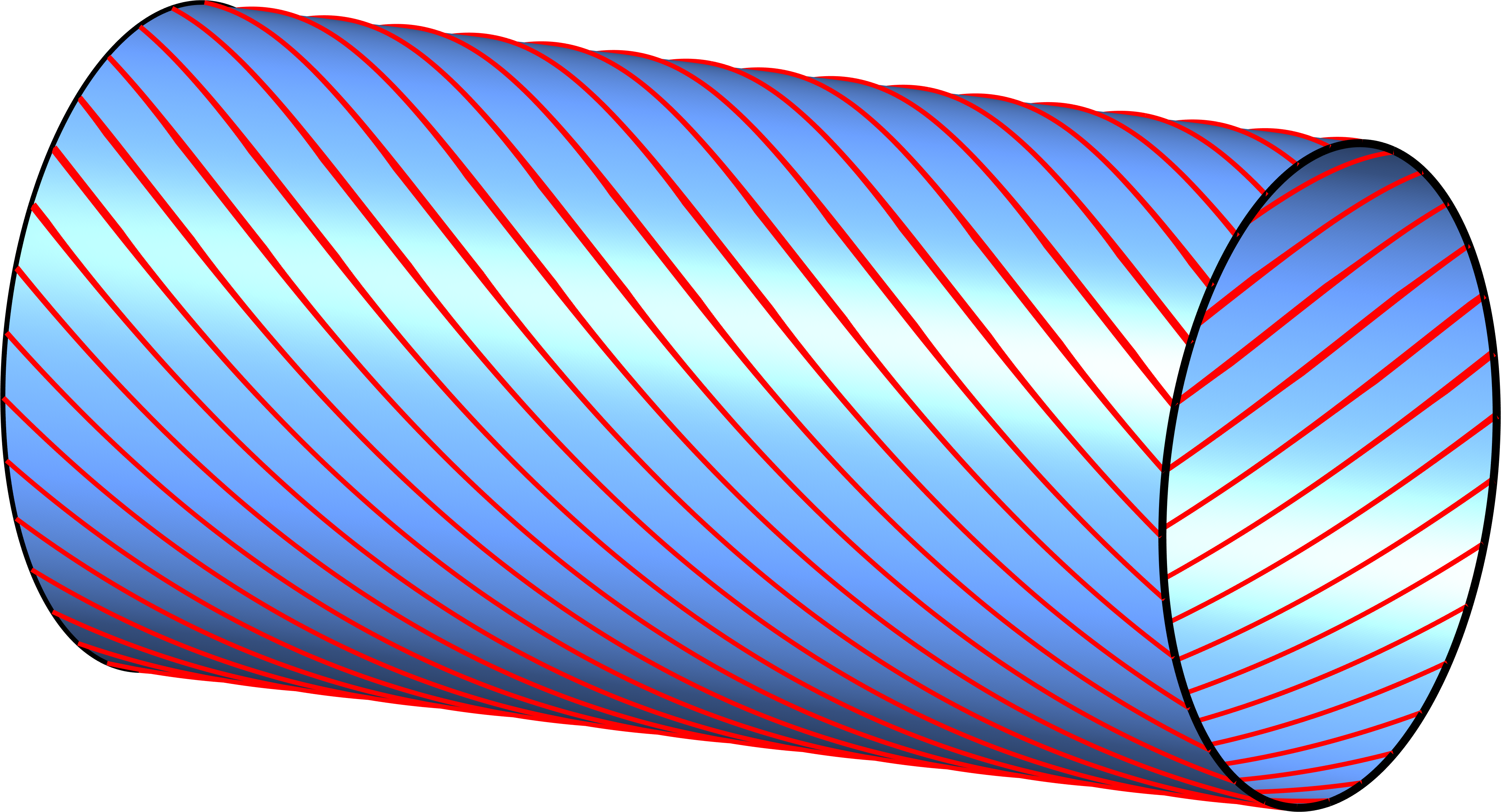}}\hfill
	\subfigure[Surface FEM discretization]{\includegraphics[width=0.4\textwidth]{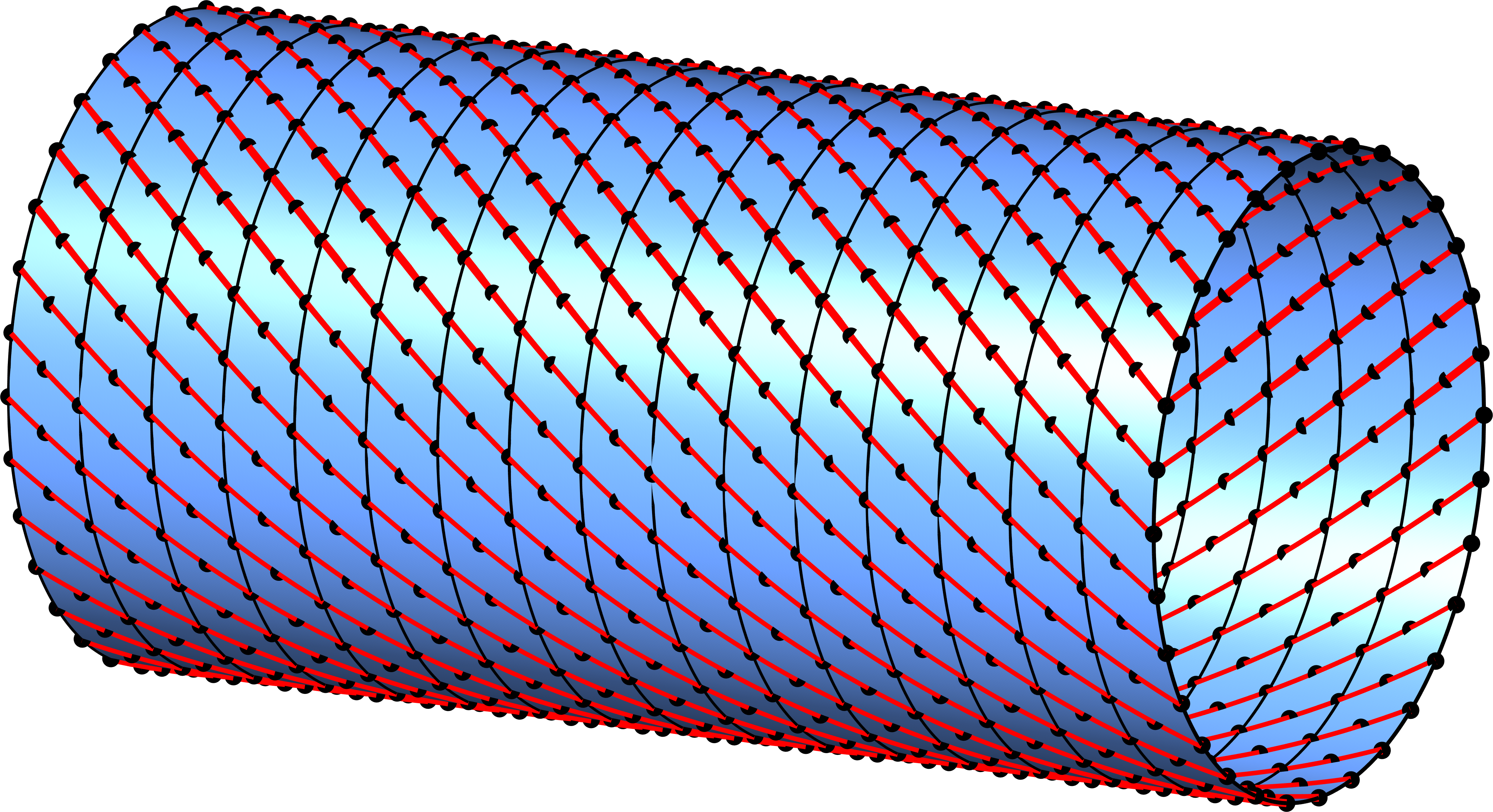}}
	\caption{Figure (a) shows the cylindrical surface in blue and some arbitrarily selected fibers in red. The disretized surface and discretized fibers as employed in the Surface FEM are shown in Figure (b).}
	\label{fig:SpiralSrfMesh}
\end{figure}

\begin{figure}
	\subfigure[residual error (BTF)]{\includegraphics[width=0.3\textwidth]{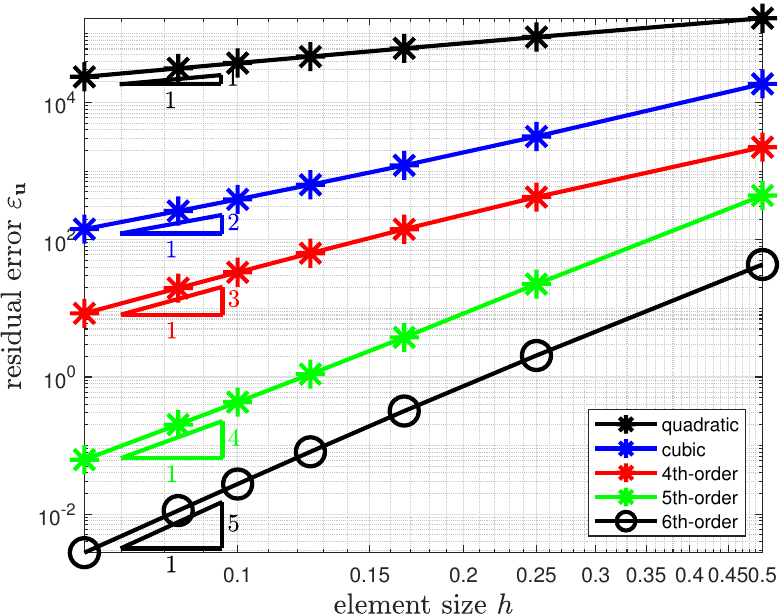}}\hfill
	\subfigure[stored-energy error (BTF)]{\includegraphics[width=0.3\textwidth]{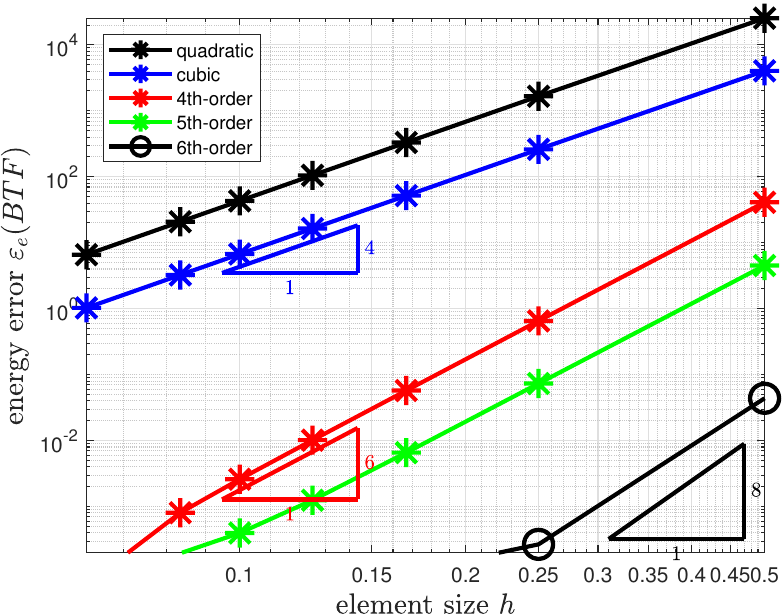}}\hfill
	\subfigure[stored-energy error (SRF)]{\includegraphics[width=0.3\textwidth]{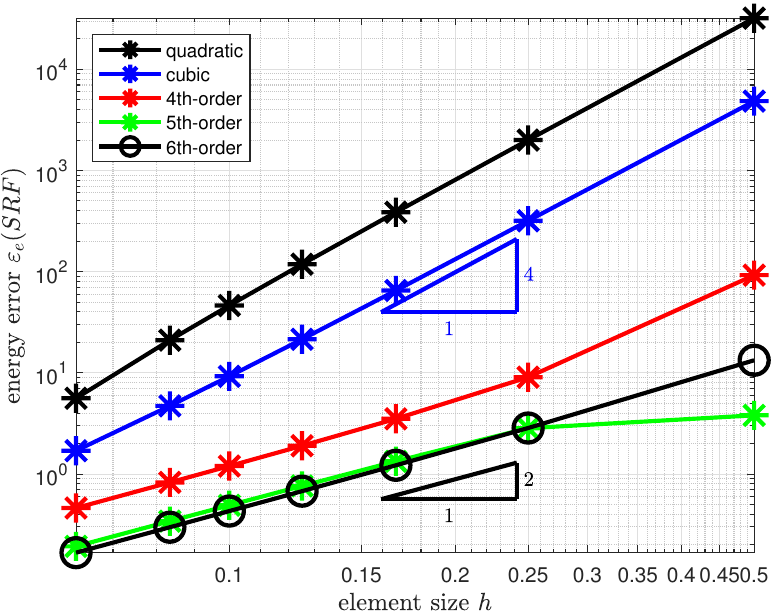}}
	\caption{Figure (a) shows the convergence results of the residual error and figure (b) the stored-energy error of the continuous embedding approach, respectively. Figure (c) shows the stored-energy error of the discrete approach.}
	\label{fig:TC4-ResConv}
\end{figure}

A cylindrical membrane is reinforced by spiraling fibers as shown in Fig.~\ref{fig:TC4-GeomVis}(a). This test case is used to compare the proposed method with an approach where the fibers are discretely described. At first, we discuss the new method (Bulk Trace FEM - BTF) and then, we compare this to the Surface FEM (SRF) with discrete fibers.\\
\\
The cylinder has a radius of $r=2.5\,$m, a length of $l = 10\,$m, parallel to the $Y$-axis (the cylinder's cross-section center points are located at $[2.5,Y,0]^{\mathrm{T}}$), and the thickness of the membrane is set to unity. Spiraling fibers are defined on the cylinder's surface. The scalar-valued function 
\begin{equation*}
	\phi = \theta_0 + Y\bigg(\frac{\\
		\pi\\
	}{l}\bigg),
\end{equation*} 
is associated with the fibers as shown in Fig.~\ref{fig:TC4-GeomVis}. Therein, $\theta_0$ is the angle of the polar coordinates for the cylinder's cross section, stored at the nodes of the employed meshes. Note that a jump in the scalar-valued function occurs when $\theta_0 = 2\pi$ changes to $\theta_0 =0$, i.e., when the circle is closed. Therefore, the correct value has to be assigned properly to each element which is adjacent to the spiral at which the jump occurs. In Fig.~\ref{fig:SpiralSrfMesh}(a), the cylindrical membrane is shown in blue and the fibers are shown in red. The incompressible Neo-Hooke material is used with $\mu_1 = 0.4225\,$MPa for the membrane and the fibers, respectively. The cylinder is loaded with a body force of $\vek{f}_{\Gamma} = [0,0,-845]^{\mathrm{T}}\,$kN and the Dirichlet boundary conditions are set to zero for the $X$- and $Z$-direction, respectively, but the cylinder surface is extended in $Y$-direction by $5\,$m at both sides in opposite directions.
Fig.~\ref{fig:TC4-ResVis} shows the undeformed cylindrical membrane in blue and the deformed one in red with some fibers plotted on the deformed surface.
\\
\\
This test case can now be used to compare it with a classical approach, i.e., where the fibers are considered as discrete structures in the finite element method using the Surface FEM (SRF); see \cite{Fries_2023a}. Using a special mesh design, and the level-sets used, the edges of the mesh in the direction of the cylinder's length are now aligned with the fibers. Furthermore, also the nodes on the other edges which are not aligned with the fibers and in the interior of the elements are located along a specific spiral fiber (i.e., as if these nodes would be connected by a `virtual' edge), see Fig.~\ref{fig:SpiralSrfMesh}(b). With that property, the fibers can be introduced as discrete structures along these (real and `virtual') edges of the surface mesh, which is used to discretize the cylindrical membrane.
For the finite element analysis, one has to make sure that the fibre-reinforced approach does not get stiffer upon increasing the number of elements and/or nodes, i.e., increasing the number of considered fibers. Therefore, an area $A$ is assigned to the fibers initially. This area is scaled with the number of fibers and the values of the level-set function as $A_{\mathrm{discr}} = \delta_f \mult A$ with $\delta_f = (\phi_{\mathrm{max}}-\phi_{\mathrm{min}}) / (n-1)$, where $n$ is the number of nodes in each circular cross section of the cylinder which is normal to the cylinder's length axis. Thereby, the stiffness of each fiber can be scaled properly and the results are comparable with the continuous case of embedded fibres from above.\\
\\
Fig.~\ref{fig:TC4-ResConv} shows convergence results of both approaches. Figs.~\ref{fig:TC4-ResConv}(a) and (b), which show the convergence results of the residual error and the stored-energy error for the newly introduced approach with continuously embedded fibers, indicate optimal convergence behaviour. The plot which shows the stored-energy error for the case with the \emph{discrete} fibers in Fig.~\ref{fig:TC4-ResConv}(c), also converges in the reference energy, however, those convergence rates are bounded by 2. This is not surprising because the geometry of the discrete fibers is considered conceptually different to the homogeneously embedded fibers being the major focus of this work. This is similar to a previous study in \cite{Fries_2023a}, where discrete and continuous fiber-reinforcements in a planar quarter annulus where studied and compared to each other with similar findings. The reference value for the stored elastic energy, obtained with the continuous embedding model, is $\mathfrak{e}_{\mathrm{ref}} = 93.70356208059$ MJ.

\section{Conclusions and outlook}\label{sec:CaO}
A mechanical model for hyperelastic, anisotropic, curved, two-dimensional membranes is introduced. The source for the anisotropy in the membranes, which act as a matrix or bulk material, are continuously embedded fibers. These one-dimensional fibers are geometrically modelled as the intersection curve of the membrane's midsurface and level sets of a scalar function defined in $\mathbb{R}^3$. With this approach, the fibers may be identified as individual level sets whereas in most standard models, the geometries of fibers are introduced by vector fields on/in the matrix material and the material characteristics of fibers are considered by modified strain energy functions. Herein, fiber families have their own strain energy functions, being independent of the membrane, i.e., the matrix material.\\
\\
Furthermore, a finite element method is formulated for the numerical analysis of the proposed mechanical model. The numerical approach is based on previous works of the authors for the simultaneous analysis of PDEs on manifolds and was labelled Bulk Trace FEM in \cite{Fries_2023a}. Herein for the first time, the Bulk Trace FEM is applied to a bulk domain which is a two-dimensional manifold itself (the membrane) into which further one-dimensional manifolds (the fibers) are continuously embedded. Therefore, the used differential operators depend on the curved geometry of both considered manifolds. For smooth physical fields, higher-order convergence rates in the residual error and the stored-energy error are achieved in the numerical results.\\
\\
The proposed mechanical model and corresponding numerical method provide important fundamentals for continuously embedded sub-structure models in the context of curved geometries, being the major goal of this paper. In future research, the model will be employed in real-world applications such as in the mechanical modelling of biological tissues, e.g., artery walls, or woven textiles and paper materials. Further research topics are how standard features of classical models for anisotropic materials, e.g., the sliding of fibers in the matrix material, the consideration of the tension-compression switch, and \emph{dispersed} fiber modeling, can be included in the proposed model. Furthermore, especially for biological materials, an extension to (geometrically nonlinear) \emph{shells} (rather than membranes) as matrix materials, probably even with embedded beams, may be highly interesting and suggested towards realistic simulations of biological tissues.


\section*{Acknowledgements}
This research was funded in part by the Austrian Science Fund (FWF) 10.55776/PAT6420824.
\printbibliography	
 
\end{document}